\PassOptionsToPackage{pdfpagelabels=false}{hyperref}   % Fix warning pdfpagelabels is turned off
\expandafter\def\csname ver@fixltx2e.sty\endcsname{}   % Fix warning fixltx2e is not required

\documentclass[fleqn,usenatbib]{mnras}

\usepackage{graphicx}	% Including figure files
\usepackage{amsmath, nccmath}	% Advanced maths commands
\usepackage{amssymb}	% Extra maths symbols
\usepackage{siunitx}
\usepackage{makecell}
\usepackage{accents}
\usepackage{empheq}
\usepackage{upgreek}

\renewcommand{\vec}[1]{\underaccent{\tilde}{#1}}
\defcitealias{deVaucouleurs1976SecondGalaxies.}{RC2}

\usepackage{fancyhdr}
\numberwithin{equation}{section}

\DeclareSIUnit\angstrom{\text{Å}}

\usepackage{txfonts}

\usepackage[T1]{fontenc}
\usepackage{ae,aecompl}

\title[Evolution of Stars and Gas in Galaxies]{Evolution of Stars and Gas in Galaxies}

\author[Beatrice M. Tinsley]{
Beatrice M. Tinsley\\
Yale University Observatory\\
}

\date{First published 1980, in Fundamentals of Cosmic Physics, Volume 5, pp. 287--388}

\pubyear{1980}

\begin{document}
\label{firstpage}
\pagerange{\pageref{firstpage}--\pageref{lastpage}}
\maketitle

% Don't make up new ones

%%%%%%%%%%%%%%%%%%%%%%%%%%%%%%%%%%%%%%%%%%%%%%%%%%

%%%%%%%%%%%%%%%%% BODY OF PAPER %%%%%%%%%%%%%%%%%%

\thispagestyle{plain}

\section{Overview}
\label{sec:Overview}

Essentially everything of astronomical interest is either part of a galaxy, or from a galaxy, or otherwise relevant to the origin or evolution of galaxies. Diverse examples are that the isotopic composition of meteorites provides clues to the history of star formation billions of years ago, and cosmological tests for the deceleration of the Universe are strongly affected by changes in the luminosities of galaxies during the lookback time sampled. The aim of this article is to review some of the vital connections that galaxy evolution makes among many astronomical phenomena.

The evolution of galaxies can be broadly divided into three areas: dynamical evolution, chemical evolution, and the evolution of photometric properties. Chemical and photometric evolution are the main topics of this review, and dynamical processes will be considered only where they directly affect the others. The article is intended to be self-contained for readers with a general knowledge of astronomy and astrophysics but with no special expertise on galaxies. Because the background and literature relevant to the evolution of stars and gas in galaxies are so extensive, complete coverage cannot be attempted. Instead, comprehensive reviews and recent papers are emphasized in the references; these include detailed works in closely related fields, such as galaxy formation and nucleosynthesis, that are treated very briefly here. The history of the subject has been discussed in several recent reviews \citep[e.g.][]{Sandage1975GalaxiesUniverse, Spinrad1975TheSpectra, Trimble1975TheElements, vandenBergh1975StellarGalaxies}, so here the emphasis will be on current ideas.

\medskip

This first Section gives an overview of relevant properties of galaxies and theoretical ideas on how they formed and evolved, followed by an outline of the rest of the article.

\subsection{Stellar Populations and Chemical Composition of Galaxies}
\label{subsec:Stellar_Populations}

Galaxies are made of stars and interstellar matter (ISM), with a variety of properties that correlate remarkably with the forms of galaxies as they appear on the sky. Introductory references to supplement the following brief outline include the classic lectures of \citet{Baade1963EvolutionGalaxies} and subsequent reviews by \citet{Morgan1969OnGalaxies}, \citet{King1971StellarGalaxies, King1977Introduction:Day}, \citet{vandenBergh1975StellarGalaxies}, and \citet{Sandage1975GalaxiesUniverse}. The \emph{Hubble Atlas of Galaxies} \citep{Sandage1961c} is an essential source book for photographs and descriptions of typical galaxies, while the \emph{Atlas of Peculiar Galaxies} \citep{Arp1966AtlasGalaxies} illustrates a great variety of unusual forms.

\medskip

Most galaxies can be arranged in a natural sequence of morphological types, the Hubble sequence. The ``earliest'' are elliptical (E) galaxies, with apparently smooth distributions of yellow--red stars and seldom any signs of internal structure (except that individual stars and globular clusters can be resolved in the closest ones); and the ``latest'' are irregulars (Irr I) with no symmetry and with disorganized patches of blue stars, hot gas, and dust throughout\footnote{By a historical misfortune, the early-type galaxies are dominated by late-type (red) stars, and the late-type galaxies are dominated by early-type (blue) stars!}. In between are the spirals, in order from Sa types with large nuclear bulges and tightly wound arms to Sc types with small bulges and open arms; the nuclear bulges closely resemble elliptical galaxies, while spiral arms, or more often bits of spiral arms, are traced by patchy blue stars and gas and by dark dust lanes, superposed on an underlying smoother disk of redder stars. The galaxies classified as S0 have disks but lack spiral structure and its associated young stars and gas; Hubble put them between types E and Sa, but \citet{vandenBergh1976a} pictures them as a sequence parallel to the spirals, ordered by bulge-to-disk ratio. Another ``dimension'' of the Hubble sequence is based on the presence or absence of a bar in the center, from the ends of which the spiral arms seem to emerge; there are barred spirals of types SBa with tight arms through SBc with loosely wound arms. The Hubble sequence has been elaborated by \citet{deVaucouleurs1959ClassificationGalaxies} into a more detailed system indicating intermediate forms and additional structural features.

Although the sequence E--S0--Sa--Sb--Sc--Irr I was defined by the shapes of galaxies, it is very much a sequence of stellar populations. The earliest galaxies show no signs of hot young stars and (usually) have undetectably little gas or dust, while the latest galaxies are undergoing active star formation and have significant amounts of ISM. The Yerkes classification system of galaxies \citep{Morgan1957AGalaxies, Morgan1969OnGalaxies} emphasizes the correlation between forms and stellar populations of galaxies. For spirals, this system is based simply on the central concentration of light, and it is found that the highly concentrated galaxies have late-type nuclear spectra (indicating a mixture of G, K, and M stars), while those with little central concentration have early-type spectra (dominated by B stars) even in the nucleus. The Yerkes types of galaxies are correlated loosely with Hubble types, which is not surprising since the Hubble classification is based partly on the amount of light concentrated into a nuclear bulge and partly on the tightness of spiral structure. The increasing prevalence of blue, young stars as one goes from elliptical through irregular galaxies is reflected in a general progression from redder to bluer integrated colors.

From studies of the integrated spectra and colors of galaxies, it is concluded that nearly all galaxies have some very old stars, and that their different proportions of red and blue stars in the light are due mainly to different current rates of star formation. Most (if not all) galaxies appear to be many billions of years old, but whereas significant star formation ceased billions of years ago in ellipticals, it has occurred at a roughly constant rate in the latest types of galaxies. Theories of galaxy formation, and of star formation within galaxies, are challenged to explain this close correlation between the forms of galaxies and their histories of star formation.

Some galaxies fit nowhere into a tidy sequence of types, and they are classified as simply ``peculiar'' or as one of the special types of peculiars such as ring galaxies or Irr~II (which are early-type galaxies in form but strewn with dust, young stars etc.). Many of the peculiar systems are interacting galaxies, distorted by tidal effects \citep{Toomre1972GalacticTails}. Sometimes the colors of peculiar galaxies are so blue that any underlying old population is completely outshone by newly-formed stars, and sometimes they are so red that most of the activity must be hidden in dust clouds. Although these very peculiar galaxies are rare, they are of special interest since some of them may be undergoing the kinds of violent changes and rapid star formation that characterized normal galaxies in their youth \citep{Larson1976c}.

\medskip

The chemical compositions of galaxies in general show much less diversity than their stellar populations. Few instances are known of interstellar gas that is deficient or overabundant in heavy elements by more than a factor of four relative to the Sun, and although stellar metallicities range over more than two orders of magnitude, they are usually within a factor of three of Solar\footnote{The term ``metals'' in this field generally includes all elements heavier than helium. True metals, especially iron, are most easily detected in the spectra of stars, while the common non-metallic elements, especially oxygen, are most accessible in the ISM. Another loose usage in this field is that the word ``element'' frequently means a particular nuclide (e.g., the ``element'' $^{13}$C).}. Many of the differences among chemical compositions in galaxies are systematic: metallicities of both stars and gas tend to decrease outward from the centers of galaxies, and average metallicities tend to increase with galaxy luminosity.

Models for chemical evolution of galaxies aim to account for their compositions in terms of the production of elements by stars (mainly) and the mixing of stellar ejecta with interstellar gas. Gas flows often play important roles in chemical evolution, diluting the products of nucleosynthesis with unenriched material from outside the galaxy, and carrying metals from one part of the galaxy to another.

Many more details are known about the stellar population in the Solar vicinity of our own galaxy than anywhere else, and some of these details contain clues to large-scale processes of galaxy formation. For example, stars with high space velocities that make them members of the halo\footnote{The word ``halo'' is used in the astronomical literature in two ways: (i) its classical meaning, a spheroidal population of ordinary stars, typified by globular clusters; and (ii) an invisible spheroidal component that provides much more galactic mass than can be ascribed to known stars and gas. The first meaning is implied throughout this article, unless otherwise stated.} population are metal-poor by factors of ten or more relative to the Sun, while stars with disk motions have metallicities almost entirely within a factor of three of Solar. Since the classic paper of \citet{Eggen1962EvidenceCollapsed.}, this difference has been interpreted as evidence that the halo stars formed first, before much chemical enrichment by deaths of massive stars had taken place. Those authors also used the kinematics of halo stars, and a model for the collapse of the Galaxy from particular initial conditions, to infer a free-fall timescale, ${\sim} 2 \times 10^{8} \: \rm yr$, for the collapse and formation of halo stars. Ages that have recently been determined for halo and disk stars now point toward a slower collapse, however; globular clusters in the Galaxy appear to have a range of ages from ${\sim} 12 - 16 \: \rm Gyr$, while very few disk stars near the Sun are older than $10 \: \rm Gyr$ \citep{Demarque1977StellarGalaxy, Saio1977AgesII, Saio1977OnStars, Twarog1980AnNeighborhood}. The picture presented by these values is of a rather slow collapse, and a very long timescale for star formation to get underway in the outer disk of the Galaxy. A very young star cluster at least $10 \: \rm kpc$ from the Sun in the Galactic anticenter direction has recently been assigned a metallicity as low as those of halo stars \citep{Christian1979UBV21}, so it seems that the outermost disk is still in a very early stage of chemical evolution.

\medskip

In general, composition differences among the old stars of galaxies serve as frozen-in tracers of their early chemical evolution, while abundances in young stars and the ISM indicate the progress of continuing enrichment. These effects are closely tied to the history of star formation in various regions of galaxies, which relates them to factors determining photometric properties. It will be seen later in this review that the aspects of galactic evolution of greatest importance for photometric evolution and for chemical evolution, respectively, are to a large extent complementary, so the two fields of study tend to yield different types of information about the evolution of galaxies.

\subsection{Galaxy Formation and Evolution}
\label{subsec:Galaxy_Formation}

The striking correlations between the forms and contents of galaxies, and systematic changes of content with position in galaxies, point to the importance of dynamical processes in chemical and photometric evolution. Many theoretical ideas on the formation and dynamical evolution of galaxies have been motivated by these correlations, and dynamical models in turn have led to further understanding of such properties. A brief sketch of some current theoretical views of galaxy formation and dynamical evolution will therefore be given here. A fuller non-technical introduction to the field is given by \citet{Larson1977a}, and more technical reviews are by \citet{Doroshkevich1978SpatialGalaxies}, \citet{Gott1977RecentFormation}, \citet{Jones1976TheObservation}, \citet{Larson1976b}, and \citet{Rees1977CosmologyFormation}.

\medskip

In the conventional cosmological picture, the primeval gas emerging from the Big Bang some $10 - 20 \: \rm Gyr$ ago consisted almost entirely of hydrogen and helium. There must have been large enough density perturbations in this gas for some regions to be locally bound and to collapse despite the expansion of the Universe, and the existence today of galaxies and bound clusters with a $10^{10}$-fold range of masses shows that there was a very wide spread of density perturbations. A favored mass scale is ${\sim} 10^{6} \: \rm M_{\odot}$, which is the Jeans mass in the Universe after (re-)combination of the primeval plasma about $10^{6} \: \rm yr$ after the Big Bang. Maybe star formation first began in lumps of this mass, which could themselves have been part of larger-scale perturbations destined to collapse eventually into a massive galaxy. Although models for galaxy formation often start from an idealized smooth density distribution, gas in this state would be Jeans unstable on much smaller scales than the protogalaxy, and one would expect star formation to occur in regions of much higher density than average.

Independently of cosmological details, a protogalaxy can be pictured as a lumpy gas cloud governed by self-gravity (and perhaps by perturbations due to neighbors) rather than by the expansion of the Universe. If no stars formed, the gaseous protogalaxy would collapse, collisions between gas clouds would dissipate the energy parallel to the net angular momentum vector, and the system would become a flat disk. (The process of ``dissipation'' in this context usually implies that some of the kinetic energy of the gas is lost by collisionally induced radiation). If, on the other hand, stars form in much less than the collapse time, there would be no dissipation and a spheroidal system of stars would be produced; subclustering due to the initially lumpy gas distribution would be wiped out except in regions with about the mass and density of globular clusters. A spheroidal system can also be formed in much longer than the free-fall time if dissipation takes much longer, perhaps because the gas is in clouds that collide infrequently.

In general, the formation of spheroidal stellar systems -- elliptical galaxies and the bulges of spirals -- requires that stars form on a timescale less than that of gaseous dissipation, while the formation of a disk requires the opposite situation. The amount of disk possessed by a galaxy therefore depends on the amount of gas that remained to collapse dissipatively, after efficient star formation in the spheroidal component. This dynamical picture is consistent with the great ages, ${\sim} 10^{10} \: \rm yr$, inferred for stars in spheroidal systems, and the contrasting range of ages (including ongoing star formation) in disks.

In detail, the differences between spheroidal and disk systems are not quite so clear-cut, because the formation of condensed nuclei in elliptical galaxies and the central bulges of spirals probably involved some dissipation. The enhanced metallicities of nuclear stars are thought to be due to an inward concentration of metals, as gas enriched by massive stars further out dissipated its energy and condensed toward the center.

S0 galaxies appear to be intermediate systems, having disks but no young stars. One class of theories of their origin suggests that these properties are intrinsic, in that star formation efficiently used up the gas after it had collapsed to a disk. In another class of theories, S0 galaxies are former spirals that are devoid of young stars because their ISM was swept away, either in a collision with another galaxy or by an intergalactic wind due to motion of the galaxy through intergalactic gas. The high frequency of S0 galaxies in dense clusters, where intergalactic gas has been detected, suggests that at least some were formed in the latter way, but other S0 galaxies are quite isolated and may be of the ``intrinsic'' type.

Sweeping by collisions and intracluster gas are examples of interactions with the environment that probably affect the evolution of many galaxies, including apparently normal ones and very peculiar objects. Other effects include mergers of galaxies in the centers of clusters and in close binaries, tidal disruption in close passages, and bursts of star formation due to such violent interactions among gas-rich galaxies. A spiral galaxy may also accrete gas from its surroundings during most of its life, gradually forming a more massive and more extended disk and acquiring new gas to fuel continued star formation. Various aspects of interactions between galaxies and their environment are reviewed by \citet{Ostriker1977a}, \citet{Saar1978InteractionGalaxies}, and \citet{Toomre1977MergersConsequences}.

\subsection{Plan of this Review}
\label{subsec:Review_plan}

From the preceding outline, it should be clear that attempts to understand the evolution of stars and gas in galaxies inevitably get involved in very diverse aspects of astronomical theory and observation. This is not a field in which one can hope to develop a complete theory from a simple set of assumptions, because many relevant data are unavailable or ambiguous, and because galactic evolution depends on many complicated dynamical, atomic, and nuclear processes which themselves are incompletely understood. Because a logical development is inappropriate, this article treats different aspects of galactic evolution as pieces of a jigsaw puzzle that may some day be put together.

\begin{figure}
	\includegraphics[width=0.47\textwidth]{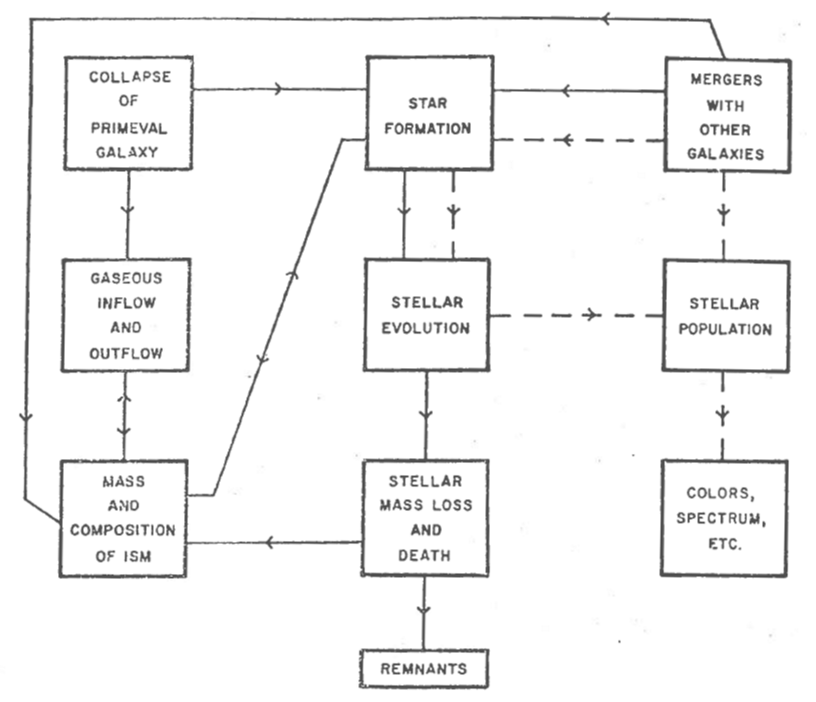}
    \caption{A schematic outline of the links between various constituents of galaxies and evolutionary processes. \emph{Solid lines} indicate the main factors in chemical evolution, and \emph{dashed lines} are for photometric evolution.}
    \label{fig:fig1}
\end{figure}

Figure~\ref{fig:fig1} is a very schematic diagram of relations among some of the most important factors in galactic evolution. It serves to summarize parts of the preceding outline, to suggest how a final jigsaw might be linked, and to show how the topics discussed below are related to each other. Two main sets of processes and constituents are indicated.

\begin{enumerate}
  \item Chemical evolution depends mainly on the areas connected by solid lines, which indicate that stars form from the ISM and return some of their mass, including new products of nucleosynthesis, to the ISM at death; the composition of the ISM also depends on gas flows to and from the region under consideration.
  
  \vspace{1mm}
  
  \item Photometric properties depend mainly on the areas on the areas connected by dashed lines, indicating that the history of star formation and the evolution of individual stars determine the population at any time.
\end{enumerate}

\medskip

Sections~\ref{sec:Aims and Methods 1},~\ref{sec:Evolution solar neighborhood},~and~\ref{sec:Chemical Evolution of Galaxies} below are on chemical evolution and Sections~\ref{sec:Approaches to Photometric Evolution}~and~\ref{sec:Colors and Star Formation Rates} are on the evolution of photometric properties. Since the formation and evolution of stars are vital to all aspects of galactic evolution, these processes are reviewed first in Section~\ref{sec:Star_formation}.

\section{The Formation and Evolution of Stars}
\label{sec:Star_formation}

The evolution of galaxies depends critically on properties of stars as a function of mass: their formation rates, evolution in the Hertzsprung--Russell (HR) diagram, lifetimes, and the mass and composition of material returned to the ISM. This Section reviews the properties of stars that are most relevant to later discussion of galaxy evolution.

\subsection{Basic Physical Properties of Stars}
\label{subsec:Physical_Properties}

Table~\ref{tab:table1} lists for reference some properties of main sequence (MS) stars with chemical compositions typical of the nearby disk population, i.e. with Solar or slightly lower metallicities. Because stars evolve in the HR diagram while still on the MS, this Table only applies to a particular stage of evolution for each mass: stars with total MS lifetimes less than $10 \: \rm Gyr$ (masses $ \gtrsim 1 \: \rm M_{\odot}$) are listed with the properties they have in the middle of this lifetime; less massive stars are listed at age $5 \: \rm Gyr$ or with mean empirical properties. The theoretical properties of MS stars of a given mass and composition differ slightly in different series of calculations, and they are sensitive to the helium and heavy-element abundances, even within the range of values applicable to disk stars, so the tabulated values should not be accepted too literally. This Table is intended only to provide approximate relations between such quantities as spectral type and MS lifetime, for quick reference.

\begin{table}
	\centering
	\caption{Representative data for main sequence stars. Quantities tabulated are typical values for stars of approximately Solar composition, from a variety of sources including mainly the following. Masses $m \leq 0.6 \: \rm M_{\odot}$: data from \citet{Veeder1974LuminositiesPhotometry}; $0.8 \leq m \leq 10 \: \rm M_{\odot}$: theoretical tracks of \citet{Mengel1979StellarSequence} and \citet{Alcock1978TheStars}; $m \geq 20 \: \rm M_{\odot}$: theoretical tracks of \citet{Chiosi1978MassiveMass-loss.} with maximum mass loss rates. For $m \leq 0.6 \: \rm M_{\odot}$, the mean empirical main sequence is given; for $0.8 \leq m \leq 1.0 \: \rm M_{\odot}$, points are for an age of $5 \: \rm Gyr$; and for more massive stars, points are for the middle of the main sequence lifetime ($\tau_{\rm ms} / 2$).
	}
	\label{tab:table1}
	\begin{tabular}{ccccccc} % three columns, alignment for each
	    \hline
	    $m \: \rm \left( M_{\odot} \right)$ & $\tau_{\rm ms}$ & $\log \left( L / \rm L_{\odot} \right)$ & $M_{\rm V}$ & $\log \ T_{\rm eff}$ & $B - V$ & Sp\\
		\hline
		0.15 & -- & -2.5 & 14.2 & 3.48 & 1.80 & M7\\
		0.25 & -- & -2.0 & 12.0 & 3.52 & 1.60 & M5\\
		0.4 & -- & -1.4 & 10.0 & 3.57 & 1.48 & M1\\
		0.6 & -- & -0.9 & 7.6 & 3.64 & 1.18 & K5\\
		0.8 & 25 & -0.4 & 6.0 & 3.70 & 0.88 & K1\\
		0.9 & 15 & -0.2 & 5.4 & 3.73 & 0.76 & G8\\
		1.0 & 10 & 0.0 & 4.9 & 3.76 & 0.64 & G2\\
		1.1 & 6.4 & 0.2 & 4.3 & 3.79 & 0.56 & F8\\
		1.2 & 4.5 & 0.4 & 3.7 & 3.82 & 0.47 & F6\\
		1.3 & 3.2 & 0.5 & 3.5 & 3.84 & 0.42 & F5\\
		1.4 & 2.5 & 0.7 & 3.0 & 3.86 & 0.36 & F2\\
		1.5 & 2.0 & 0.8 & 2.8 & 3.88 & 0.30 & F0\\
		2 & 0.75 & 1.3 & 1.4 & 3.98 & 0.00 & A0\\
		3 & 0.25 & 2.1 & -0.2 & 4.10 & -0.12 & B7\\
		4 & 0.12 & 2.6 & -0.6 & 4.18 & -0.17 & B5\\
		6 & 0.05 & 3.2 & -1.5 & 4.30 & -0.22 & B3\\
		8 & 0.03 & 3.6 & -2.2 & 4.35 & -0.25 & B1\\
		10 & 0.02 & 3.9 & -2.7 & 4.40 & -0.27 & B0.5\\
		15 & 0.01 & 4.4 & -3.7 & 4.45 & -0.29 & B0.5\\
		20 & 0.008 & 4.7 & -4.3 & 4.48 & -0.30 & B0\\
		30 & 0.006 & 5.1 & -5.1 & 4.51 & -0.31 & O9.5\\
		40 & 0.004 & 5.4 & -5.7 & 4.53 & -0.31 & O9\\
		60 & 0.003 & 5.7 & -6.2 & 4.58 & -0.32 & O5\\
		\hline
	\end{tabular}
\end{table}

\begin{figure}
	\includegraphics[width=0.47\textwidth]{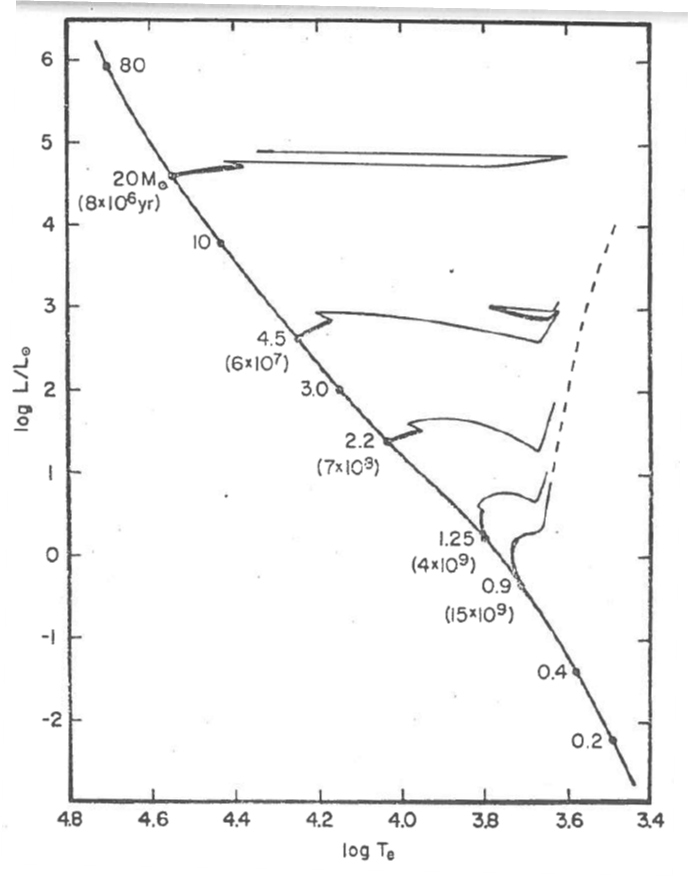}
    \caption{Theoretical HR diagram indicating main sequence lifetimes and evolutionary tracks of stars with approximately Solar composition. The zero-age main sequence (ZAMS) is shown, with a few stellar masses ($\rm M_{\odot}$) and lifetimes (yr). Also shown are representative evolutionary tracks, shaded heavily where a star evolves slowly relative to adjacent points. \emph{Sources}: ZAMS and tracks for $m \leq 3 \: \rm M_{\odot}$: \citet{Mengel1979StellarSequence}, $Z = 0.01$, $Y = 0.20$; ZAMS for $3 < m \leq 10 \: \rm M_{\odot}$: \citet{Alcock1978TheStars}, $Z = 0.01$, $Y = 0.29$; track for $m = 4.5 \rm \: M_{\odot}$: \citet{Harris1976EffectsClouds}, $Z = 0.01$, $Y = 0.28$; ZAMS and tracks for $m \geq 20 \rm \: M_{\odot}$: \citet{Chiosi1978MassiveMass-loss.}, case with maximum mass loss rate, $Z = 0.02$, $Y = 0.28$.}
    \label{fig:fig2}
\end{figure}

\begin{figure}
	\includegraphics[width=0.47\textwidth]{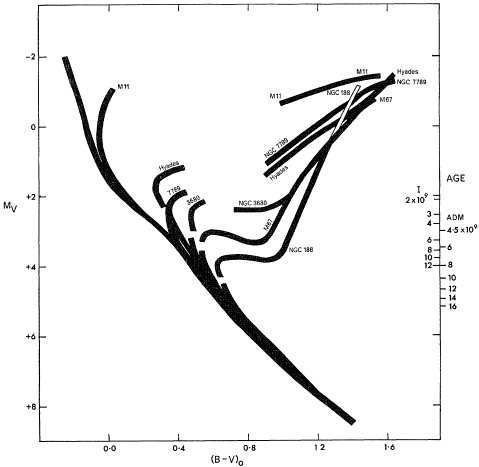}
    \caption{Composite color--magnitude diagram for open clusters \citep{Sandage1969IsochronesClusters}. Ages at the right refer to the blue point for a given cluster. Modern distances and age scales correspond to younger ages, e.g., $5 \times 10^{9} \: \rm yr$ for NGC~188 \citep{Demarque1977StellarGalaxy}. This diagram is an empirical counterpart of Figure~\ref{fig:fig2}.}
    \label{fig:fig3}
\end{figure}

An overview of stellar evolution in the HR diagram is provided by Figures~\ref{fig:fig2}~and~\ref{fig:fig3}. Figure~\ref{fig:fig2} is a theoretical HR diagram with the zero-age MS and evolutionary tracks for stars of a few masses; the tracks are drawn lightly in regions where stellar evolution is rapid (relative to neighboring points on the same track) since few stars are observed at those stages. A given star spends only about $10\%$ of its MS lifetime in stages of evolution beyond the MS, so the supergiant and giant regions of the HR diagram contain a relatively ephemeral population of stars. Much the same information is given empirically in Figure~\ref{fig:fig3}, which is a reproduction of the composite color--magnitude diagram for open clusters of \citet{Sandage1969IsochronesClusters}. The clusters have too few stars for the faster stages of evolution to be well represented; not only are there gaps between the MS and the giant branch, but the latest giants are absent from all but one or two clusters. Despite this problem, the diagram of \citet{Sandage1969IsochronesClusters} gives a vivid picture of stellar evolution.

\medskip

Using these Figures, one can envisage the evolution of a single generation of stars in a galaxy, starting with their populations on the zero-age MS. As time passes, the top of the MS peels away and short-lived supergiants appear; the MS turnoff evolves down to mid-B spectral type in $10^{8} \: \rm yr$, and by $10^{9} \: \rm yr$ late A and later stars remain on the MS; between $10^{9}$ and $10^{10} \: \rm yr$, the turnoff evolves from late A to early G, and there is a population of evolving K and M giants. An elliptical galaxy, for example, that had no star formation for the last many billions of years, would be expected to consist only of K and M giants and the MS up to a late F or early G turnoff. In a galaxy like our own, where star formation has presumably been continuous for billions of years, the whole array of MS stars, turnoff, and giant branches can be found.

\subsection{The Initial Mass Function}
\label{subsec:IMF}

Because the properties of stars depend strongly on their masses, the distribution of stellar masses at birth is a function vital to the photometric and chemical evolution of galaxies. Various definitions and notations for this function are used in the literature, so care is needed to avoid confusion in comparing different papers. The following notation is used here: the number of stars formed in the mass interval ($m$, $m + dm$) and in the time interval ($t$, $t + dt$) is
\begin{equation}
    \phi \left( m \right) \psi \left( t \right) dm \ dt,
	\label{eq:eq2.1}
\end{equation}
where $\psi \left( t \right)$ is the total mass of stars formed per unit time, and $\phi \left( m \right)$, which may itself be a time-dependent function, is therefore normalized so that
\begin{equation}
    \int_{0}^{\infty} m \phi \left( m \right) dm = 1.
	\label{eq:eq2.2}
\end{equation}
The function $\psi \left( t \right)$ is called the star formation rate (SFR) and $\phi \left( m \right)$ is called the initial mass function (IMF). It is often useful to approximate $\phi \left( m \right)$ by a power law, at least over some range of masses, and in these cases the notation used here is
\begin{equation}
    \phi \left( m \right) \propto m^{-\left( 1 + x \right)},
	\label{eq:eq2.3}
\end{equation}
where $x$ is called the slope of the IMF\footnote{Quantities equivalent to $x$, $-x$, $\left( 1 + x \right)$, and $-\left( 1 + x \right)$ have been called ``the slope of the IMF'' by different authors.}.

A comprehensive review of the IMF in the Solar neighborhood and elsewhere has recently been written by \citet{Scalo1978TheSpectrum}, so the following account includes only those aspects that are most relevant to the rest of this article.

\subsubsection{The local IMF}
\label{subsubsec:Local IMF}

The IMF in the Solar neighborhood can be derived from counts of field stars, using principles originally derived by \citet{Salpeter1955TheEvolution.}, and generalized to allow for a time-dependent SFR by \citet{Schmidt1959TheFormation., Schmidt1963TheMass.}, among others. A recent thorough rediscussion of the data and methods is given by \citet{Miller1979TheNeighborhood}.

Stars are counted as a function of absolute magnitude, and the counts must first be reduced to the present mass distribution of stars, $n(m)$, where $n(m) \ dm$ will denote the present number of MS stars in the mass interval ($m$, $m + dm$). Several non-trivial problems arise at this stage, including the following:

\begin{enumerate}
  \item The number of stars in each absolute magnitude interval must be corrected for post-MS stars, which requires a knowledge of color or spectral type; these corrections can be large -- e.g., near $M_{\rm V} = +1$ there are about equal numbers per unit volume of A dwarfs and K giants.
  
  \vspace{1mm}
  
  \item Field stars of different masses have very different distributions perpendicular to the Galactic plane, the scaleheight decreasing with increasing mass. The function $n(m)$ obtained for stars in a unit volume therefore differs systematically from the function for a unit column perpendicular to the plane. Because one is normally interested in the whole population of stars formed at all heights (or formed in the plane and accelerated to more extended orbits), near the Sun's galactocentric distance, the star counts should be reduced to the \emph{average number per square parsec} at this distance from the Galactic center.
  
  \vspace{1mm}
  
  \item The transformation between magnitudes and masses of MS stars depends on their chemical composition.
  
  \vspace{1mm}
  
  \item Stars evolve in the HR diagram while on the MS, so there is not a unique mass--luminosity relation even for a given composition.
\end{enumerate}

The papers cited above discuss how these problems can be handled, and the resulting uncertainties in $n(m)$.

Let us assume that we know the present mass distribution of MS stars per square parsec in the Solar neighborhood, $n(m)$. This function can be used to estimate the IMF and SFR as follows. Stars with lifetimes $\left( \tau_{\rm m} \right)$ greater than the age of the Galaxy $\left( t_{1} \right)$ have accumulated since star formation began ($t \equiv 0$), so Equation~(\ref{eq:eq2.1}) gives directly
\begin{equation}
    n(m) = \int_{0}^{t_{1}} \phi \left( m \right) \psi \left( t \right) dt, \: \: \tau_{\rm m} \geq t_{1}.
	\label{eq:eq2.4}
\end{equation}
If the IMF is assumed constant, this expression simplifies to
\begin{equation}
    n(m) = \phi \left( m \right) \overline{\psi}_{1} t_{1}, \: \: \tau_{\rm m} \geq t_{1},
	\label{eq:eq2.5}
\end{equation}
where $\overline{\psi}_{1}$ is the average past SFR; Equation~(\ref{eq:eq2.5}) can be used in any case if $\phi \left( m \right)$ is interpreted as the average past IMF. The mass with $\tau_{\rm m} = t_{1}$ is called the present turnoff mass ($m_{1}$) since it defines the lowest MS turnoff point in the HR diagram for local field stars. In practice, one must consider how the mass--lifetime relation depends on metallicity, but to illustrate the principles here these effects will be ignored; approximate round numbers are $m_{1} \simeq 1 \: \rm M_{\odot}$, $t_{1} \simeq 10 \: \rm Gyr$.

The present population of stars with $\tau_{\rm m} < t_{1}$ includes only those that were formed at times less than $\tau_{\rm m}$ ago, so from Equation~(\ref{eq:eq2.1}) their mass distribution is
\begin{equation}
    n(m) = \int_{t_{1} - \tau_{\rm m}}^{t_{1}} \phi \left( m \right) \psi \left( t \right) dt, \: \: \tau_{\rm m} < t_{1}.
	\label{eq:eq2.6}
\end{equation}
If $\phi(m)$ is constant, it can be taken out of the integral, but $n(m)$ still depends on details of the past SFR. A simpler equation can be written for stars with lifetimes shorter than the present timescale for changes in $\psi \left( t \right)$:
\begin{equation}
    n(m) = \phi \left( m \right) \psi_{1} \tau_{\rm m}, \: \: \tau_{\rm m} \ll t_{1},
	\label{eq:eq2.7}
\end{equation}
where $\psi_{1}$ is the present SFR. The function $n(m)$ is supposed to be averaged over the patchy distribution of the youngest stars in the Solar neighborhood, so Equation~(\ref{eq:eq2.7}) is probably a good approximation for $m \gtrsim 2 \: \rm M_{\odot}$, i.e., for stars with lifetimes $\lesssim 1 \: \rm Gyr$ and MS spectral types earlier than about A0; it would hold for all $m > m_{1}$ if the SFR were constant.

From the empirical function $n(m)$, therefore, one can use Equation~(\ref{eq:eq2.5}) to derive the shape of the (average) IMF for stars below about $1 \: \rm M_{\odot}$ and Equation~(\ref{eq:eq2.7}) to derive the shape of the IMF for stars above about $2 \: \rm M_{\odot}$. These two pieces of the IMF are not determined with the same multiplicative factors, since the quantities actually determined are $\phi \left( m \right) \overline{\psi}_{1} t_{1}$ and $\phi \left( m \right) \psi_{1}$, respectively. In particular, the relative values of the IMF for $1 \: \rm M_{\odot}$ and $2 \: \rm M_{\odot}$ depend on the ratio
\begin{equation}
    T_{1} \equiv \frac{\overline{\psi}_{1} t_{1}}{\psi_{1}},
	\label{eq:eq2.8}
\end{equation}
which is evidently a timescale for star formation in the Solar neighborhood.

There is an intermediate mass interval, $1 \lesssim m / \rm M_{\odot} \lesssim 2$, for which Equation~(\ref{eq:eq2.6}) cannot be simplified, so the shape of the IMF cannot be derived independently of details of $\psi(t)$. This gap is usually filled by assuming that the functions for higher and lower masses can be smoothly interpolated, as illustrated, for example, by \citet[][Figure~1]{Schmidt1963TheMass.}. In this way, $\phi(m)$ for the intermediate mass range is determined without too much ambiguity and constraints are set on the quantity $T_{1}$.

\begin{figure}
	\includegraphics[width=0.47\textwidth]{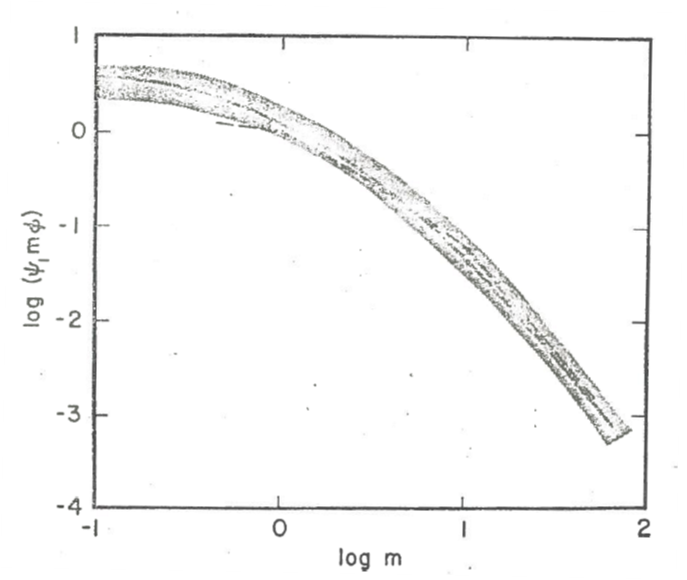}
    \caption{The initial mass function in the Solar neighborhood, shown in the form $\log \left[ \psi_{1} m \phi(m) \right]$ versus $\log \ m$, where $\phi(m)$ is the IMF itself, normalized as in Equation~(\ref{eq:eq2.2}), and $\psi_{1}$ is the present star formation rate in $\rm M_{\odot} \ pc^{-2} \ Gyr^{-1}$. Masses are in solar units. \emph{Solid line and shaded area}: range of functions derived by \citet{Miller1979TheNeighborhood}, allowing for uncertainties in the basic data and in the history of the star formation rate. \emph{Dashed line}: power-law segments, given by Equation~(\ref{eq:eq2.9}); the slopes of these lines are $-x$ in the notation of Equation~(\ref{eq:eq2.3}). (The dashed line and the shaded area are discrepant for $-0.4 < \log \ m < 0.0$ because different stellar luminosity functions were used).}
    \label{fig:fig4}
\end{figure}

\medskip

The IMF derived along the preceding lines by \citet{Miller1979TheNeighborhood} is illustrated in Figure~\ref{fig:fig4}, where the quantity plotted (logarithmically) is $m \phi(m) \psi_{1}$. The solid line is their analytical relation for the case of a constant SFR and a Galactic age $t_{1} = 12 \: \rm Gyr$: $m \phi(m) \psi_{1} = 3.83 \exp [ -1.09 ( \log m + 1.02 )^{2} ] \: \rm pc^{-2} \ Gyr^{-1}$ (with $m$ in solar units). The shaded area shows the range of uncertainty allowed by their analytic fits to limiting cases, based on uncertainties in the basic data and the limits for smoothly joining the low- and high-mass ends, as explained above. \citet{Miller1979TheNeighborhood} stress that, if one is making models for the Solar neighborhood, the choice of an IMF within the shaded area should be made consistently with the adopted SFR function, for otherwise the two functions together would not reproduce the observed star counts.

Power laws are very tractable analytically, so it is tempting to fit straight lines to the IMF derived from the star counts. \citet{Salpeter1955TheEvolution.} originally derived a famous slope $x = 1.35$ for the local IMF, but subsequent star counts and stellar lifetimes show that a single power law for the whole mass spectrum is a poor approximation; serious systematic errors can be introduced in some applications by ignoring the increase of slope with mass in the local IMF. The four dashed straight lines in Figure~\ref{fig:fig4} are intended as a compromise between the data and analytical convenience. (The discrepancy with the shaded area between $0.4$ and $1 \: \rm M_{\odot}$ arises because the dashed line for those masses is based on the luminosity function of \citet{Wielen1974TheCatalogue.}, which is flatter than the function adopted by \citealp[][Figure~1]{Miller1979TheNeighborhood}). The equations of these four lines, which have slopes $-x$ in the logarithmic plot (see Equation~\ref{eq:eq2.3}), are:
\begin{equation}
\begin{aligned}
    m \phi(m) \psi_{1} = & \ 1.00 m^{-0.25}, \: \: 0.4 < m \leq 1.0, \\
                         & \ 1.00 m^{-1.0}, \: \: 1.0 < m \leq 2.0, \\
                         & \ 1.23 m^{-1.3}, \: \: 2.0 < m \leq 10.0, \\
                         & \ 12.3 m^{-2.3}, \: \: 10.0 < m \leq 50.0,
	\label{eq:eq2.9}
\end{aligned}
\end{equation}
in units of $\rm pc^{-2} \ Gyr^{-1}$, with masses in solar units. (It is fortuitous that the coefficient 1.00 occurs in these equations). A slope is not given for $m < 0.4 \: \rm M_{\odot}$ because M dwarfs contribute little to the light of any galaxy and nothing directly to chemical evolution; their contribution to the total mass can be included with the invisible objects below $0.1 \: \rm M_{\odot}$.

Integrating Equation~(\ref{eq:eq2.9}) with respect to mass, we find that the present SFR for masses $0.4 - 50 \: \rm M_{\odot}$ is $3.0 \: \rm M_{\odot} \ pc^{-2} \ Gyr^{-1}$. The total SFR, $\psi_{1}$, could be derived if we knew the contribution from objects below $0.4 \: \rm M_{\odot}$, and $\psi_{1}$ in turn would allow the IMF to be normalized as in Equation~(\ref{eq:eq2.2}). As described in Section~\ref{subsubsec:Local SFR}, \citet{Miller1979TheNeighborhood} find $\psi_{1}$ in this way by defining as ``stars'' only objects above $0.1 \: \rm M_{\odot}$, for which there are star counts giving the shape of the IMF. Alternatively, $\psi_{1}$ can be estimated indirectly (see Section~\ref{subsubsec:Local SFR}). The values found are within a factor of ${\sim} 2$ of $10 \: \rm M_{\odot} \ pc^{-2} \ Gyr^{-1}$, which is several times greater than the rate for objects above $0.4 \: \rm M_{\odot}$.

\subsubsection{The IMF at other times and places}
\label{subsubsec:Other IMF}

Several questions can be asked about the constancy of the IMF. Is the IMF the same in regions other than the Solar neighborhood? Is it constant in time, locally and elsewhere? If variations occur, do they involve the shape of the IMF for stars above say $0.4$ or $1 \: \rm M_{\odot}$, or do they involve only the mass fraction in ``inert'' low-mass objects?

\medskip

Star clusters in the Galaxy and the Magellanic Clouds have a variety of MS luminosity functions indicating significant variations in the IMF on the scale of clusters (\citealp{DaCosta1977TheClusters, Freeman1977StarGalaxies}; others reviewed by \citealp{Scalo1978TheSpectrum}); also, there are regions of the Milky Way where the only newborn stars seem to be T Tauri stars, with masses $\lesssim 2 \: \rm M_{\odot}$, and other regions abounding in young OB stars. These observations, however, do not necessarily imply variations on galaxy-wide scales in averages over many sites of star formation. In fact, counts of field stars in nearby galaxies have not revealed any significant deviations from the local IMF \citep{Butcher1977ACloud, Hardy1977TheRegions, Lequeux1979ComparisonGroup}. The clearest evidence for large-scale variations is the mass function for nearby halo stars of \citet{Schmidt1975TheStars}; this function has $x \simeq 2$ for $0.25 - 0.75 \: \rm M_{\odot}$, whereas local disk stars have $x < 1$ in the same mass range. There are also some early-type spiral galaxies with widespread patches of young blue stars but a deficiency of \textsc{H~ii} regions, suggesting that the IMF is deficient in O stars; examples are the Sombrero galaxy, M104 \citep{vandenBergh1976b, Schweizer1978GalaxiesTails}, and NGC~2841 \citep{Kormendy1977}.

The upper part of the IMF $\left( m > 0.4 \: \rm M_{\odot} \right)$ can be studied indirectly by comparing colors of galaxies with those of models based on various IMFs. As discussed in Section~\ref{sec:Colors and Star Formation Rates}, the results are consistent with the local IMF holding (on large scales) everywhere, but in some respects the test is rather insensitive. For example, a galaxy like the Sombrero is so dominated by old yellow--red stars that the \emph{integrated} colors at optical wavelengths would be changed negligibly by the presence or absence of stars above $10 \: \rm M_{\odot}$ in the IMF. The red--infrared spectra of galaxies show that giant stars, rather than late dwarfs, provide most of the light at these wavelengths, and this result sets some constraints on the IMF for stars of ${\sim} 0.4 - 1 \: \rm M_{\odot}$: if the IMF were much steeper than the local function (say if $x > 2$), low-mass dwarfs would contribute much of the infrared light, contrary to the spectroscopic information.

Because of this giant dominance in the infrared light, photometric observations are very insensitive to the mass fraction in stars below $0.4 \: \rm M_{\odot}$, i.e. M dwarfs; these stars could be numerous enough to dominate the mass of the system while contributing very little light! A handle on the mass fraction in invisible stars is given by the mass-to-luminosity ($M / L$) ratio, which is crudely a ratio of very low-mass to more massive stars. In the Solar neighborhood, the integrated luminosity of known stars (per $\rm pc^{3}$), including giants, divided by the dynamically-estimated density of mass, yields a ratio $M / L_{\rm B} \simeq 3 \: \rm M_{\odot} / L_{B \odot}$ \citep{Faber1979MassesGalaxies}. If the IMF and proportion of invisible matter were invariant, all galaxies should have this $M / L$ ratio, apart from a systematic increase by a factor of ${\sim} 10$ from the bluest galaxies to the reddest (because of the decreasing fraction of young blue stars), and apart from a scatter ${\sim} 10\%$ due to different mass fractions of ISM. In fact, observed $M / L$ values show a \emph{scatter} by a factor of 10 or more at a given color, and a tendency to increase from values of a few in the central parts of galaxies to ${\sim} 100$ in the outer regions \citep{Faber1979MassesGalaxies}. It appears that different proportions of hidden matter exist in different galaxies, and that the invisible mass fraction increases with radius. If this matter condensed before ordinary star formation began in galaxies, as in some pictures of the formation of heavy invisible halos \citep[e.g.][]{White1978CoreClustering}, then it would have no direct effect on chemical or photometric evolution. (It would have indirect effects, because the motions of gas and stars in the galaxy would be influenced by the hidden mass). However, if there are variations in the low-mass part of the stellar IMF, chemical evolution would be affected significantly, because variable proportions of stellar matter would be returned to the ISM by evolving stars.

Certain chemical properties of galaxies can be explained by invoking variations in the shape of the IMF for massive stars. For example, the extreme paucity of metal-poor dwarfs in the Solar neighborhood could be due to an initial burst of massive metal-producing stars; and variations in relative abundances of heavy elements could be due to variable proportions of the stars that synthesize such elements. But in all cases, other explanations not involving variations in the IMF are available, as discussed in Sections~\ref{sec:Evolution solar neighborhood}~and~\ref{sec:Chemical Evolution of Galaxies}.

\medskip

In summary, significant variations in the IMF may be rare enough for the assumption of a universal function (the local IMF) to be useful for many contexts. However, it is equally relevant to ask how galactic evolution would be affected by changes in both the form of the IMF for visible stars and the relative weight of very low-mass objects.

\subsection{Rates of Star Formation}
\label{subsec:SFRs}

The rate of star formation is one of the main factors in galactic evolution. The shape of a galaxy depends on the timescale for star formation relative to collapse and dissipation timescales, the colors and luminosity depend on the age distribution of stars, and chemical evolution depends on the SFR relative to the gas flow rates and the gas mass. Various ways of estimating SFRs in galaxies are reviewed here.

\subsubsection{The local SFR}
\label{subsubsec:Local SFR}

There are several ways of estimating the present SFR in the Solar neighborhood, which is the quantity in $\rm M_{\odot} \ pc^{-2} \ Gyr^{-1}$ denoted $\psi_{1}$ above. Stars above ${\sim} 2 \: \rm M_{\odot}$ are no problem, because they are so short-lived that their numbers scale directly with $\psi_{1}$ (Equation~\ref{eq:eq2.7}), but the ages of less massive stars are unknown, as is the mass fraction contained in objects below $0.1 \: \rm M_{\odot}$. Most of the methods that have been suggested for overcoming these problems reduce in principle to estimating the average past SFR ($\overline{\psi}_{1}$) and then the ratio $\overline{\psi}_{1} / \psi_{1}$ \footnote{Another approach in the literature is to find the formation rate of relatively massive stars (above $2 \: \rm M_{\odot}$ or a greater limit), and to scale to the total SFR via an \emph{assumed} IMF. Obviously, the problems mentioned above are sidestepped rather than solved, so the conclusions should be regarded as correspondingly uncertain.}. The following two methods are typical.

\begin{enumerate}
  \item A dynamical estimate of the total surface density in the Solar neighborhood, called the ``Oort limit'', can be derived from motions of stars perpendicular to the Galactic plane; its value is probably ${\sim} 80 - 100 \: \rm M_{\odot} \ pc^{-2}$ \citep{Oort1965StellarDynamics}. Of the total surface density, ISM in various forms accounts for $5 - 10 \: \rm M_{\odot} \ pc^{-2}$ \citep{Gordon1976CarbonNucleons}, so stars are presumed to contribute ${\sim} 70 - 95 \: \rm M_{\odot} \ pc^{-2}$ (neglecting any possible non-stellar condensed objects). This is not the total mass of stars ever formed, because some fraction $R$ has been ejected back to the ISM by evolving and dying stars; $R$ depends on the IMF, whose normalization is not known without $\psi_{1}$ itself, but all reasonable normalizations give $R$ in the range ${\sim} 0.1 - 0.3$ (as shown in Section~\ref{subsec:Assumptions}). The mass of stars ever formed is thus $10\% - 30\%$ greater than the mass now present; the mean past SFR is thus given by $75 < \overline{\psi}_{1} t_{1} < 125 \: \rm M_{\odot} \ pc^{-2}$.
  
  The next step is to use the ratio $T_{1}$ (Equation~\ref{eq:eq2.8}), derived by joining smoothly the IMFs with different multiplicative factors for stars above $2 \: \rm M_{\odot}$ and below $1 \: \rm M_{\odot}$, respectively. Plausible values of $T_{1}$ lie between 5 and $20 \: \rm Gyr$ \citep{Tinsley1976ChemicalInhomogeneities, Miller1979TheNeighborhood}, so we are left with the following range of values for $\psi_{1}$ itself: $1.5 < \psi_{1} < 25 \: \rm M_{\odot} \ pc^{-2} \ Gyr^{-1}$. The uncertainties at various stages of the derivation are mutually independent, and the $20\%$ uncertainty in $R$ is relatively unimportant. A more stringent upper limit could be obtained if one could assume that the SFR has not increased since the time the oldest disk stars formed, ${\sim} 12 \: \rm Gyr$ ago; then $T_{1} > 12 \: \rm Gyr$ and $\psi_{1} < 10 \: \rm M_{\odot} \ pc^{-2} \ Gyr^{-1}$. However, although the assumption of a monotonically decreasing SFR is common, recent work on the age distribution of disk stars \citep{Twarog1980AnNeighborhood} shows that the SFR more likely \emph{increased} between 12 and $5 \: \rm Gyr$ ago, and was approximately constant in the last ${\sim} 5 \: \rm Gyr$. In this case, $T_{1}$ is probably between about 5 and $10 \: \rm Gyr$, and the above limits on $\overline{\psi}_{1} t_{1}$ lead to $8 < \psi_{1} < 25 \: \rm M_{\odot} \ pc^{-2} \ Gyr^{-1}$. Values only ${\sim} 10 - 20\%$ of these have usually been derived in the literature, on the assumption that $\psi (t)$ has been a decreasing function for at least $10 \: \rm Gyr$.
  
  \vspace{1mm}
  
  \item Instead of using the Oort limit to estimate the total mass of stars ever formed, one can use star counts to obtain the existing population (and to allow for the total mass of stars that have evolved away in the time since they formed) down to ${\sim} 0.1 \: \rm M_{\odot}$, and neglect any less massive objects. In this way, \citet{Miller1979TheNeighborhood} derive a range of values $43 < \overline{\psi}_{1} t_{1} < 144 \: \rm M_{\odot} \ pc^{-2}$. Combining these values with appropriate limiting values of $T_{1}$, they find self-consistent values in the range $3 < \psi_{1} < 7 \: \rm M_{\odot} \ pc^{-2} \ Gyr^{-1}$.
\end{enumerate}

Method (i) above gave larger values than those of \citet{Miller1979TheNeighborhood} for two reasons: the estimated mass of stars now present above $0.1 \: \rm M_{\odot}$ falls short of estimates based on the Oort limit (although the discrepancy may not be significant), and \citet{Miller1979TheNeighborhood} considered both decreasing and increasing SFRs. Altogether, it is plausible to suggest that $\psi_{1}$ lies within a factor of about two of $10 \: \rm M_{\odot} \ pc^{-2} \ Gyr^{-1}$.

\subsubsection{The SFR elsewhere}
\label{subsubsec:Elsewhere SFR}

Estimates of the SFR in other places use various indicators of the numbers of fairly massive stars, and convert these to a total SFR by means of an assumed IMF, which is usually the local function (or an assumed power-law approximation). Methods that have been used include the following.

\begin{enumerate}
  \item The brightest individual stars, supergiants, and occasionally upper-MS stars, can be counted in nearby galaxies. Apart from complications due to differences in chemical composition, the numbers of massive stars in given stages of evolution should be directly proportional to the SFR, if the IMF is the same everywhere; their numbers relative to counts of similar stars in the Solar neighborhood thus scale fairly directly with the SFR. \citet{Lequeux1979ComparisonGroup} has applied this method to galaxies of the Local Group, using various types of stars, and several authors \citep[e.g.][]{Searle1973TheGalaxies, Larson1974PhotometricGalaxies} have used such rare objects as supernovae in the same way for more distant galaxies.
  
  \vspace{1mm}
  
  \item If hot young stars are present in a region, they generally emit most of the Lyman continuum photons that are available for ionizing interstellar hydrogen. The flux of $\rm H \alpha$, $\rm H \beta$, or free--free radio emission is therefore approximately proportional to the SFR. This method has been applied to external galaxies by \citet{Cohen1976H-alphaGalaxies} using integrated $\rm H \alpha$ equivalent widths, and by \citet{Huchra1977StarGalaxies} using $\rm H \beta$; \citet{Smith1978StarGalaxy.} have studied SFRs in the Galaxy using radio emission from \textsc{H~ii} regions.
  
  \vspace{1mm}
  
  \item The integrated colors of galaxies depend strongly on their relative proportions of young and old stars; as a result, UBV colors in general provide a rough estimate of the ratio (SFR) / (mass of old stars). This approach will be discussed in detail in Section~\ref{sec:Colors and Star Formation Rates}.
  
  \vspace{1mm}
  
  \item Regions of star formation are often heavily obscured by dust, but even so the infrared luminosity (starlight re-radiated by dust grains) gives a measure of the SFR, as discussed in Section~\ref{subsubsec:Highly reddened galaxies}.
\end{enumerate}

Most of these methods rely strongly on the assumption of a universal IMF, since they sample only the very massive stars; integrated colors can reveal large departures from an assumed IMF (Section~\ref{subsubsec:Other IMF}), but even these are completely blind to stars below ${\sim} 0.5 \: \rm M_{\odot}$. Even granted the assumption of universality, important uncertainties arise in scaling from the observations to a total SFR, because the statistics of very massive stars are poor. Large systematic errors can arise from the use of a single power law for the whole IMF (a common practice), as can be clearly seen from Figure~\ref{fig:fig4}.

Despite the uncertainties, the results of various studies have led to a coherent picture. In the Galaxy, some $10\%$ of the current star formation is occurring in the innermost $1 \: \rm kpc$, and most of the remainder is concentrated in a ring between 5 and $8 \: \rm kpc$ from the center, which is the site of most of the Galaxy's giant molecular clouds, infrared emission, and other signs of intense star formation. (There are several relevant reviews in the symposium volume edited by \citealp{Burton1979TheGalaxy}). Morphologically normal galaxies show a systematic trend of SFR with type, which can be described as a progression from a very short timescale for star formation in the earliest types to very long in the latest. Some peculiar galaxies appear to be undergoing intense bursts of star formation. These properties will be discussed later in Section~\ref{sec:Colors and Star Formation Rates}.

\subsubsection{Factors affecting the SFR}
\label{subsubsec:SFR factors}

An obvious question to ask is \emph{why} stars form at various rates in various places. Factors affecting SFRs have been reviewed by \citet{Larson1977b}, and some salient points will be mentioned here.

\medskip

A popular assumption, following \citet{Schmidt1959TheFormation.}, is that the SFR varies as a power $n$ of the gas density. However, attempts to determine $n$ empirically run into two problems: $n$ appears to vary within and among galaxies, indicating that the gas density is not the only relevant quantity; and the very definition of ``gas density'' is ambiguous since it depends strongly on the spatial resolution of the observations. Consequently, a power-law formula for the SFR cannot be given the status of a physical law, but it can be regarded as a useful parametrization in some circumstances (as in the later study of \citealp{Schmidt1963TheMass.}).

The physical environment needed for star formation is evidently cool, gravitationally bound gas clouds, so the main question is: under what conditions does interstellar gas clump into clouds? \citet{Larson1977b} argues that gas compression is the key requirement for star formation, and that both large- and small-scale dynamical processes are important. On small scales, interstellar shock and ionization fronts have been widely discussed as causes of star formation. These processes can be self-sustaining, since supernovae and \textsc{H~ii} regions arising from recently-formed stars may induce further star formation nearby \citep[e.g.][]{Elmegreen1977SequentialAssociations}. Large-scale compression mechanisms that may lead to star formation include gravitational settling of gas into a thin layer, density waves of bar or spiral form, high-velocity collisions between gas streams in young or interacting galaxies, and accretion of intergalactic clouds. If star formation can be self-propagating over long distances, differential rotation can spread the active regions out to produce spiral structure \citep{Gerola1978StochasticGalaxies}. Large-scale processes are especially relevant to the long-term evolution of galaxies, and since some of them involve interactions with external matter, galaxies do not necessarily evolve as independent systems. This point has been emphasized in a review by \citet{Saar1978InteractionGalaxies}. Further discussion of star formation can be found in the reviews cited, many references therein, and in a symposium volume edited by \citet{DeJong1977StarFormation}.

\medskip

Although some understanding has developed of the conditions under which stars form, no formula for the SFR has emerged that can make a useful \emph{prediction} for any galaxy. \citet{Lynden-Bell1977OnFormation} has pointed out that a proper expression for the SFR would contain so many unknown parameters as to be useless. He wrote, ``This rate $R$ (the SFR) probably depends on gas density $\rho_{\rm g}$, gas sound speed $c_{\rm s}$, shock frequency $\omega_{\rm s}$, shock strength $V_{\rm T} / c_{\rm s}$, gas rotation $\Omega$ and shearing rate $A$, the magnetic field strength $\lvert \vec{B} \rvert$, the gas metal abundance $Z$, and possibly the background star density $\rho_{*}$. Thus
\begin{equation*}
    R = R \left( \rho_{\rm g}, \ c_{\rm s}, \ \omega_{\rm s}, \ V_{\rm T} / c_{\rm s}, \ \Omega, \ A, \ \lvert \vec{B} \rvert, \ Z, \ \rho_{*} \right).
	\label{eq:eq2.Lynden-Bell}
\end{equation*}
However, if we knew the true functional form of $R$ and offered it to a galaxy builder he would probably tell us `Oh, go and jump in the lake, that's far too complicated'.''

The only feasible approach to galaxy building is to be less ambitious than to want a physical formula for the SFR. Instead, schematic expressions, guided by the above ideas on relevant factors, can be tested to see how galactic evolution is affected by various parameters. Some examples already mentioned just above and in Section~\ref{sec:Overview} are that self-propagating star formation can lead naturally to spiral structure, and that the final shape of a galaxy depends on the ratios of the timescales for star formation, collapse, and gaseous dissipation in the protogalaxy. Many examples of schematic expressions for the SFR, in models for chemical and photometric evolution, will be considered later.

\subsection{Stellar Evolution Beyond the Main Sequence}
\label{subsec:Beyond MS}

Late stages of stellar evolution are important to galaxies because stars burn much of their nuclear fuel after leaving the main sequence. The products of this late fuel guzzling provide most of the chemical enrichment of galaxies, and the energy liberated provides most of their integrated light. The following outline of stellar evolution and nucleosynthesis emphasizes aspects that will be relevant later in the discussion of galaxies, and necessarily only skims the surface of these topics. Further details and references can be found in the comprehensive review by \citet{Trimble1975TheElements}, and in other papers cited below.

\subsubsection{Stars near Solar mass}
\label{subsubsec:Solar mass}

Giant stars of approximately Solar metallicity and mass are especially interesting because they provide most of the light of elliptical and S0 galaxies and the nuclear bulges of spirals. The dominant old-disk population of the Solar neighborhood provides a convenient sample of such stars for detailed study; these may be somewhat younger than most stars in spheroidal systems, but evolution on the giant branch is thought to be rather insensitive to the initial stellar mass in the small range whose lifetimes are ${\sim} 5 - 15 \: \rm Gyr$.

\medskip

\begin{figure}
	\includegraphics[width=0.47\textwidth]{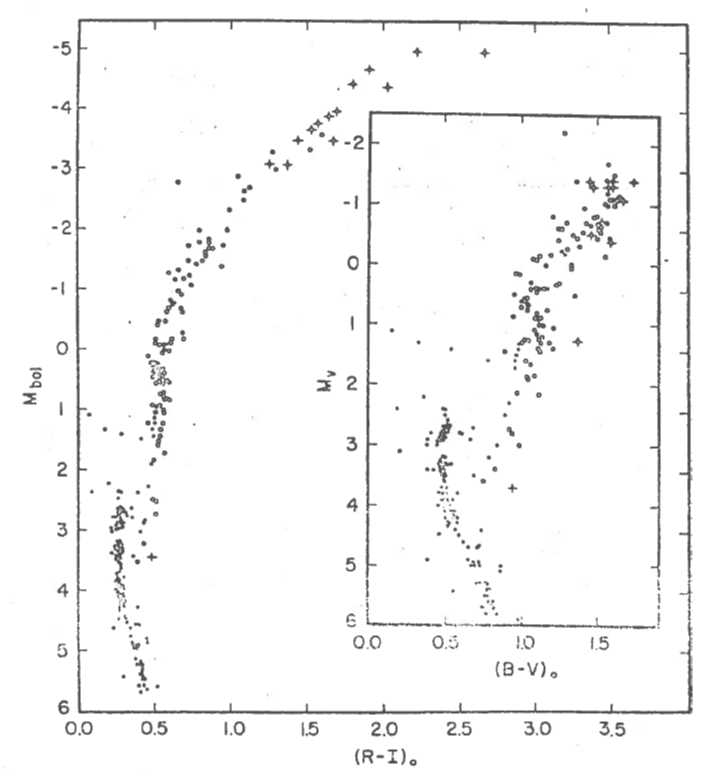}
    \caption{Color--magnitude diagrams for the old open cluster M67 (\citealp{Racine1971Photometry+12}; \emph{filled circles}) and for red giants in old-disk groups (\citealp{Eggen1962EvidenceCollapsed., TinsleyGunn1976b}; \emph{open circles}); \emph{crosses} denote variable stars. The M67 data have been reduce to $M_{\rm bol}$ versus $R-I$, as in \citet{TinsleyGunn1976b}. This diagram illustrates the types of stars that apparently contribute most of the light in old, metal-rich systems, such as giant elliptical galaxies and the nuclear bulges of spirals. Because of the small total cluster population, there are no late giants in M67, and because of selection effects the visually brightest stars are over-represented in the groups.}
    \label{fig:fig5}
\end{figure}

Figure~\ref{fig:fig5} gives color--magnitude diagrams of the old open cluster M67 and a sample of field giants, which are necessary for completeness because the cluster is so poor that it contains none of the short-lived M giants. Selection effects in the field sample mean that earlier giants are badly under-represented relative to later types, and an unbiased luminosity function would show roughly a factor of two decrease in the number of stars per bolometric magnitude interval, for each magnitude increase in luminosity up the giant branch. Some relevant aspects of the stellar evolution responsible for this distribution of stars are as follows \citep[e.g.][]{Iben1974PostStars, TinsleyGunn1976b}:

\begin{enumerate}
  \item When the central $10\%$ of hydrogen has been converted to helium, the star leaves the MS and its site of hydrogen burning moves from the center to a shell around the helium core; as the shell is established, the star moves along the subgiant branch (at $M_{\rm bol} \simeq 2.5$ in Figure~\ref{fig:fig5}) to the base of the giant branch, at which point the helium core mass is ${\sim} 0.2 \: \rm M_{\odot}$.
  
  \vspace{1mm}
  
  \item During the first ascent of the giant branch to the so-called ``tip'' at $M_{\rm bol} \simeq -3$, the helium core grows to ${\sim} 0.45 \: \rm M_{\odot}$. Although this ascent occupies only about $10\%$ of the preceding lifetime, the energy emitted, which is nearly proportional to the mass of hydrogen burned, is greater.
  
  \vspace{1mm}
  
  \item Helium ignition in the degenerate core -- the ``helium flash'' -- halts the ascent of the giant branch, and the star very quickly drops to a lower luminosity, $M_{\rm bol} \simeq 0.5$, where it remains for about $10^{8} \: \rm yr$ while the core helium is burned to carbon (and a little oxygen). This stage appears as the horizontal branch to the left of the giant branch in HR diagrams of globular clusters, but most metal-rich stars form a clump of early K giants superposed on the ascending giant branch; some metal-rich blue horizontal-branch stars are found in the old-disk population, and these may represent a minority that lost a large fraction of their envelopes at the helium flash \citep[e.g.][]{Butler1976MetalWindow}. Although the total energy emitted by core-helium-burning stars is relatively small, blue horizontal-branch stars can make an important contribution to the light of galaxies whose bluest stars otherwise are at the turnoff.
  
  \vspace{1mm}
  
  \item The star then ascends the giant branch for a second time (called the asymptotic giant branch stage because it appears as an asymptote to the left of the first giant branch in the HR diagrams of globular clusters), with two shells burning hydrogen and helium respectively. The star can become brighter and cooler than the first giant branch ``tip'', as seen in Figure~\ref{fig:fig5} but not in all HR diagrams for globular clusters.
  
  \vspace{1mm}
  
  \item The end of this stage of evolution is determined by mass loss, without which the whole star would become a carbon-oxygen white dwarf. In fact, there is much evidence that several tenths of a solar mass of envelope are lost before the burning shells reach out to consume the whole initial mass of hydrogen. For example, the average mass of white dwarfs is ${\sim} 0.7 \: \rm M_{\odot}$, well below the turnoff mass (${\sim} 1 \: \rm M_{\odot}$) for field stars; planetary nebulae, most of which belong to the old-disk population, represent ${\sim} 0.2 \: \rm M_{\odot}$ of ejected matter; and spectroscopic observations of M giants show that they are losing mass in a wind. The existence of both very late M giants and blue horizontal-branch stars in the old-disk population suggests that mass loss occurs at a variety of rates among stars of a given initial mass and composition. Since the average final core mass of solar-mass stars must be the white dwarf mass, ${\sim} 0.7 \: \rm M_{\odot}$, the energy liberated during the second ascent of the giant branch is comparable to that liberated on the first ascent, with the difference that most of it now appears in the near infrared from very cool stars.
\end{enumerate}

Current stellar models do not predict post-MS evolution accurately enough to be used directly in synthetic galaxies. The main uncertainties are the depth of the surface convection zone, which determines the effective temperature at a given luminosity, and mass loss; a related problem is the outward mixing of products of nucleosynthesis, which are observed on the surfaces of many giants but cannot be predicted in detail. Consequently, galaxy models must rely heavily on empirical studies of giant populations. The implications of these problems for understanding photometric properties of elliptical galaxies are reviewed by \citet{Faber1977ThePopulations}. Mass loss and mixing also mean that even these low-mass stars contribute significant amounts of newly synthesized elements to the ISM, including for example carbon and nitrogen, which are often overabundant at the surfaces of giants and in planetary nebulae. The isotopes $^{13}$C and $^{14}$N are by-products of hydrogen burning by the CNO cycle, while $^{12}$C must be a product of core helium burning \citep[e.g.][]{Trimble1975TheElements, IbenI.1978OnMedium}.

\subsubsection{Stars of $1 - 4 \: \rm M_\odot$}
\label{subsubsec:1 - 4 Msun}

%\begin{figure*}
%	\includegraphics[width=0.97\textwidth]{Figures/Fig6.png}
%    \caption{Composite HR diagrams for (mainly) post-main-sequence stars in galactic clusters grouped by age \citep{Harris1976EvolvedClusters.}. Ages ($\tau$ in yr) are indicated. These diagrams illustrate the distribution of supergiants expected in galaxies containing stars of the appropriate ages.}
%    \label{fig:fig6}
%\end{figure*}

\begin{figure*}
\centering
	\includegraphics[width=0.49\textwidth]{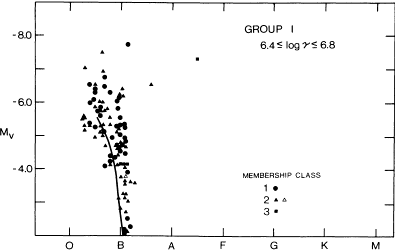}
	\includegraphics[width=0.49\textwidth]{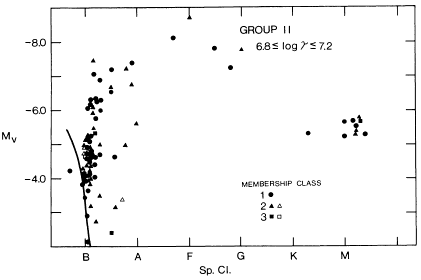}
	\includegraphics[width=0.49\textwidth]{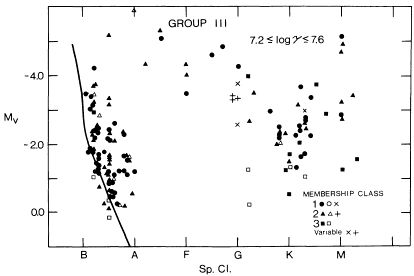}
	\includegraphics[width=0.49\textwidth]{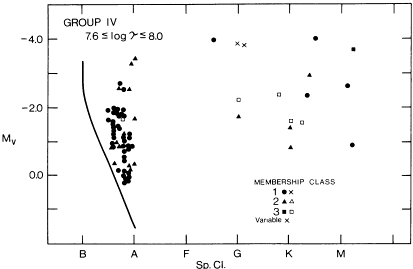}
    \caption{Composite HR diagrams for (mainly) post-main-sequence stars in galactic clusters grouped by age \citep{Harris1976EvolvedClusters.}. Ages ($\tau$ in yr) are indicated. These diagrams illustrate the distribution of supergiants expected in galaxies containing stars of the appropriate ages.}
    \label{fig:fig6}
\end{figure*}

White dwarfs appear in open clusters where their progenitors must have been stars with initial masses up to ${\sim} 4 \: \rm M_{\odot}$ \citep[e.g.][]{Tinsley1975b}. The substantial mass loss required is in agreement with stellar models that allow for winds \citep{Fusi-Pecci1976OnSun., Mengel1976SteadyIgnition.}; these models suggest that stars up to some limit ${\sim} 3 - 6 \: \rm M_{\odot}$ lose so much mass that they fizzle before the point of carbon ignition, which would otherwise occur (possibly explosively) when the degenerate core reached $1.4 \: \rm M_{\odot}$. Therefore, stars with MS lifetimes in the whole range from ${\sim} 10^{8}$ to ${\sim} 10^{10} \: \rm yr$, from mid-B to early G on the MS, evolve internally similarly to the solar-mass stars discussed above, and they make similar contributions to nucleosynthesis.

\subsubsection{Stars above $8 \: \rm M_\odot$}
\label{subsubsec:8 Msun}

Stars initially more massive than ${\sim} 8 \: \rm M_{\odot}$, which are early B and O stars on the MS, are believed to ignite carbon (quietly) in a non-degenerate core and to proceed through many stages of nuclear burning. Models acquire an onion-skin structure of successive elements, with unaltered envelope material (mostly hydrogen) on the outside, then layers of helium, carbon, oxygen, neon, magnesium, etc., with the ashes of each nuclear reaction providing fuel for the next. These stars and their contributions to nucleosynthesis are discussed by \citet{Arnett1978OnStars}, \citet{Weaver1978PresupernovaStars}, and \citet{Chiosi1979TheGalaxy}, the last stressing the very important effects of mass loss. Theoretical models predict that the stars undergo core collapse after the last exoergic reaction, formation of iron-peak elements, and the event is expected to send a shock wave through the star leading to explosive nucleosynthesis of most of the less abundant elements. Because Type~II supernovae are associated with the most massive stars in galaxies, they are believed to represent these explosive stellar deaths. The lower initial mass limit for stars with this fate is somewhere in the range $6 - 10 \: \rm M_{\odot}$, depending on the structure of stellar cores at carbon ignition; it is sometimes suggested that there is also an upper limit for such deaths, stars above the higher limit collapsing to black holes that swallow most of the new elements \citep[e.g.][]{Wheeler1978a}.

\medskip

The evolution of massive stars in the HR diagram (as well as internally) must be strongly affected by mass loss, since winds of ${\sim} 10^{6} \: \rm M_{\odot} \ yr^{-1}$ are inferred from the spectra of MS O stars and much higher mass-loss rates are common in supergiants \citep{Conti1978MassStars}. These rates mean that the stellar mass is changing significantly on an evolutionary timescale. The evolutionary tracks of such stars in the HR diagram cannot be predicted reliably \citep[e.g.][]{Chiosi1978MassiveMass-loss., Stothers1978StellarStars}, so models for galaxies with young stars are subject to some uncertainty. Empirical illustrations of the evolutionary tracks of massive stars are provided by compilations of HR diagrams of open clusters in age groups \citep{Harris1976EvolvedClusters.}, of which the youngest few are reproduced in Figure~\ref{fig:fig6}. The general trends with age are clear, but there are obviously many uncertainties due to the small numbers of supergiants, and a scatter that may be due partly to the presence of clusters with different metallicities.

\subsubsection{Stars of intermediate mass}
\label{subsubsec:Intermediate mass}

There may be some stars of mid-B type on the MS, with masses in the interval ${\sim} 4 - 8 \: \rm M_{\odot}$, that neither die quietly as white dwarfs nor follow the sequence of nuclear burning stages through core collapse and (Type~II) supernova explosion. Such stars would be characterized by carbon ignition in a degenerate core, and it is possible that they do not exist if the upper limit for white dwarf formation exceeds the lower limit for non-degeneracy. If they do exist, their final evolution is very uncertain, since it is not known whether degenerate carbon ignition would blow a star apart or lead quietly to further burning; this problem is reviewed by \citet{Wheeler1978a}.

Regardless of their final fate, stars of about $4 - 8 \: \rm M_{\odot}$\footnote{Values of stellar masses will from now on be understood to be initial values, unless otherwise stated.} make important contributions to galactic enrichment in elements that are produced during hydrogen and helium burning and are mixed up to the surface. Overabundances of carbon, nitrogen, barium and other elements are seen in the spectra of many giants in this mass range, and their theoretical origin is discussed by \citet{IbenI.1978OnMedium}.

The evolution of stars between 4 and $8 \: \rm M_{\odot}$ in the HR diagram is subject to the same uncertainties that beset more and less massive stars. The groups of clusters in \citet{Harris1976EvolvedClusters.} again give a useful guide, but they are rather sparse, as seen in Figure~\ref{fig:fig6}.

\medskip

About half of all supernovae in spiral galaxies are Type~I supernovae (SNI), which, because of their sites in galaxies are associated with stars below ${\sim} 6 \: \rm M_{\odot}$. Their origin has been controversial for some time, as reviewed by \citet{Tammann1974StatisticsSupernovae, Tammann1977AStatistics}, \citet{Tinsley1975b, Tinsley1977a}, and \citet{Wheeler1978a}. This leaves a serious gap in our understanding of stellar deaths and nucleosynthesis, especially since SNI appear to eject significant quantities of iron \citep{Kirshner1975Supernova5253}. Alternative theories include the following.

\begin{enumerate}
  \item SNI are exploding white dwarfs. This has long been the standard theory, because SNI occur in elliptical galaxies where only stars ${\sim} 1 \: \rm M_{\odot}$ are conventionally believed to be dying. However, the white dwarf theory conflicts with the large envelopes (of supergiant size and $\gtrsim 1 \: \rm M_{\odot}$) that are invoked to explain the early spectra and light curves of SNI, and with the clear association of SNI with a young stellar population in spiral and irregular galaxies \citep{Oemler1979TypeStars}.
  
  \vspace{1mm}
  
  \item SNI are stars ${\sim} 2 \: \rm M_{\odot}$ that live for ten times longer than normal by mixing completely on the MS. \citet{Wheeler1978b} proposed this model to be consistent with the envelope masses, populations in elliptical galaxies, the extreme deficiency of hydrogen in SNI spectra, and some properties of the Crab nebula. Its problems are that occasional complete mixing of MS stars must be postulated ad hoc, and that often SNI are associated with a much younger stellar population.
  
  \vspace{1mm}
  
  \item SNI are stars of intermediate mass, ${\sim} 4 - 6 \: \rm M_{\odot}$. This was suggested by \citet{Renzini1976OnSupergiants} on the grounds that these stars could blow away their hydrogen-rich envelopes before dying, and by \citet{Oemler1979TypeStars} to account for the association of SNI with ongoing star formation in spiral and irregular galaxies. This hypothesis implies a rate of star formation in elliptical galaxies that should be marginally detectable.
  
  \vspace{1mm}
  
  \item There are two kinds of SNI, those from low-mass stars that appear in elliptical galaxies, and those from more massive stars that account for the frequency in spiral and irregular galaxies \citep{Dallaporta1973OnSupernovae}. While this hypothesis avoids the problems of the preceding ones, the properties of SNI (while not exactly homogeneous) do not divide naturally into two classes suggestive of very different stellar origins.
\end{enumerate}

In any case, the production of iron by SNI and the red-giant nucleosynthesis mentioned above show that stars with a very wide range of masses contribute to galactic enrichment in various elements.

\subsubsection{Effects of initial composition}
\label{subsubsec:Initial composition}

Stars now forming and evolving in different regions of galaxies have at least a ten-fold range of metallicities, and the first stars everywhere were presumably very metal-poor. It is therefore important to ask how stellar evolution is affected by initial composition.

\medskip

The evolution of stars from ${\sim} 0.7 - 1 \: \rm M_{\odot}$ has been widely studied as a function of metallicity ($Z$) and helium abundance ($Y$), for the interpretation of HR diagrams of globular clusters \citep[e.g.][]{Iben1974PostStars, Mengel1979StellarSequence, Sweigart1978EvolutionaryStars}. Much less attention has been paid to more massive stars. In the largest systematic survey to date, stars up to $7 \: \rm M_{\odot}$ with a wide range of $Z$ and $Y$ have been followed to the base of the giant branch \citep{Mengel1979StellarSequence}; \citet{Sweigart1978EvolutionaryStars} have followed some of these up the giant branch. A few more massive models have been studied by \citet{Trimble1973EvolutionStars} and \citet{Harris1976EffectsClouds}, and \citet{Alcock1978TheStars} have followed the evolution of stars up to $10 \: \rm M_{\odot}$, with various $Z$, through core helium exhaustion. The results of these calculations suggest that important systematic differences in nucleosynthesis may occur as a result of initial $Z$ differences, but they do not go far enough to provide quantitative estimates. Models for chemical evolution usually assume that the production of primary elements -- those that are made from hydrogen and helium that were initially in the star -- is a function of stellar mass alone, but this assumption may be leading to systematic errors. (An exception is the study undertaken by \citealp{Arnett1971GalacticNucleosynthesis} of explosive nucleosynthesis of certain neutron-rich primary elements which may depend systematically on the star's initial $Z$; \citealp{Arnett1971GalacticNucleosynthesis} showed how such effects could be tested using relative abundances of metals in metal-poor stars).

\medskip

Some elements are believed to be synthesized in amounts that depend \emph{directly} on the initial composition of the star, because they are made not only from hydrogen and helium but also from heavier elements that were present initially. These are called secondary elements, and some examples are:

\begin{enumerate}
  \item $^{13}$C and $^{14}$N are made in stellar envelopes as by-products of hydrogen burning in the CNO cycle, mainly from $^{12}$C that was present initially\footnote{If any fresh $^{12}$C, a product of helium-burning in the star's own core, gets mixed to the hydrogen-burning region and processed in this way, the resulting $^{13}$C and $^{14}$N would be called primary products.}.
  
  \vspace{1mm}
  
  \item Barium is made from the addition of neutrons to iron nuclei, probably in the intershell region of asymptotic-branch giants \citep[e.g.][]{IbenI.1978OnMedium}.
\end{enumerate}

Because the structure of stars at different stages of evolution depends on their initial composition, these and other elements may also be affected \emph{indirectly} by initial composition.

Extensive calculations of stellar models lie ahead before the dependence of both primary and secondary nucleosynthesis on composition is adequately understood. Almost all calculations to date have used Solar relative abundances of the heavy elements (all scaled with the single initial metallicity parameter $Z$), and it will be important also to find out how stellar evolution and nucleosynthesis depend on variations among these relative abundances.

Another set of questions concerns changes in the HR diagram that would affect the colors of galaxies with stars of different compositions. A population of low $Z$ is in general bluer than normal for two reasons: the effective temperature of a star of a given mass and evolutionary stage is (in most cases of interest) higher at lower $Z$, because metal-poor stars tend to have smaller radii; and most colors at a given effective temperature are bluer at lower $Z$, because the reduced line-blanketing restores more flux at short wavelengths than at long wavelengths. These effects are partially offset when one considers populations of a given age, because the turnoff mass is lower at lower $Z$, but this factor is relatively unimportant. Some implications of these changes will be discussed in Section~\ref{subsec:Data2} and Section~\ref{subsubsec:Variations in metallicity}.

\section{Aims and Methods of Chemical Evolution}
\label{sec:Aims and Methods 1}

Studies of chemical evolution aim to account for abundance distributions of the elements, including the variation of stellar metallicities with age and position in the Galaxy, abundance gradients in galaxies, variations in relative abundances of elements heavier than helium, and related observations. The main processes governing chemical evolution are indicated in Figure~\ref{fig:fig1}: star formation, nucleosynthesis, mass loss from evolving and dying stars, and gas flows. Because of the importance of dynamical processes, another motivation for studying chemical evolution is that it leads to clues to the dynamical evolution of galaxies, as examples in Sections~\ref{sec:Evolution solar neighborhood}~and~\ref{sec:Chemical Evolution of Galaxies} will show. Details of the basic processes have been extensively reviewed elsewhere, so the emphasis here will be on methods of modeling chemical evolution, and selected applications. Reviews for further background material include \citet{Audouze1976ChemicalGalaxies.}, \citet{Faber1977ThePopulations}, \citet{Pagel1978a, Pagel1978b}, \citet{Trimble1975TheElements}, \citet{vandenBergh1975StellarGalaxies}, and others cited in Section~\ref{sec:Star_formation}. There is no clearer introduction to chemical evolution than the seminal paper by \citet{Schmidt1963TheMass.} on the subject.

\subsection{Basic Assumptions and Equations}
\label{subsec:Assumptions}

Models for chemical evolution follow, analytically or numerically, abundance changes in the ISM of a region and the resulting abundance distributions in stars. For most purposes, the region under study in a particular model can be assumed to have ISM of uniform composition and to gain or lose mass through gas flows only; complicated models for whole galaxies are generally divided up, for computation, into zones with these properties. The basic equations are then easy to derive.

The total mass $M$ of the system changes according to the net inflow rate $f$ of gas:
\begin{equation}
    \frac{dM}{dt} = f.
	\label{eq:eq3.1}
\end{equation}
($f$ is often called the infall or accretion rate). The mass of stars $M_{\rm s}$ changes via star formation and mass loss from evolving stars:
\begin{equation}
    \frac{dM_{\rm s}}{dt} = \psi - E,
	\label{eq:eq3.2}
\end{equation}
where $\psi$ is the SFR and $E$ is the total ejection rate from stars of all masses and ages. Similarly, the mass of ISM, called ``gas'' ($M_{\rm g}$), changes through star formation, ejection, and net inflow into the region:
\begin{equation}
    \frac{dM_{\rm g}}{dt} = -\psi + E + f,
	\label{eq:eq3.3}
\end{equation}
Since $M = M_{\rm s} + M_{\rm g}$, the sum of Equations~(\ref{eq:eq3.2})~and~(\ref{eq:eq3.3}) is simply (\ref{eq:eq3.1}).

The fraction of gas in the system is written
\begin{equation}
    \mu \equiv \frac{M_{\rm g}}{M},
	\label{eq:eq3.4}
\end{equation}
so the mass of stars is obviously
\begin{equation}
    M_{\rm s} = \left( 1 - \mu \right) M.
	\label{eq:eq3.5}
\end{equation}

The ejection rate can be written in terms of the IMF and SFR, Equation~(\ref{eq:eq2.1}), if one makes the usual approximation that each star undergoes its entire mass loss after a well-defined lifetime. Let $w_{\rm m}$ be the remnant mass and $\tau_{\rm m}$ the lifetime of a star of (initial) mass $m$; the star was therefore formed at time ($t - \tau_{\rm m}$) if it dies at time $t$. Then the ejection rate at time $t$ is
\begin{equation}
    E(t) = \int_{m_{\rm t}}^{\infty} \left( m - w_{\rm m} \right) \psi \left( t - \tau_{\rm m} \right) \phi(m) \ dm,
	\label{eq:eq3.6}
\end{equation}
where $m_{\rm t}$ is the mass with $\tau_{\rm m} = t$ (the turnoff mass) and $\phi(m)$ is the IMF at time ($t - \tau_{\rm m}$). The approximation of sudden mass loss at the end of each star's lifetime is usually a reasonable one, since for most stars nearly all of the mass loss is thought to occur in a final small fraction of their lives; mass loss on the MS is important for O stars, however, so the approximation fails if timescales $< 10^{7} \: \rm yr$ are of interest.

\medskip

For illustration, let us consider the evolution of a single ``metal'' abundance parameter $Z$, where ``metals'' are one or all of the common primary elements (e.g., C, O, Fe), and let us assume that their production is a function only of stellar mass, not of initial composition. (The methods for calculating this schematic $Z$ are readily adapted to more realistic cases). Let us also assume for simplicity that material ejected from stars is mixed instantly throughout the ISM in the region under study. Then the mass of metals in the gas, which is $ZM_{\rm g}$, evolves via star formation (putting metals from the ISM into stars), ejection, and gas flows, according to the equation
\begin{equation}
    \frac{dZM_{\rm g}}{dt} = -Z \psi + E_{Z} + Z_{\rm f} f,
	\label{eq:eq3.7}
\end{equation}
where $E_{Z}$ is the total ejection rate of metals from stars and $Z_{\rm f}$ is the mean metal abundance in infalling gas. $E_{Z}$ includes both newly synthesized metals and those that were in the star from birth and re-ejected. Let $p_{\rm zm}$ be the mass fraction of a star of mass $m$ that is converted to metals and ejected; then the rate of ejection of new metals from stars at time $t$ is
\begin{equation}
    \int_{m_{\rm t}}^{\infty} m p_{\rm zm} \psi \left( t - \tau_{\rm m} \right) \phi(m) \ dm.
	\label{eq:eq3.8}
\end{equation}
Unprocessed material occupies a mass ($m - w_{\rm m} - m p_{\rm zm}$) of the ejected part of a star of mass $m$, and the metal abundance in this region is $Z(t - \tau_{\rm m})$. Thus the total ejection rate of old and new metals is obtained by adding the ejection rate from all masses of these unprocessed metals to the rate (Equation~\ref{eq:eq3.8}):
\begin{equation}
\begin{medsize}
    E_{Z}(t) = \int\limits_{m_{\rm t}}^{\infty} \left[ \left( m - w_{\rm m} - m p_{\rm zm} \right) Z \left( t - \tau_{\rm m} \right) + m p_{\rm zm} \right] \psi \left( t - \tau_{\rm m} \right) \phi(m) \ dm.
	\label{eq:eq3.9}
\end{medsize}
\end{equation}

The mean metallicity of stars ever formed, denoted $Z_{\rm s}$, can be obtained by writing an equation for the conservation of metals: a mass $Z_{\rm s} M_{\rm s}$ of metals is stored in stars, a mass $Z M_{\rm g}$ is in the gas, and their sum is the mass of new metals ever ejected, which is obtained by integrating Equation~(\ref{eq:eq3.8}) over time. Thus $Z_{\rm s}$ is given by
\begin{equation}
    Z_{\rm s} M_{\rm s} + Z M_{\rm g} = \int_{0}^{t} \int_{m_{\rm t^{\prime}}}^{\infty} m p_{\rm zm} \psi \left( t^{\prime} - \tau_{\rm m} \right) \phi(m) \ dt^{\prime} dm.
	\label{eq:eq3.10}
\end{equation}

\medskip

It is convenient to define some integrals that depend on the IMF and parameters of stellar evolution but not on $\psi(t)$. The \emph{returned fraction} $R$ is defined by the integral
\begin{equation}
    R \equiv \int_{m_{1}}^{\infty} \left( m - w_{\rm m} \right) \phi(m) \ dm,
	\label{eq:eq3.11}
\end{equation}
where $m_{1}$ is the present turnoff mass, and is usually taken to be $1 \: \rm M_{\odot}$ in evaluating $R$. Because the IMF is normalized as in Equation~(\ref{eq:eq2.2}), $R$ is the fraction of mass put into stars at a given time that is thereafter returned to the ISM in the lifetime of a solar-mass star; for example, if the stars in a galaxy were all formed in a single initial burst with a total mass $M$, then ${\sim} 10^{10} \: \rm yr$ later the mass in stellar form would be $M(1 - R)$. An estimate of $R$ can be obtained using the local IMF as approximated in Equation~(\ref{eq:eq2.9}), with a white dwarf mass of $0.7 \: \rm M_{\odot}$ as the remnant mass $m_{\rm w}$ for $m \leq 4 \: \rm M_{\odot}$ and a neutron star mass of $1.4 \: \rm M_{\odot}$ for $m > 4 \: \rm M_{\odot}$; the result is, $R = 0.17 \times (10 \: \rm M_{\odot} \ pc^{-2} \ Gyr^{-1} / \psi_{1})$.

A second useful integral is the \emph{yield} $y$, which is the mass of new metals ejected (eventually) when unit mass of matter is locked into stars. The yield can be defined for any element of interest, and for the ``metals'' considered above it is given by
\begin{equation}
    y \equiv \frac{1}{1 - R} \int_{m_{1}}^{\infty} m p_{\rm zm} \phi(m) \ dm.
	\label{eq:eq3.12}
\end{equation}
Again, if a mass $M$ of stars were formed in a single initial burst, then ${\sim} 10^{10} \: \rm yr$ later, the mass of new metals ejected would be $y(1 - R)$. Sometimes the quantities $p_{\rm zm}$ are called stellar yields, and $y$ is called the net yield.

Although $R$ and $y$ are defined with a specific lower mass limit, their values do not depend strongly on $m_{1}$, so they turn out to be useful parameters in many situations. For example, $R$ has been used in Section~\ref{subsubsec:Local SFR} to relate the mass of stars now in the Solar neighborhood to the mass ever formed, giving a reasonable estimate even though star formation did not occur in a single burst; and the yield will prove to be a first approximation to the abundance of metals resulting from a given IMF and stellar production parameters (e.g., $p_{\rm zm}$), independently of the detailed history of the system.

Many insights into chemical evolution can be obtained using analytical approximations to the above equations, while for detailed modeling they can be cast into forms suitable for numerical computation. These techniques will be outlined in turn.

\subsection{Analytical Approximations}
\label{subsec:Approximations}

The equations of chemical evolution are especially easy to handle if one makes an approximation known as \emph{instantaneous recycling}: stars are divided into two classes, those that live forever (masses less than some value $m_{1}$) and those that die as soon as they are born ($m > m_{1}$). Thus the lower mass limits in the integrals in Equations~(\ref{eq:eq3.6})~--~(\ref{eq:eq3.10}) are taken to be the time-independent $m_{1}$, and the assumed instant deaths allow the arguments ($t - \tau_{\rm m}$) to be replaced by simply ($t$). (Note that the assumption now made of death immediately after birth, so $\tau_{\rm m} \equiv 0$, is a much stronger and less realistic assumption than the earlier one of sudden mass loss, which simply allowed $\tau_{\rm m}$ to be well-defined for each star). Equations~(\ref{eq:eq3.6})~and~(\ref{eq:eq3.9}) reduce at once to the much simpler relations,
\begin{equation}
    E(t) = R \psi(t),
	\label{eq:eq3.13}
\end{equation}
\begin{equation}
    E_{Z}(t) = R Z(t) \psi(t) + y(1 - R) \left[ 1 - Z(t) \right] \psi(t),
	\label{eq:eq3.14}
\end{equation}
in which all time-dependent quantities are evaluated at the current time $t$ only. Since $Z \ll 1$ in all situations of interest, Equation~(\ref{eq:eq3.14}) can be written more simply
\begin{equation}
    E_{Z}(t) = R Z(t) \psi(t) + y(1 - R) \psi(t).
	\label{eq:eq3.15}
\end{equation}
Substituting in Equations~(\ref{eq:eq3.2}),~(\ref{eq:eq3.3}),~and~(\ref{eq:eq3.7}), we have now
\begin{equation}
    \frac{dM_{\rm s}}{dt} = (1 - R) \psi,
	\label{eq:eq3.16}
\end{equation}
\begin{equation}
    \frac{dM_{\rm g}}{dt} = -(1 - R) \psi + f,
	\label{eq:eq3.17}
\end{equation}
\begin{equation}
    \frac{dZM_{\rm g}}{dt} = -Z(1 - R) \psi + y(1 - R) \psi + Z_{\rm f} f.
	\label{eq:eq3.18}
\end{equation}
An equation for the metal abundance $Z$ of the gas, rather than the mass of metals it contains, is obtained by substituting Equation~(\ref{eq:eq3.17}) into (\ref{eq:eq3.18}), using the identity $M_{\rm g} dZ / dt = d(ZM_{\rm g}) / dt - Z dM_{\rm g} / dt$; the result is
\begin{equation}
    M_{\rm g} \frac{dZ}{dt} = y(1 - R) \psi + \left( Z_{\rm f} - Z \right) f.
	\label{eq:eq3.19}
\end{equation}

For simplicity, let us assume that the IMF is constant. (The methods developed for a constant IMF can be adapted when necessary to more complicated cases). The quantities $R$ and $y$ are therefore constants.

Equation~(\ref{eq:eq3.16}) now gives an expression for the mass of stars at time $t$,
\begin{equation}
    M_{\rm s} (t) = (1 - R) \int_{0}^{t} \psi \left( t^{\prime} \right) \ dt^{\prime} \equiv (1 - R) \overline{\psi} t.
	\label{eq:eq3.20}
\end{equation}
The mean metallicity of stars can be obtained from Equation~(\ref{eq:eq3.10}), using the instantaneous recycling approximation and the assumption of a constant IMF: the integrals with respect to time and mass separate, and are simply $M_{\rm s} / (1 - R)$ and $y(1 - R)$ respectively, so the result is $Z_{\rm s} M_{\rm s} + Z M_{\rm g} = y M_{\rm s}$, from which
\begin{equation}
    Z_{\rm s} = y - \frac{\mu}{1 - \mu} Z.
	\label{eq:eq3.21}
\end{equation}
This result shows that $Z \rightarrow y$ as $\mu \rightarrow 0$, which simply states that, when there is no gas, all the metals ever made and ejected ($y M_{\rm s}$) are incorporated into later generations of stars. An estimate of the yield for the local IMF is therefore the mean metallicity of stars in the Solar neighborhood (where the gas fraction is small, $\mu \sim 0.05$), i.e., $y \simeq 0.8 \: \rm Z_{\odot}$. The approximation $y \sim \: \rm Z_{\odot}$ will be useful below.

Further solutions to the above equations depend on the assumptions made about gas flows, so two extreme cases will be considered for illustration.

\subsubsection{A closed system, initially unenriched gas}
\label{subsubsec:closed system}

In this case, we set $f = 0$, and the initial values are $M_{\rm g0} = M$, $M_{\rm s0} = 0$, $Z_{0} = Z_{\rm s0} = 0$. Obviously, $M = \rm constant$. Time can be eliminated as an explicit variable if Equation~(\ref{eq:eq3.19}) is divided by (\ref{eq:eq3.17}), with the result
\begin{equation}
    M_{\rm g} \frac{dZ}{dM_{\rm g}} = -y,
	\label{eq:eq3.22}
\end{equation}
i.e.,
\begin{equation}
    Z = y \ln \left( \frac{M_{\rm g0}}{M_{\rm g}} \right) = y \ln \left( \frac{M}{M_{\rm g}} \right) = y \ln \left( \frac{1}{\mu} \right).
	\label{eq:eq3.23}
\end{equation}
This solution is valid as long as $Z \ll 1$. If not, Equation~(\ref{eq:eq3.14}) must be used instead of (\ref{eq:eq3.15}); the result is then $Z = 1 - \mu^{y}$, which reduces to (\ref{eq:eq3.23}) if $Z \ll 1$, and has the limit $Z \rightarrow 1$ as $\mu \rightarrow 0$. (This formally correct limit is of course absurd, and is a consequence of neglecting the effects of metallicity on stellar nucleosynthesis). Original derivations of these results were given by \citet{Talbot1971TheModel}, and Equation~(\ref{eq:eq3.23}) was derived independently by \citet{Searle1972InferencesGalaxies}. Through Equations~(\ref{eq:eq3.21})~and~(\ref{eq:eq3.23}), $Z_{\rm s}$ can also be expressed as a function of $\mu$.

\subsubsection{A system with infall balanced by star formation}
\label{subsubsec:balanced system}

Here it is assumed that star formation just keeps up with the rate of infall plus stellar gas loss:
\begin{equation}
    \psi = f + R \psi,
	\label{eq:eq3.24}
\end{equation}
so that, from Equation~(\ref{eq:eq3.17}),
\begin{equation}
    M_{\rm g} = {\rm constant} = M_{0}.
	\label{eq:eq3.25}
\end{equation}
The picture is that the total mass grows by infall, while star formation maintains a constant gas mass. Other initial conditions are $M_{\rm s0} = 0$ and $Z_{0} = Z_{\rm s0} = 0$, and the infalling gas will be assumed to be unenriched so that $Z_{\rm f} = 0$. To eliminate time in this case, we divide Equation~(\ref{eq:eq3.17}) by (\ref{eq:eq3.1}), with the result
\begin{equation}
    M_{\rm g} \frac{dZ}{dM} = y - Z.
	\label{eq:eq3.26}
\end{equation}
It is useful to solve this Equation in terms of a parameter $\nu$ that gives the ratio of mass accreted to the initial mass:
\begin{equation}
    \nu \equiv \frac{M - M_{0}}{M_{0}} = \frac{M - M_{\rm g}}{M_{\rm g}} = \frac{1}{\mu} - 1.
	\label{eq:eq3.27}
\end{equation}
Then Equation~(\ref{eq:eq3.26}) has the solution
\begin{equation}
    Z = y \left(1 - e^{-\nu} \right),
	\label{eq:eq3.28}
\end{equation}
a result due originally to \citet{Larson1972b}. Equation~(\ref{eq:eq3.21}) again gives $Z_{\rm s}$. An important difference between this infall model and the preceding closed model is that here $Z \rightarrow y$ as $\mu \rightarrow 0$; an equilibrium is set up between the rate of infall of metal-free gas and the rate of enrichment by evolving stars.

\subsubsection{Generalities}
\label{subsubsec:generalities}

Although the above two models are extremely schematic, their Equations~(\ref{eq:eq3.23})~and~(\ref{eq:eq3.28}) for the ``metal'' abundance of the gas have some common properties that are found in a wide variety of models for chemical evolution. The following generalizations are approximately true even when instantaneous recycling is a poor approximation, but they break down if the IMF is time-dependent.

\begin{enumerate}
  \item \emph{$Z$ is proportional to the net yield, $y$.} Thus any primary elements whose stellar production parameters are independent of composition will have abundances in proportion to their respective yields. The yields in turn can be calculated from Equation~(\ref{eq:eq3.12}), for a given IMF, independently of any model for star formation or gas flows in the system. The utility of this result is that theories of nucleosynthesis can be tested by comparing predicted relative yields of various elements with their observed relative abundances.
  
  \vspace{1mm}
  
  \item \emph{$Z / y$ depends chiefly on current properties of the system}, and is insensitive to its past history. This statement is exactly true in the above models because of the assumption of instantaneous recycling (which led to the cancellation of time as an explicit variable). In numerical models that allow for finite stellar lifetimes, the same result holds  approximately, and relevant current properties include the gas fraction ($\mu$) and the ratio of SFR to gas flow rate ($\psi / f$).
  
  \vspace{1mm}
  
  \item \emph{$Z / y$ depends rather weakly on model-dependent quantities} such as $\mu$ and $\psi / f$. Consequently, regions of galaxies with gas fractions differing by orders of magnitude may have interstellar abundances differing by only small factors. This prediction allows one to test whether theories of nucleosynthesis give the right absolute amounts of elements, as well as the right relative amounts; in particular, the yield of an element, calculated for the local IMF, should be in order of magnitude equal to its abundance in the Solar System or nearby stars.
\end{enumerate}

\subsection{Numerical Models}
\label{subsec:Numerical Models}

Chemical evolution is best studied numerically in cases where approximations that make the analytical approach transparent (or possible) break down. Situations in which instantaneous recycling is no longer a useful approximation include the following:

\begin{enumerate}
  \item If the SFR is a strongly decreasing function of time, systematic errors result from setting $\psi (t - \tau_{\rm m}) = \psi(t)$; in particular, low-mass stars formed early die later in much greater numbers (relative to massive stars) than would be predicted on the basis of the current SFR.
  
  \vspace{1mm}
  
  \item If one is interested in time-dependent abundance ratios arising from nucleosynthesis in stars of different lifetimes, the effects would be entirely lost by neglecting finite stellar lifetimes. Even if instantaneous recycling can be assumed, the analytical approach becomes intractable in models with time-dependent IMFs, or in considerations of radioactive elements with lifetimes of billions of years, except in very special cases.
\end{enumerate}

For numerical computation, equations such as (\ref{eq:eq3.1})~--~(\ref{eq:eq3.10}) are expressed as differences in finite steps of time and sums over a grid of stellar masses. The programming required is little more than careful book-keeping, so the effort in models for chemical evolution goes not into techniques but into the astrophysical input. Many examples of numerical models will be found in the references cited in Sections~\ref{sec:Evolution solar neighborhood}~and~\ref{sec:Chemical Evolution of Galaxies}, in the discussion of particular problems.

\section{Chemical Evolution in the Solar Neighborhood}
\label{sec:Evolution solar neighborhood}

The extensive data available on chemical compositions of nearby stars and \textsc{H~ii} regions, and details of the isotopic composition of Solar System material, have led to many efforts to build consistent models for chemical evolution in the Solar neighborhood. In trying to explain the abundance distributions, these models also shed light on dynamical processes that have affected the local region of the Galaxy, and they set some constraints on theories of nucleosynthesis.

The ``neighborhood'' must be carefully defined to include the range of environments occupied by nearby stars and their predecessors in their motions around the Galaxy, and to include stars that were formed near the plane but now occupy a thicker disk. A small ``local swimming hole'' around the Sun (in Baade's words) will not do. Instead, the Sun's ``nucleogenetic pool'' is usefully defined as a region about a kpc wide in the Galactic plane (to include most stellar epicyclic motions and periodic gas motions), extended around the Galaxy at the Sun's galactocentric distance (to include passages of local material through spiral arms and related small-scale patchiness), and extending about a kpc out of the plane (to include stars whose periodic motions carry them far from the plane). This region is shaped like a cylindrical shell, and quantities like the mass of stars, SFR, etc. are expressed in units per square parsec, referring to \emph{an average column at the Solar galactocentric distance}. Spatial averages over this pool can be interpreted as short-term time averages during the history of a given small amount of material, and models generally consider explicitly only timescales longer than passages between spiral arms, etc. (Some special studies, such as the interpretation of elements with radioactive lifetimes ${\sim} 10^{6} - 10^{9} \: \rm yr$, do need to consider short-term changes). Although the region has been defined to include most relevant periodic motions, it cannot be treated as a closed system, because it is probably affected by systematic radial gas flows in the disk and by infall from outside, especially at early times.

Halo stars are sometimes treated as part of the ``neighborhood'' (in which case its extent out of the plane must be many kpc), but more often the halo is treated as a separate subsystem of the Galaxy, whose effects on the evolution of the disk are perturbations from outside. Properties of halo stars are in any case relevant to the Solar neighborhood, so they will be discussed along with the properties of disk stars.

\subsection{Outline of Relevant Data}
\label{subsec:Data1}

Chemical abundances in the Solar neighborhood and in the Solar System have been reviewed in detail by \citet{Trimble1975TheElements} and several authors at a recent conference at Yale University. An outline of the points most relevant to this article will be given here. Because of systematic radial gradients in chemical abundances in the Galaxy (Section~\ref{subsec:Data2}), the following remarks apply only to stars within a few kpc of the Sun's galactocentric distance.

\begin{figure}
	\includegraphics[width=0.47\textwidth]{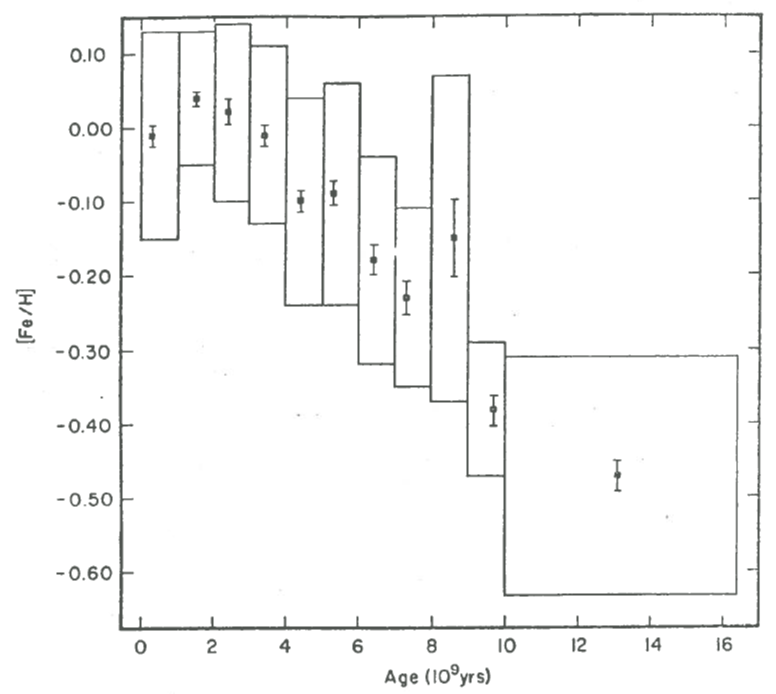}
    \caption{The relation between age and metallicity for disk stars in the Solar neighborhood \citep{Twarog1980AnNeighborhood}. Boxes denote the range of ages referred to by each point, and the $\pm 1 \sigma$ spread in $\rm [Fe / H]$ for the age group. Error bars attached to the points denote the standard error in the mean. Note that the mean metallicity increases by a factor of less than four during the lifetime of the disk.}
    \label{fig:fig7}
\end{figure}

Stars with halo kinematics are all metal-poor, with $\rm [Fe / H] \lesssim -1$, whereas stars with disk motions nearly all have $\rm [Fe / H] \gtrsim -0.5$. (As usual, abundance ratios in square brackets are logarithms of the ratios relative to their Solar values). Since most halo stars are probably older than most disk stars (Section~\ref{subsec:Stellar_Populations}), their composition differences indicate a systematic increase of metallicity with time of formation. The disk stars alone show a systematic increase of metallicity with time of formation, illustrated in Figure~\ref{fig:fig7}, but the trend is only about a factor of three over the whole lifetime of the disk. Open clusters show a very weak dependence of metallicity on age, if any, because they are found at such a wide range of galactocentric distances that the radial gradient in metallicity obscures the time-dependence.

A powerful tool for chemical evolution is the metallicity distribution of all stars ever formed, which is given by samples of G--M dwarfs since these stars have remained on the MS since they formed. The distribution can be roughly approximated by a log-normal function, with a mean $\langle \left[ \rm Fe / H \right] \rangle = -0.1$ and standard deviation $\sigma \left( \left[ \rm Fe / H \right] \right) = 0.2$, allowing for observational errors (\citealp{Pagel1975MetalNeighbourhood, Pagel1978a}; and references therein). Its striking characteristic is the extreme paucity of metal-poor stars: only a few percent of stars in the local cylinder (including halo stars) have $\rm [Fe / H] < -1$. The lack of metal-poor disk stars proves to be one of the most powerful constraints on models for the Solar neighborhood, as discussed below.

Stellar ``metallicities'' are commonly denoted $\rm [Fe / H]$ because most photometric indices of the abundance of heavy elements are sensitive to blanketing by numerous iron lines. Detailed abundance analyses show that other elements heavier than helium usually occur in fixed proportions, so a single metallicity index is often adequate. However, the exceptions to this uniformity are often of great interest. Certain abundance variations in \emph{evolved} stars are believed to show products of the stars' own nucleosynthesis that are mixed to the surface and lost in a stellar wind (Section~\ref{subsec:Beyond MS}) -- direct observational evidence for enrichment of the ISM. Some elements also show relative abundance variations in \emph{unevolved} stars, and in these cases the anomalies must have been present in the ISM from which the star formed.

\medskip

Besides these interesting abundance variations, there are others that should be ignored or ``corrected'' for in the context of chemical evolution. Extraneous effects include depletion of metals in the diffuse interstellar gas due to their incorporation into grains; anomalous interstellar abundance ratios, especially between isotopes of the same element, caused by chemical fractionation\footnote{This is a case of real chemistry, not in the special sense of ``chemical'' evolution of galaxies. Other terminology has been suggested to avoid the term ``chemical evolution'' now that real chemistry of the ISM is becoming an important topic in astronomy, but the alternative suggestions have their own problems. For example, ``nuclear evolution'' can be confused with the evolution of the nuclei (centers) of galaxies, and it obscures the great importance of gas flows, in addition to nucleosynthesis, in determining the evolution of chemical abundances in galaxies.}; and large abundance variations of various elements on the surfaces of Ap and related stars, produced by differential diffusion of elements in the stellar atmospheres \citep{Trimble1975TheElements}. Unfortunately, it is not always clear whether a particular composition anomaly is due to some such extraneous effect, to nucleosynthesis in the star observed, or to significant composition changes in the ISM.

\subsection{The ``G-Dwarf Problem''}
\label{subsec:G-Dwarf}

The metallicity distribution of long-lived disk dwarfs leads to a problem originally discovered by \citet{vandenBergh1962TheAbundances.} and \citet{Schmidt1963TheMass.}: there are far fewer metal-poor stars than one predicts using the most straightforward assumptions. This contradiction is called the ``G-dwarf problem'', although a stellar metallicity distribution has now been confirmed for M dwarfs \citep{Mould1978InfraredDwarfs}. The straightforward assumptions involved are the following:

\begin{enumerate}
  \item the Solar neighborhood can be modeled as a closed system;
  
  \vspace{1mm}
  
  \item it started as $100 \%$ metal-free gas;
  
  \vspace{1mm}
  
  \item the IMF is constant; and
  
  \vspace{1mm}
  
  \item the gas is chemically homogenous at any time.
\end{enumerate}

Any model based on these assumptions is called a \emph{``simple'' model}. It is now recognized that assumptions (i) and (ii) are almost certainly false and that more realistic alternatives lead to a natural solution of the G-dwarf problem; assumptions (iii) and (iv) are probably also unrealistic, but there is less evidence for believing that they are violated sufficiently to solve the problem entirely. Although the G-dwarf problem is no longer regarded as a difficulty in chemical evolution, it will be reviewed here because of its earlier importance in motivating many models. Moreover, a review of this problem serves to emphasize that simple models cannot be used consistently in any arguments relating to chemical evolution in the Solar neighborhood -- the G-dwarf metallicities are so much at variance with such models that they must ignore very important effects, and could lead to highly misleading results in other contexts. This conclusion also leads to suspicion of simple models for other regions of galaxies.

\medskip

The problem itself can be derived analytically, following \citet{Pagel1975MetalNeighbourhood}, using the instantaneous recycling approximation and the results for a simple model in Section~\ref{subsubsec:closed system}. Let us consider the cumulative metallicity distribution of stars ever formed (which is represented observationally by dwarfs of mid-G type and later). From Equation~(\ref{eq:eq3.5}), the fraction of all stars that had been made while the gas fraction was $\geq \mu$ is
\begin{equation}
    \frac{M_{\rm s}}{M_{\rm s1}} = \frac{1 - \mu}{1 - \mu_{1}},
	\label{eq:eq4.1}
\end{equation}
where subscripts 1 denote present values. From Equation~(\ref{eq:eq3.23}), these stars were formed while the metallicity of the gas was $\leq y \ln \ \mu^{-1}$. The fraction of stars with metallicities $\leq Z$, denoted $S(Z)$, is therefore
\begin{equation}
    S(Z) = \frac{M_{\rm s}}{M_{\rm s1}} = \frac{1 - \exp(-Z / y)}{1 - \mu_{1}}.
	\label{eq:eq4.2}
\end{equation}
It is useful to eliminate the yield using the relation $y = Z_{1} / (\ln \ \mu_{1}^{-1})$ (again Equation~\ref{eq:eq3.23}), so Equation~(\ref{eq:eq4.2}) can finally be written in terms of the metallicity ratio $Z / Z_{1}$ and the present gas fraction $\mu_{1}$:
\begin{equation}
    S(Z) = \frac{1 - \mu_{1}^{Z / Z_{1}}}{1 - \mu_{1}}.
	\label{eq:eq4.3}
\end{equation}

\begin{figure}
	\includegraphics[width=0.47\textwidth]{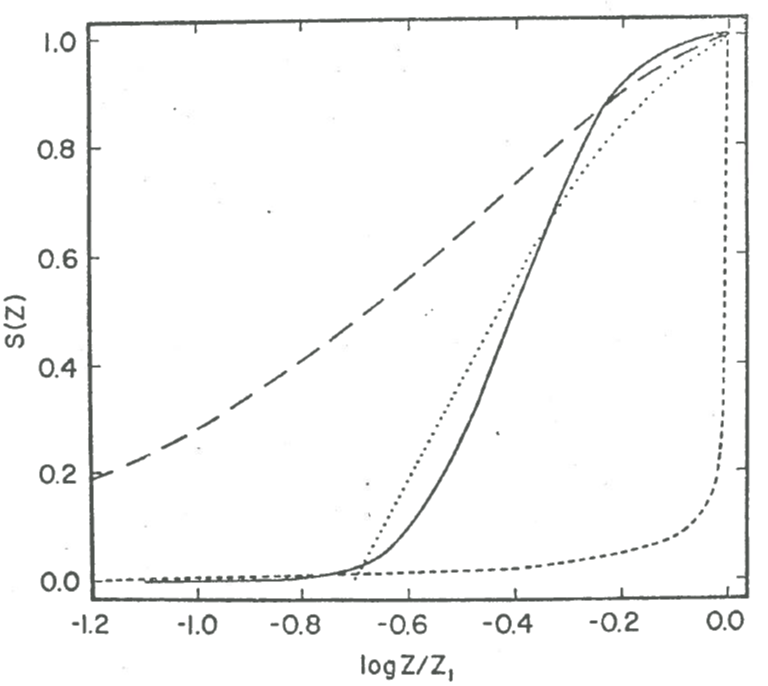}
    \caption{Cumulative stellar metallicity distribution functions. $S(Z)$ is the fraction of stars that have metallicities $\leq Z$, with a maximum value $Z_{1}$. \emph{Solid line}: log-normal representation of the data for stars in the Solar neighborhood; $Z_{1}$ is taken as $2 \: \rm Z_{\odot}$ since more metal-rich stars are very rare; $\langle \log \ Z / \rm Z_{\odot} \rangle = -0.1$, and $\sigma( \log \ Z / \rm Z_{\odot}) = 0.15$, allowing for observational errors \citep[e.g.][]{Pagel1978a}. \emph{Long dashes}: the ``simple'' model for chemical evolution (Section~\ref{subsec:G-Dwarf}). \emph{Short dashes}: an extreme infall model (Section~\ref{subsubsec:infall}). \emph{Dots}: a model with a finite initial metallicity (Section~\ref{subsubsec:pre-enrichment}).}
    \label{fig:fig8}
\end{figure}

This distribution is compared in Figure~\ref{fig:fig8} with a log-normal approximation to the data. (The model illustrated has $\mu_{1} = 0.05$, and the shape of the curve is insensitive to $\mu_{1}$ within the range of likely values). Obviously, there are far fewer metal-poor stars in the Solar neighborhood than predicted by this formula. The assumptions defining the simple model allow only two significant quantities to define a specific model: the IMF and the SFR. However, the IMF affects only the yield, which does not appear in Equation~(\ref{eq:eq4.3}) because metallicities have been scaled to the present value; the SFR was eliminated in the derivation of Equation~(\ref{eq:eq3.23}), and numerical models show that very nearly the same result is obtained if the approximation of instantaneous recycling is dropped. Thus the failure of the ``simple'' class of models is independent of the particular IMF or SFR used. It may seem at first sight that one could eliminate the problem by making the metals rapidly using a large SFR at early times; however, since there is initially unenriched gas and a constant IMF, a fixed quota of metal-poor G dwarfs must form along with the massive stars that produce metals -- no matter how rapidly it all happens.

Attempts to solve the G-dwarf problem have motivated numerous alternative models for the Solar neighborhood, in which one or more of the four straightforward assumptions are dropped. The following outline considers each assumption in turn, and some types of model will be developed from a different viewpoint in Section~\ref{subsec:Effects of Galaxy Formation}. Further details and references, on these and other models, can be found in reviews by \citet{Audouze1976ChemicalGalaxies.}, \citet{Pagel1975MetalNeighbourhood}, \citet{Tinsley1974ConstraintsNeighborhood} and \citet{Trimble1975TheElements}.

\subsubsection{Infall}
\label{subsubsec:infall}

It is unlikely that the region called the ``Solar neighborhood'' has evolved as a closed system. In particular, the disk could have grown gradually by accretion, on a timescale similar to that for star formation, as proposed by \citet{Larson1972a} and \citet{Sciama1972OnChromosphere}. The effect on the stellar metallicity distribution is then as follows \citep{Larson1972b}. Suppose the disk consists initially of unenriched gas, as in the simple case, but with a very small mass (surface density). The first stars formed are metal-poor, but their numbers are proportional only to the initial disk mass; star formation substantially enriches the gas while the disk is still far below its final mass. Then further star formation continually enriches the new gas that is accreted, so ultimately the only very metal-poor stars are the few initial ones.

The model in Section~\ref{subsubsec:balanced system} illustrates this behavior analytically. According to Equation~(\ref{eq:eq3.28}), $Z$ remains essentially constant at the yield value after the disk mass has increased to several times its initial value. The stellar metallicity distribution, $S(Z)$, can be derived analytically for this model \citep{Tinsley1975a}, and the result is a function strongly peaked at $Z = y$, as shown in Figure~\ref{fig:fig8}. Here we have a problem opposite to that of the simple model: there are too few stars in the model with $Z < Z_{1}$! The problem is alleviated if the width of the observed distribution is due more to errors than has been thought, but in any case this particular model is very schematic, showing effects of infall in a most extreme way, and more realistic models with infall (Section~\ref{subsec:Effects of Galaxy Formation}) predict metallicity distributions intermediate between the extreme and the simple cases.

\subsubsection{Pre-enrichment of the disk gas}
\label{subsubsec:pre-enrichment}

If the gas that formed the disk had a significant initial metal abundance, the number of more metal-poor disk stars would obviously be zero. This idea was introduced by \citet{Truran1971EvolutionaryGalaxy}, who postulated a pregalactic burst of massive stars. \citet{Ostriker1975GalacticHalos} considered the effects of massive stars in the young halo, whose ejecta enriched the gas that was condensing into a disk; the likely importance of this effect will be reviewed below (Section~\ref{subsubsec:Metals from the young halo}).

The effects of pre-enrichment can be seen analytically by simply adding a finite initial metal abundance $Z_{0}$ to the solution of Equation~(\ref{eq:eq3.22}). Then (\ref{eq:eq3.23}) becomes $Z = Z_{0} + y \ln \ \mu^{-1}$, and the exponent $Z / Z_{1}$ in (\ref{eq:eq4.3}) is replaced by $(Z - Z_{0}) / (Z_{1} - Z_{0})$. Figure~\ref{fig:fig8} shows the resulting stellar metallicity distribution in the case $Z_{0} = 0.2 Z_{1}$, and there is clearly no excess of metal-poor stars. A more realistic picture will be discussed in Section~\ref{subsec:Effects of Galaxy Formation}.

\subsubsection{Variable IMF}
\label{subsubsec:Variable IMF}

The original solution to the G-dwarf problem, proposed by \citet{Schmidt1963TheMass.}, was to postulate an IMF richer in massive stars at early times; the number of long-lived stars formed before the gas reaches a given metallicity is therefore reduced. (An analytical solution illustrating this effect is given by \citealp{Pagel1975MetalNeighbourhood}). Many alternative models with variable IMFs have been considered subsequently, but they all suffer from being ad hoc: although arguments can be offered for massive stars forming more easily at low metallicity, they are not compelling, and there is no independent way of verifying suggested changes in the IMF in the distant past.

\subsubsection{Metal-enhanced star formation}
\label{subsubsec:Metal-enhanced star formation}

\citet{TalbotRaymondJ.1973TheFormation} suggested that there are always large chemical inhomogeneities in the ISM, and that stars form preferentially in regions with above-average metallicity. \citet{TalbotRaymondJ.1974SensitivityElements} has developed this idea into a physical model, in which the enhanced cooling by heavy elements favors star formation in metal-rich regions. For this process alone to solve the G-dwarf problem, it would require both very large chemical inhomogeneities and very inefficient star formation in metal-poor regions; more plausibly, metal-enhanced star formation leads only to some reduction in the proportion of metal-poor stars.

\medskip

In summary, none of the four assumptions underlying the simple model is likely to be strictly true, and departures from any or all of them could alter the predicted metallicity distribution in the required sense of including fewer metal-poor stars.

\subsection{Effects of Galaxy Formation}
\label{subsec:Effects of Galaxy Formation}

In retrospect, the failure of the simple model is not surprising, because the Solar neighborhood was treated as a closed box, whereas gas flows during the formation of the Galaxy would affect chemical evolution strongly in two ways:

\begin{enumerate}
  \item some pre-enrichment of the initial disk gas by massive stars in the young halo, and
  
  \vspace{1mm}
  
  \item later accretion of metal-poor gas.
\end{enumerate}

The ``formation'' of the Galaxy could in the latter sense be continuing today.

\subsubsection{Metals from the young halo}
\label{subsubsec:Metals from the young halo}

The amount of initial enrichment by massive halo stars can be estimated from the present mass of long-lived stars in the halo. Let $y_{\rm h}$ be the yield during the time of halo formation. Then, by definition of the yield, a mass $M_{\rm h}$ of surviving halo stars implies that a mass ${\sim} y_{\rm h} M_{\rm h}$ of metals was made and ejected at an early time; of these metals, a mass $Z_{\rm h} M_{\rm h}$ remains locked in halo stars (where $Z_{\rm h}$ is their mean metallicity), and the remaining $(y_{\rm h} - Z_{\rm h}) M_{\rm h}$ either fell to the disk or was lost from the Galaxy. \citet{Ostriker1975GalacticHalos} suggest on geometrical grounds that about half of those metals were retained by the disk. Now the halo yield, $y_{\rm h}$, may differ from the disk yield because the IMF for low-mass stars appears to differ between the two regions (Section~\ref{subsubsec:Other IMF}), but for a very rough estimate we can consider $y_{\rm h} \sim {\rm Z_{\odot}} \gg Z_{\rm h}$. The mass of metals falling on the disk is therefore ${\sim} 0.5 \ {\rm Z_{\odot}} M_{\rm h}$, and the initial metallicity of the disk gas is this quantity divided by the initial disk mass. The appropriate value of $M_{\rm h}$ is not the mass of halo stars in a local cylinder, but the mass (mostly further out) from which ejected gas would fall to this part of the disk; since the surface density of halo stars is locally only a few percent of the disk value, and it decreases with increasing radius \citep{Schmidt1975TheStars}, $M_{\rm h}$ is probably only a small fraction of the present mass of the disk in the local cylinder. However, the initial disk mass may also have been small. Altogether, it is possible that metals captured from the young halo could have been enough to have enriched the initial disk gas significantly. A value of $Z_{0} / \rm Z_{\odot} \sim 0.2$ would completely solve the G-dwarf problem, as illustrated above.

Early enrichment by halo stars appears as an important effect in the dynamical models for the formation of disk galaxies of \citet{Larson1976a}, whose chemical evolution has been studied further by \citet{Tinsley1978ChemicalDisks}. At model radii corresponding to the Sun's position in the Galaxy, $Z$ reaches about a third of its final value before stars with disk kinematics begin to form, and the initial disk mass in these models is about $10\%$ of the final value. As a result, the fraction of local stars with $Z < \rm Z_{\odot} / 3$ is no greater than observed.

It may be concluded that the disk in the Solar neighborhood plausibly started with a small fraction of its present mass, in the form of gas that had been significantly enriched by massive stars in the young halo.

\subsubsection{Later metal-poor infall}
\label{subsubsec:Later metal-poor infall}

In the dynamical models just mentioned, the outer disk grows on a timescale of billions of years by accretion of metal-poor gas; star formation occurs on a similar timescale, so the metallicity tends to reach a roughly constant value near the yield, as in the schematic infall model of Section~\ref{subsubsec:balanced system}.

A significant additional factor, first suggested by \citet{Lynden-Bell1975TheGalaxies}, is that the ratio of SFR to infall rate ($\psi / f$) is unlikely to remain strictly constant in a realistic picture; its value is more likely to increase as the surrounding gas is used up, as indeed occurs in the dynamical models. An asymptotic relation between $\psi / f$ and $Z$ can be derived using the Equations of Section~\ref{subsec:Approximations} with the instantaneous recycling approximation. Let
\begin{equation}
    (1 - R) \psi = kf,
	\label{eq:eq4.4}
\end{equation}
where $k$ is a constant, so Equation~(\ref{eq:eq3.19}) becomes
\begin{equation}
    M_{\rm g} = \frac{dZ}{dt} = \left( ky + Z_{\rm f} - Z \right) f.
	\label{eq:eq4.5}
\end{equation}
According to this Equation, there is an asymptotic value of $Z$,
\begin{equation}
    Z \rightarrow ky + Z_{\rm f},
	\label{eq:eq4.6}
\end{equation}
so the metal abundance of the gas tends toward a value $ky$ if $Z_{\rm f} = 0$. Thus if $\psi / f$ increases, so does $Z$, and there is steady enrichment rather than the constant $Z$ obtained in Section~\ref{subsubsec:balanced system}. Examples of this behavior are seen in the numerical models of \citet{Tinsley1978ChemicalDisks}, in particular in the regions between ${\sim} 7$ and $15 \: \rm kpc$ which have general properties resembling the Solar neighborhood. The slow increase of $Z$ leads to a less peaked stellar metallicity distribution than that in the ``extreme'' infall model shown in Figure~\ref{fig:fig8}.

\subsubsection{Timescales for chemical evolution}
\label{subsubsec:Timescales for chemical evolution}

The original evidence for infall of extragalactic gas was the explanation of the high-velocity clouds of neutral hydrogen \citep{Oort1970TheHydrogen.}, which required an infall rate comparable to the SFR. Further evidence is given by the timescale for star formation to consume the ISM in the Solar neighborhood, as pointed out by \citet{Larson1972a}: the local surface density of ISM is ${\sim} 8 \: \rm M_{\odot} \ pc^{-2}$ while the SFR is ${\sim} 10 \: \rm M_{\odot} \ pc^{-2} \ Gyr^{-1}$ (Section~\ref{subsubsec:Local SFR}), so the ISM is being used up on a timescale
\begin{equation}
    \tau_{*} \equiv \frac{M_{\rm g}}{(1 - R) \psi} \sim 1 \: \rm Gyr.
	\label{eq:eq4.7}
\end{equation}
If there is not a source of new gas, flowing in at about the SFR, we must be living very near the end of active star formation in this part of the disk. Various constraints related to interstellar pressure balance and Galactic X-ray emission allow an infall rate up to ${\sim} 1 \: \rm M_{\odot} \ pc^{-2} \ Gyr^{-1}$ in the Solar neighborhood \citep{Cox1976AccretionRate}. This is enough to extend the lifetime of the gas against total consumption by only ${\sim} 10\%$, but all of the above estimates are rather uncertain, and it is possible that $f \sim \psi$ in the Solar neighborhood, so the gas is continually replenished.

A related timescale is that for chemical enrichment, which can be written
\begin{equation}
    \tau_{\rm Z} \equiv \frac{Z}{\left| dZ / dt \right|} = \frac{Z M_{\rm g}}{\left| y(1 - R) \psi + \left( Z_{\rm f} - Z \right) f \right|},
	\label{eq:eq4.8}
\end{equation}
using Equation~(\ref{eq:eq3.19}). If infall were negligible, this Equation would give $\tau_{\rm Z} = \tau_{*} (Z / y) \simeq \tau_{*} \sim 1 \: \rm Gyr$; we would then expect to see $Z$ increasing on a timescale of $1 \: \rm Gyr$ now, which is not consistent with the data in Figure~\ref{fig:fig7}. However, if gaseous inflow is occurring at a rate comparable to the SFR, $\tau_{\rm Z}$ could be much longer.

Other processes than infall have been suggested to account for the slow enrichment rate in the Solar neighborhood. For example, various authors beginning with \citet{Truran1971EvolutionaryGalaxy} have suggested a time-dependent IMF with a greater yield in the past; thus at present $Z \gg y$, so $\tau_{\rm Z} \gg \tau_{*}$.

Whatever such additional processes may affect the evolution and distribution of metal abundance, it seems almost inevitable that gas flows related to galaxy formation have a strong influence on chemical evolution. The dynamical models discussed above were designed to reproduce the structural properties of galaxies \citep{Larson1976a}, so it is significant that they predict both pre-enrichment of the disk gas and later metal-poor infall, which thus appear to be the most natural solutions to the G-dwarf and timescale problems of the Solar neighborhood. From these results, it is to be expected that the chemical evolution of other regions of galaxies is affected by gas flows; examples will be reviewed in Section~\ref{sec:Chemical Evolution of Galaxies}.

\subsection{Relative Abundances of the Elements}
\label{subsec:Relative Abundances of the Elements}

Relative abundances of elements heavier than helium provide information on both nucleosynthesis and galactic evolution, in a variety of ways that are discussed extensively in the literature (see reviews by \citealp{Audouze1976ChemicalGalaxies.}, \citealp{Pagel1978a}, and \citealp{Trimble1975TheElements}). Here the emphasis will be on points relevant to galactic evolution.

\subsubsection{Primary elements from different stars}
\label{subsubsec:Primary elements from different stars}

Stars of different masses eject new primary elements in different proportions, so if the ISM is not well mixed we expect to see some variations in relative abundances on the surfaces of unevolved stars. (Variations would arise similarly from any dependence of primary nucleosynthesis on the initial compositions of stars, Section~\ref{subsubsec:Initial composition}). Until recently, most reported variations were within the uncertainties of measurement and analysis, but some trends are now emerging.

The relative abundances of carbon, oxygen, and iron (for which the data are reviewed by \citealp{Pagel1978a}) are an interesting example. All unevolved stars with reliable carbon abundances have $\rm [C / Fe] = 0$ within the errors, i.e. a Solar abundance ratio\footnote{This statement is tentative, since very few metal-poor stars have quantitative carbon abundances and some globular clusters show preliminary evidence for carbon excesses and/or deficiencies \citep{McClure1979ObservationalPopulations}.}, but oxygen is systematically overabundant in metal-poor stars, a typical value being $\rm [O / Fe] = 0.5$ in stars with $\rm [Fe / H] < -1$. Since these stars have halo kinematics, it appears that the yield of oxygen was relatively high during the early life of the Galaxy. Oxygen is the most abundant element after H and He, so it would be more appropriate to say that the yield of iron (and perhaps carbon) was relatively low.

One possible explanation is that the IMF was different, favoring massive stars, which have a higher oxygen/carbon ratio in their ejecta according to current stellar models \citep[e.g.][]{Arnett1978OnStars}. Another possibility is that significant fractions of iron and carbon are ejected by stars with such long lives that they die too late to enrich the halo stars. This idea agrees qualitatively with the mass dependence of stellar nucleosynthesis outlined in Section~\ref{subsec:Beyond MS}: carbon stars (and planetary nebulae) with initial masses from ${\sim} 1 - 8 \: \rm M_{\odot}$ indicate that carbon production takes place partly in very long-lived stars; Type~I supernovae, with iron-enriched envelopes, show that iron is ejected at least partly by stars of $\lesssim 6 \: \rm M_{\odot}$; but there is no evidence for oxygen production outside the massive stars ($\gtrsim 10 \: \rm M_{\odot}$) that are theoretically predicted to be its main source. Models showing the effects of nucleosynthesis in stars with such a wide range of lifetimes have been studied by \citet{Tinsley1979StellarEvolution}. It is found that systematic variations in relative abundances arise if star formation occurs on timescales less than the lifetime of the least massive star producing significant amounts of an element of interest. For example, iron produced by stars of intermediate mass (${\sim} 4 - 6 \: \rm M_{\odot}$) is underabundant relative to oxygen if stars form on a timescale ${\sim} 10^{8} \: \rm yr$. Even if the whole time for halo formation is several billion years, as indicated by the age spread of globular clusters (Section~\ref{subsec:Stellar_Populations}), such variations will appear in halo stars if they were formed in a series of short bursts. A very inhomogeneous protogalaxy, with stars forming in dense lumps, has been suggested on dynamical grounds (Section~\ref{subsec:Galaxy_Formation}), so it is interesting to note how chemical abundances could be affected.

\subsubsection{Secondary elements}
\label{subsubsec:Secondary elements}

Secondary elements are those synthesized from ``seed'' elements heavier than helium that were already present in the star at birth; some examples have been mentioned in Section~\ref{subsubsec:Initial composition}. The stellar yields of these elements increase with metallicity, so their abundances are expected to increase more rapidly (and to be relatively lower in metal-poor stars) than those of primary elements. As reviewed by \citet{Pagel1978a}, this prediction is borne out qualitatively by relative underabundances of nitrogen and barium in metal-poor stars, but the data are in quantitative disagreement with theoretical models.

\begin{figure}
	\includegraphics[width=0.47\textwidth]{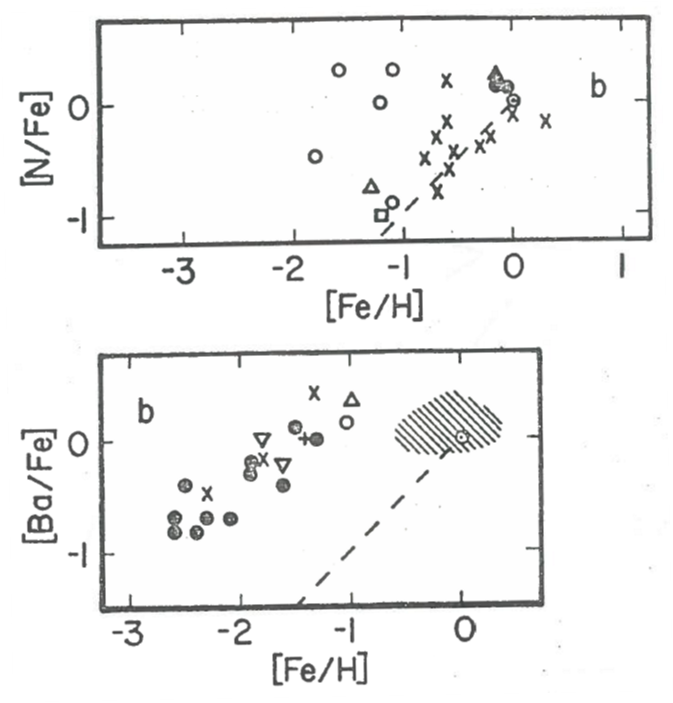}
    \caption{Ratios of primary to secondary abundances versus metallicity. \emph{Data points} (and shaded area with many points) are from sources cited by \citet{Tinsley1979StellarEvolution}, and \emph{dashed lines} correspond to a simple scenario for secondary nucleosynthesis (Section~\ref{subsubsec:Secondary elements}). The nitrogen data are all for dwarfs; barium data include giants, but these do not give a different picture from the dwarfs alone, so there is probably no significant contamination by the stars' own nucleosynthesis.}
    \label{fig:fig9}
\end{figure}

The canonical prediction for secondary abundances is that they increase as the square of primary abundances, e.g. $\rm Ba / H \propto (Fe / H)^{2}$ or $\rm Ba / Fe \propto Fe / H$. This prediction arises, as follows, from the conventional assumption that a star of given mass produces and ejects an amount of secondary element in direct proportion to the star's initial primary seed abundance. Let us consider a secondary element ``$s$'', with primary seed $Z$, and let $p_{\rm sm}$ be the mass of new $s$ ejected by a star of mass $m$ with initial metallicity $\rm Z_{\odot}$; with another initial metallicity $Z$, the amount of new $s$ ejected is $p_{\rm sm} (Z / \rm Z_{\odot})$. Thus the expression like (\ref{eq:eq3.8}) for the rate of ejection of new $s$ is
\begin{equation}
    \int_{m_{\rm t}}^{\infty} mp_{\rm sm} \frac{Z \left( t - \tau_{\rm m} \right)}{\rm Z_{\odot}} \psi \left( t - \tau_{\rm m} \right) \phi(m) \ dm.
	\label{eq:eq4.9}
\end{equation}
The instantaneous recycling approximation to this expression can be written $y_{\rm s} (Z / {\rm Z_{\odot}}) (1 - R) \psi$, where $Z$ and $\psi$ are current values. The equation analogous to (\ref{eq:eq3.19}) for the abundance $X_{\rm s}$ of $s$ is therefore
\begin{equation}
    M_{\rm g} \frac{dX_{\rm s}}{dt} = y_{\rm s} \left( \frac{Z}{\rm Z_{\odot}} \right) (1 - R) \psi + \left( X_{\rm sf} - X_{\rm s} \right) f.
	\label{eq:eq4.10}
\end{equation}
In the simple model of Section~\ref{subsubsec:closed system}, the equation analogous to (\ref{eq:eq3.22}) is
\begin{equation}
    M_{\rm g} \frac{dX_{\rm s}}{dM_{\rm g}} = -y_{\rm s} \left( \frac{Z}{\rm Z_{\odot}} \right).
	\label{eq:eq4.11}
\end{equation}
Now $Z$ is given by Equation~(\ref{eq:eq3.23}), so (\ref{eq:eq4.11}) can be solved for $X_{\rm s}$, with the result
\begin{equation}
    X_{\rm s} = \frac{1}{2} \left( \frac{y_{\rm s}}{y \rm Z_{\odot}} \right) Z^{2},
	\label{eq:eq4.12}
\end{equation}
which is the canonical prediction $X_{\rm s} \propto Z^{2}$ mentioned above. Since Solar abundances are near the mean disk values, they must also be approximately the yield values, so the prediction is equivalent to $X_{\rm s} / Z \propto Z$ with a zero-point near the Solar values.

Stellar nitrogen and barium abundances are plotted in Figure~\ref{fig:fig9}, where the canonical relation would correspond in each case to the straight line at $45^{\circ}$ through about the Solar values (zeros in the logarithmic notation). It can be seen that the nitrogen abundances scatter greatly about the predicted relation, mainly exceeding the predicted value in metal-poor stars; and barium abundances are almost all greater than predicted, no matter where one places this zero-point among the data for stars near Solar composition.

\medskip

What is wrong? There is a logical flaw in making predictions with the simple model because it is known to be untenable on account of the stellar metallicity distributions (Section~\ref{subsec:G-Dwarf}); furthermore, instantaneous recycling may be a bad approximation for secondary elements that are made in red giants down to initial masses ${\sim} 1 \: \rm M_{\odot}$. Plausible modifications to the above theory lead to no better agreement, however, as the following examples show \citep{Tinsley1979StellarEvolution}.

\begin{enumerate}
  \item Numerical calculations allowing for delayed ejection by long-lived stars make the situation worse, because they predict even smaller secondary abundances in the most metal-poor stars.
  
  \vspace{1mm}
  
  \item Infall makes the situation even worse, because it predicts a steeper rise than $X_{\rm s} / Z \propto Z^{2}$ at the metal-rich end.
\end{enumerate}

The latter result can be derived simply in the extreme infall model of Section~\ref{subsubsec:balanced system}. Use of Equation~(\ref{eq:eq4.10}), with $X_{\rm sf} = 0$ and (\ref{eq:eq3.28}) for $Z$, leads to a result analogous to (\ref{eq:eq3.28}),
\begin{equation}
    X_{\rm s} = \frac{y_{\rm s} y}{\rm Z_{\odot}} \left( 1 - e^{-\nu} -\nu e^{-\nu} \right).
	\label{eq:eq4.13}
\end{equation}
This formula reduces to Equation~(\ref{eq:eq4.12}) in the limit of low abundances, but its asymptotic limit, $X_{\rm s} \rightarrow y_{\rm s} y / \rm Z_{\odot}$, is twice the value obtained by extrapolating (\ref{eq:eq4.12}) to the asymptotic value $Z \rightarrow y$. Thus a plot of $X_{\rm s} / Z$ versus $Z$ for this model would have a slope of $45^{\circ}$ at first, but would become steeper at the top, obviously in disagreement with the data in Figure~\ref{fig:fig9}.

From these results, it appears that the problem in Figure~\ref{fig:fig9} is not due to a simplistic choice of model for chemical evolution, but to wrong assumptions about nucleosynthesis. For nitrogen, it has often been suggested that a primary source of nucleosynthesis could contribute significantly \citep[e.g.][]{Truran1971EvolutionaryGalaxy}, and this component of the nitrogen abundance would make the ratio $\rm N / Fe$ more nearly constant. A possible scenario for primary nitrogen production is that some carbon produced by helium burning in stars of intermediate mass gets mixed into the hydrogen-burning shell region and there converted to nitrogen; the models in \citet{IbenI.1976FurtherStar} illustrate this possibility. Nitrogen abundances in \textsc{H~ii} regions, in our own and other galaxies, show a scatter similar to the stellar data, and primary production has again been proposed as an explanation \citep[e.g.][]{Smith1975SpectrophotometricGalaxies, Edmunds1978NitrogenGalaxies}.

For barium there is no obvious solution to the discrepancy in Figure~\ref{fig:fig9}, since a primary source is unlikely in terms of nuclear physics. The calculations of stellar structure and nucleosynthesis used to predict barium production are so complicated that results for Solar initial metallicities cannot be extrapolated to others \citep{IbenI.1978OnMedium}; it is therefore likely that the conventional assumption, that production scales directly with the initial $\rm Fe / H$ of a star of given mass, is wrong.

\medskip

Other aspects of secondary nucleosynthesis relevant to galactic evolution are discussed in the references cited at the beginning of Section~\ref{subsec:Relative Abundances of the Elements} and in the proceedings of a meeting on CNO isotopes in astrophysics \citep{Audouze1977CNOAstrophysics}.

\subsubsection{Radioactive elements}
\label{subsubsec:Radioactive elements}

The field of nucleochronology (or cosmochronology or nucleo-cosmochronology) uses meteoritic abundances of the decay products of certain radioactive elements to derive information on the age distribution of elements in the early Solar System. This information is relevant to the rates of star formation and nucleosynthesis before the Sun formed, and to the age of the Galaxy. Nucleochronology has been reviewed recently by \citet{Fowler1979NuclearReport}, \citet{Schramm1974Nucleo-Cosmochronology}, and \citet{Trimble1975TheElements}, and its relevance to galactic evolution is discussed further by \citet{Reeves1972SpatialNucleosynthesis}, \citet{Reeves1976ThePhenomena}, and in other papers referenced below. Some general principles will be outlined here.

\medskip

Let us consider the abundance $X_{\rm i}$ of a radioactive nuclide with decay constant $\lambda_{\rm i}$ and net yield $y_{\rm i}$. The equations of chemical evolution must be modified to allow for decay in the ISM and in stellar envelopes; the equation analogous to (\ref{eq:eq3.7}) is
\begin{equation}
    \frac{d X_{\rm i} M_{\rm g}}{dt} = -X_{\rm i} \psi + E_{\rm i} + X_{\rm if} f - \lambda_{\rm i} X_{\rm i} M_{\rm g},
	\label{eq:eq4.14}
\end{equation}
where $E_{\rm i}$ is given, analogously to Equation~(\ref{eq:eq3.9}), by
\begin{equation}
\begin{medsize}
    E_{\rm i}(t) = \int\limits_{m_{\rm t}}^{\infty} \left[ \left( m - w_{\rm m} - m p_{\rm im} \right) X_{\rm i} \left( t - \tau_{\rm m} \right) e^{-\lambda_{\rm i} \tau_{\rm m}} + m p_{\rm im} \right] \psi \left( t - \tau_{\rm m} \right) \phi(m) \ dm.
	\label{eq:eq4.15}
\end{medsize}
\end{equation}
The decay timescale, $\lambda_{\rm i}^{-1}$, is of course independent of the system's timescales for star formation and infall, so straightforward analytical solutions for $X_{\rm i}$ are possible only in special cases, even if instantaneous recycling is used. In fact, the long-lived radioactive elements are useful just \emph{because} their abundances are sensitive to evolutionary details (such as the past SFR and the age) that cancel in the instantaneous recycling approximation and hardly affect other abundance parameters in detailed models.

The simplest situation would be if all the elements formed in an initial burst. Then their decay in the ISM and stellar envelopes would lead to abundances at time $t$ given by $X_{\rm i}(t) = X_{\rm i0} \exp( -\lambda_{\rm i} t)$, where $X_{\rm i0}$ is the initial abundance and is proportional to the amount of element $i$ made, i.e. to its yield $y_{\rm i}$. In particular, at the time $T$ when the Solar System formed, two such elements would have an abundance ratio
\begin{equation}
    \frac{X_{\rm i}(T)}{X_{\rm j}(T)} = \frac{X_{\rm i0}}{X_{\rm j0}} \exp \left[ - \left( \lambda_{\rm i} - \lambda_{\rm j} \right) T \right],
	\label{eq:eq4.16}
\end{equation}
where $X_{\rm i0} / X_{\rm j0} = y_{\rm i} / y_{\rm j}$. The value of $T$ could in this case be found by substituting into Equation~(\ref{eq:eq4.16}) the yield ratio (given by nucleosynthesis theory) and the abundance ratio in the early Solar System (obtained from the abundances of decay products in meteorites). The age of the Galaxy, which is simply $T$ plus the age of the Sun, could then be determined.

Of course, the elements were probably synthesized continually during the period $T$ from first star formation in the Galaxy until the Solar System condensed, so Equation~(\ref{eq:eq4.16}) is unrealistic. As first shown by \citet{Schramm1970NucleochronologiesElements}, one can define a useful age parameter, here denoted $T_{\rm ij}$, analogously to the solution of (\ref{eq:eq4.16}) for $T$:
\begin{equation}
    T_{\rm ij} \equiv \frac{1}{\lambda_{\rm i} - \lambda_{\rm j}} \ln \left[ \frac{y_{\rm i} / y_{\rm j}}{X_{\rm i}(T) / X_{\rm j} (T)} \right],
	\label{eq:eq4.17}
\end{equation}
which can, in principle, be evaluated from meteoritic and nuclear data independently of Galactic evolution. In the limit of long-lived elements, ($\lambda_{\rm i} T \ll 1, \: \lambda_{\rm j} T \ll 1$), $T_{\rm ij}$ is just the mean age of elements in the gas, $T_{Z}$, at the time when the Solar System formed \citep{Tinsley1975c}.

The relation between the mean age ($T_{Z}$) and the elapsed time ($T$) is model-dependent, so estimates of $T_{\rm ij}$ for long-lived pairs of elements do not give $T$ directly. Different possibilities include the following.

\begin{enumerate}
  \item If essentially all nucleosynthesis of the relevant elements took place in an initial burst, then $T = T_{Z}$.
  
  \vspace{1mm}
  
  \item The simple model for chemical evolution gives the intuitive result that the mean age is half the elapsed time, $T_{Z} = T / 2$; however, because this model is discrepant with stellar metallicities, one cannot assume that $T$ is given simply by $2 T_{Z}$ in reality.
  
  \vspace{1mm}
  
  \item In extreme infall models, the ISM and heavy elements in it have a mean age ${\sim} M_{\rm g} / f$ at all times greater than $M_{\rm g} / f$; so if chemical evolution in the disk was strongly affected by infall before the Solar System formed, the value of $T_{Z}$ obtained from meteoritic abundances may reflect only the timescale for infall, independently of the age $T$.
  
  \vspace{1mm}
  
  \item There are consistent models for the Solar neighborhood, with some infall and/or some early enrichment, that have values of $T_{Z} \simeq T / 2$ (as emphasized by \citealp{Hainebach1977CommentsNucleocosmochronology}), but since not all plausible models have this relation it cannot be used confidently (as emphasized by \citealp{Tinsley1977b}).
\end{enumerate}

In summary, there is a large uncertainty in any age estimate of the Galaxy derived from nucleochronology, except of course that the age of the Solar System is a reliable lower limit!

\medskip

The initial Solar System abundances of short-lived radioactive elements are sensitive to rates of nucleosynthesis at times immediately preceding the solidification of the meteorites. Their abundances suggest that the nucleosynthesis of most elements ceased ${\sim} 10^{8} \: \rm yr$ before the solidification time, but some material was produced only ${\sim} 10^{6} \: \rm yr$ earlier. Interpretations of these timescales include passage of the pre-Solar material through spiral arms at $10^{8}$-yr intervals; enrichment by fresh supernova ejecta each $10^{6} \: \rm yr$, which could result from the average supernova rate in the Solar neighborhood; a last-minute supernova that triggered the formation of the Solar System; and locking of radioactive elements, with their decay products, into grains long before the Solar System formed. These possibilities are reviewed briefly by \citet{Podosek1978IsotopicMaterials}, and discussed in detail by several authors in a conference proceedings edited by \citet{Gehrels1978ProtostarsPlanets.}. As yet there is no consensus on the interpretation of short-lived radioactivities in the early Solar System, but ultimately they should provide valuable information on star formation and interstellar chemistry.

\section{Chemical Evolution of Galaxies}
\label{sec:Chemical Evolution of Galaxies}

For other galaxies, and for regions of our own outside the Solar neighborhood, there is much less detailed information on abundance distributions, but there are some striking trends that call for explanations involving the formation and later dynamical evolution of galaxies. A few relevant properties have been mentioned in Section~\ref{subsec:Stellar_Populations}, and now further details will be described and some of the theoretical models reviewed. Other general reviews of this subject include those by \citet{Audouze1976ChemicalGalaxies.}, \citet{Pagel1978b}, and \citet{Trimble1975TheElements}.

\subsection{Outline of Relevant Data}
\label{subsec:Data2}

Abundances are very often found to decrease outward in galaxies: gradients have been observed in the \textsc{H~ii} regions of disks, in disk stars, and in the stars of spheroidal systems including elliptical galaxies and the bulge-halo components of spirals.

\medskip

In the Galactic disk, \textsc{H~ii} regions within a few kpc of the Sun have an average gradient $d {\rm [O / H]} / dr \simeq -0.1 \: \rm kpc^{-1}$ (where $r$ is the galactocentric distance), while stars of intermediate age have a gradient $d {\rm [Fe / H]} / dr \simeq -0.05 \: \rm kpc^{-1}$; an open cluster about $10 \: \rm kpc$ from the Sun in the anticenter direction, of age only $\lesssim 10^{8} \: \rm yr$, is apparently as metal-poor as $\rm [Fe / H] < -1$ \citep{Christian1979UBV21}. (Oxygen and iron abundances are quoted for the ISM and stars, respectively, because these are the best observed elements). The uncertainties are such that the apparent age dependence of the gradient may or may not be real. These data are reviewed by \citet{Janes1977ADisk}, \citet{Mayor1977ChemicalGalaxy}, and \citet{Peimbert1977ChemicalGalaxy}. \textsc{H~ii} regions in external galaxies generally show gradients of a similar magnitude (e.g. \citealp{Smith1975SpectrophotometricGalaxies, Shields1978TheM101}; and references therein). However, only a marginal gradient appears in the Large Magellanic Cloud and none in the Small Cloud \citep{Pagel1978AClouds}.

The most obvious explanation for these gradients would be that the outer regions of disks are less chemically evolved than the inner regions, in the sense of having converted a smaller fraction of their gas to stars. The simple model, for example, would predict a $Z$ gradient given by $Z = y \ln \ \mu^{-1}$ arising from a gradient in the gas fraction $\mu$ (Section~\ref{subsubsec:closed system}). However, the best studied galaxies (the Milky Way and M101) probably do not have a sufficient gradient in $\mu$ for this explanation to suffice. \citet{Gordon1976CarbonNucleons} found that the combined surface densities of atomic and molecular hydrogen lead to a nearly constant gas fraction ($\mu \sim 0.05$) at $R > 4 \: \rm kpc$ in the Galaxy, and \citet{Shields1978TheM101} noted a similar problem in M101. The amount of ISM interior to the Sun could be overestimated, since the $\rm H_{2}$ density is derived from observations of CO molecules on the assumption of a constant abundance ratio $\rm CO / H_{2}$; if in fact $\rm CO / H_{2}$ increases inward because of a $\rm C / H$ abundance gradient, then there is less $\rm H_{2}$ than had been thought at small radii \citep{Peimbert1977ChemicalGalaxy}. With this correction, the Galaxy probably has some gradient in $\mu$, which could account in part for the interstellar composition gradient. Of course, the simple model is known to be invalid in the Solar neighborhood, so we do not expect the formula $Z = y \ln \ \mu^{-1}$ to explain gradients in detail. Other ways of generating gradients will be mentioned in Section~\ref{subsec:Abundance Gradients in Disks}.

The Galactic halo stars also have a metallicity gradient. Studies of individual stars in globular clusters show that a spread of metallicities of $\rm [Fe / H] \sim -2$ to $0$ occurs in the innermost $5 \: \rm kpc$ (measured from the Galactic center, at any angle to the disk), while the upper limit drops to ${\sim} {-1}$ at greater radii \citep{Cowley1978SpectraGalaxy, Searle1978CompositionsHalo}. It is not clear whether a systematic decline in iron and/or CNO abundances persists further out \citep{McClure1979ObservationalPopulations, Kraft1978EvidenceClusters}.

\medskip

\begin{figure}
	\includegraphics[width=0.47\textwidth]{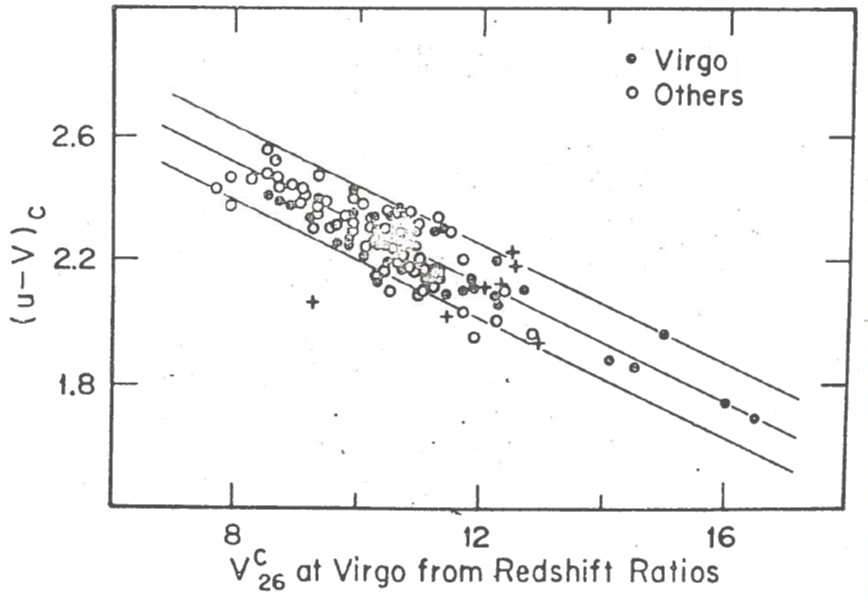}
    \caption{A color--magnitude diagram for elliptical and S0 galaxies in several clusters of galaxies \citep{Visvanathan1977TheClusters}. The color $(u - V)_{\rm c}$ is corrected for reddening, redshift, and aperture effects; magnitudes are adjusted to the distance of the Virgo cluster using redshift ratios. (\emph{Crosses} denote Virgo galaxies not used by \citealp{Visvanathan1977TheClusters} in their linear fit to the data). The \emph{straight lines} are a linear fit, with $\pm 2 \sigma$ boundaries. Despite the excellent linear fit over a large range of magnitudes, the brightest few points for the Virgo cluster alone (\emph{filled circles}) and for other clusters (\emph{open circles}) show a tendency to level off in color.}
    \label{fig:fig10}
\end{figure}

Many elliptical and S0 galaxies have gradients of integrated colors and line strengths in their spectra that are best interpreted as composition gradients. The same quantities also vary systematically with the absolute magnitudes of E and S0 galaxies, as illustrated in Figure~\ref{fig:fig10}, indicating that the brighter galaxies are more metal-rich. A thorough review of this subject is given by \citet{Faber1977ThePopulations}. The analysis and calibration of abundance effects in the integrated light of a galaxy are much more complicated than for individual stars, because line strengths are strongly affected by the mixture of stellar temperatures and luminosities, and because the whole population in the HR diagram is shifted by effects of metallicity on the interiors and atmospheres of stars (Section~\ref{subsubsec:Initial composition}). Until recently, it was not clear whether all elements heavier than helium enter the composition trends in E and S0 galaxies, or whether a few with strong spectral features (N, Mg, Na), are mainly responsible. \citet{Cohen1978AbundancesM13} has now made a detailed observational study of some lines of Ca, Na, Mg, and Fe, together with approximate theoretical predictions of how their strengths should vary with the composition of a population of stars; she finds no evidence against the abundances of all of these elements varying in the same proportions.

The color--magnitude relation for elliptical galaxies is linear over a wide range of metallicities (Figure~\ref{fig:fig10}), which suggests a power-law relation between metallicity and luminosity. A tentative calibration, subject to revision when both the data and the theoretical color--metallicity relation are more secure, is
\begin{equation}
    Z_{\rm s} \propto L_{\rm B}^{0.3}
	\label{eq:eq5.1}
\end{equation}
\citep{Tinsley1978b}, where $Z_{\rm s}$ is the average metallicity of stars in an elliptical galaxy of blue luminosity $L_{\rm B}$. This relation was derived by from population models differing only in metallicity, so that the stars in all cases had the same IMF and age (as outlined in Section~\ref{subsubsec:Evolutionary models}). Such models also predict that the mass-to-luminosity ratio should depend on metallicity, yielding a relation $M_{\rm s} / L_{\rm B} \propto L_{\rm B}^{0.13}$. A relation very similar to this has been obtained observationally by \citet{Schechter1979ObservationsGalaxies} for the cores of elliptical galaxies. Equation~(\ref{eq:eq5.1}) therefore corresponds to a tentative metallicity--mass relation,
\begin{equation}
    Z_{\rm s} \propto M_{\rm s}^{0.25},
	\label{eq:eq5.2}
\end{equation}
where $M_{\rm s}$ is the mass of stars (not any extended hidden halo material), and the main uncertainties in the exponent are due to the color--magnitude data and to the color--metallicity calibration. There is some evidence that the color--magnitude relation levels off at the magnitudes of the brightest cluster galaxies, as can be seen, for example, from the brightest few points in Figure~\ref{fig:fig10}.

\medskip

\begin{figure*}
	\includegraphics[width=0.97\textwidth]{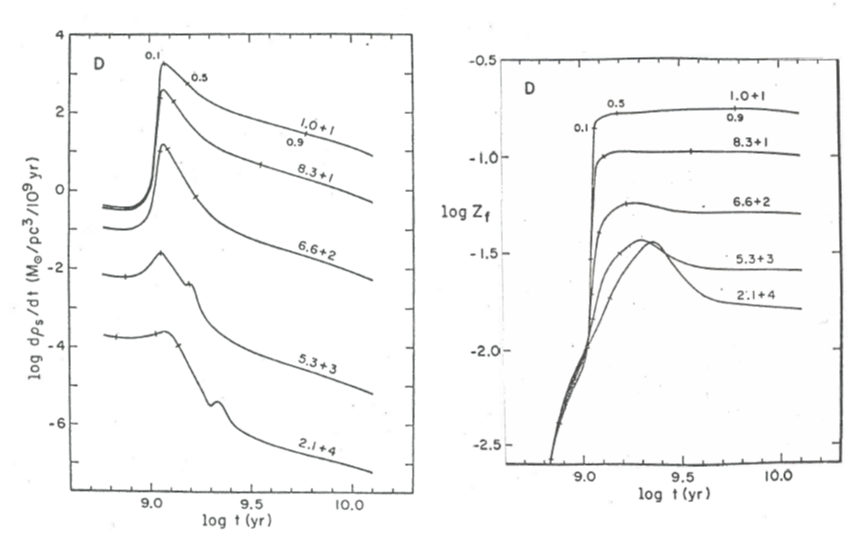}
    \caption{(a) Star formation rates and (b) metallicities of newly formed stars (i.e., $Z$ of the gas), at several radii in a collapse model for the formation of a spherical galaxy \citep{Larson1974a}. The radius in pc is marked on each curve, and the three ticks indicate the times at which star formation is $10\%$, $50\%$, and $90\%$ complete (relative to the final mass of stars) at that radius.}
    \label{fig:fig11}
\end{figure*}

Heavy elements have been detected in intergalactic gas, and since these elements almost certainly come from galactic stars they are relevant to the chemical evolution of galaxies. A feature due to iron has been observed in the diffuse X-ray emission spectra of several rich clusters of galaxies; the interpretation is that these clusters contain hot gas (${\sim} 10^{8} \: \rm K$), emitting by thermal Bremsstrahlung, with approximately Solar abundances of iron. The mass of intergalactic gas inferred from the data is model-dependent \citep[e.g.][]{Bahcall1977ParametersVirgo., Perrenod1978TheSources}, and is roughly between 1 and 30 times the total luminosity of cluster galaxies, in solar units. Now the mass of stars in galaxies is ${\sim} 10$ times their total luminosity, in solar units\footnote{This mass is not to be confused with the virial masses of clusters, which are ${\sim} 100$ times the total luminosity and which provide the main evidence for hidden non-stellar matter in association with galaxies.}, and their average metallicity is approximately Solar, so the intergalactic medium (IGM) in the rich clusters apparently contains about the same total mass of iron as do the stars themselves. These observations suggest that galaxies sometimes lose enough metal-rich gas to affect their own chemical evolution substantially.

Another striking observation of metal-rich IGM is absorption due to Ca and Mg in the spectrum of a quasar (3C~232) lying $1.9^{\prime}$ on the sky from a foreground spiral galaxy (NGC~3067); the absorption redshift matches that of the galaxy, and neutral hydrogen absorption has been detected at the same redshift in the radio spectrum of the quasar \citep{Haschick1975Neutral3067}. The line strengths and positions of the objects imply that there is gas with roughly Solar abundances at least $17 \: \rm kpc$ from the center of the galaxy \citep{Boksenberg1978The3067}.

\subsection{Abundance Gradients in Spheroidal Systems}
\label{subsec:Abundance Gradients in Spheroidal Systems}

Most stars in elliptical galaxies and in the bulge-halo components of spirals were probably formed within a few times $10^{9} \: \rm yr$ at a time ${\sim} 10^{10} \: \rm yr$ ago. The abundance gradients in these systems therefore reflect processes occurring during the time of star formation in a protogalaxy; several such processes have been suggested as possible causes of gradients.

\subsubsection{Dissipative collapse}
\label{subsubsec:Dissipative collapse}

The most extensive models exploring the effects of star formation during the collapse of an initially gaseous protogalaxy are those reviewed by \citet{Larson1976b}. In the spheroidal components of model disk galaxies, and in model ellipticals, star formation becomes increasingly concentrated toward the center as the density builds up. This effect is illustrated in Figure~\ref{fig:fig11}~(a), which shows the SFR as a function of time at several radii in a spherical collapse model. Stars formed at a given radius remain in orbits with little net inward motion, but the gas sinks further in because it is dissipative (i.e., its kinetic energy of radial motion is partly lost via collisionally induced radiation). Thus the metals ejected by evolving stars are carried inward by the gas, and an abundance gradient develops in the gas. As stars continue to form, their compositions reflect this gaseous abundance gradient. Figure~\ref{fig:fig11}~(b) shows the evolution of metallicity of newly formed stars (i.e., $Z$ of the gas) at several radii in a spherical model, and the rapid development of a gradient is clear. The same process of dissipation produces a central concentration in the gas density, which leads to a condensed nucleus of stars.

If there were no dissipation, the stars and gas would collapse together and the metals would not be concentrated inward. Thus in the outer parts of some of these models, where the protogalactic density is too low for dissipation to be effective, no stellar abundance gradient appears. The possible lack of a gradient in metallicities of globular clusters beyond ${\sim} 10 \: \rm kpc$ from the Galactic center has therefore been interpreted as showing that the collapse of the Galaxy began with the stars and gas in free-fall; conversely, the gradient at smaller radii is interpreted as showing the effects of dissipation at a later stage of the collapse \citep{Larson1977c}.

\subsubsection{A gradient in the IMF}
\label{subsubsec:A gradient in the IMF}

Aside from any effects of gas flows, negative metallicity gradients could be produced by gradients in the IMF that led to a yield decreasing outward. Since the yield (Equation~\ref{eq:eq3.12}) depends on the relative numbers of metal-producing stars, possibilities would be a steeper slope for massive stars or more low-mass ``inert'' stars, at larger radii. In the latter case, the stars that survive to the present would still have a radial gradient in their mass function, with an interesting consequence: most of the luminosity of an old population of stars comes from dwarfs near the MS turnoff and evolving giants, while most of the mass is in less massive objects that contribute little light; thus the $M / L$ ratio increases with the proportion of very low-mass stars, and the postulated gradient in the IMF would lead to an outward increase of $M / L$. Such a trend is indeed observed, although it is sometimes ascribed to an extended ``halo'' of non-stellar condensed objects that formed too soon to affect chemical evolution (Section~\ref{subsubsec:Other IMF}). \citet{vandenBergh1975StellarGalaxies} has suggested that the IMF tends to have more massive stars in regions of high density, and this view of the origin of metallicity gradients is part of his evidence. The hypothesis of a gradient in the IMF in spheroidal systems has no convincing theoretical basis, and the trends it would explain can arise in other ways, but nevertheless systematic variations in the IMF could be as important as they are hard to verify.

\subsubsection{Finite stellar lifetimes}
\label{subsubsec:Finite stellar lifetimes}

The timescale for star formation in a protogalaxy could be comparable to the lifetimes of some metal-producing stars, in which case stars formed early would be relatively unenriched. Thus if the outermost stars formed before most of the central ones, there would be a negative metallicity gradient. In the models in Section~\ref{subsubsec:Dissipative collapse} \citep{Larson1976b} it is assumed that all metals are produced by stars with lifetimes $< 3 \times 10^{7} \: \rm yr$, so this effect is negligible; but iron production by Type~I supernovae could in fact be significant on longer timescales (Section~\ref{subsubsec:Intermediate mass}). What timescales are relevant? The minimal collapse time for a protogalaxy is the free-fall time,
\begin{equation}
    t_{\rm ff} = 1.7 \times 10^{6} \left( \frac{M}{10^{11} \: \rm M_{\odot}} \right)^{-\frac{1}{2}} \left( \frac{R}{1 \: \rm kpc} \right)^{\frac{3}{2}} \: \rm yr,
	\label{eq:eq5.3}
\end{equation}
where $M$ and $R$ are the mass and radius. For example, a galaxy with $M = 2 \times 10^{11} \: \rm M_{\odot}$ and $R = 15 \: \rm kpc$ has $t_{\rm ff} = 7 \times 10^{7} \: \rm yr$; and a protogalaxy of the same mass collapsing from $R = 50 \: \rm kpc$ has $t_{\rm ff} = 4 \times 10^{8} \: \rm yr$. Much longer timescales for star formation are possible if the dissipation is slow, and the collapse time of the system can be much longer if its boundary initially expands with the Universe \citep{Gunn1972OnEvolution}. At least the outer parts of large galaxies could therefore be metal-poor partly because of the finite lifetimes of metal-producing stars.

A potential test is to look for variations in relative abundances. For example, if oxygen comes only from very massive stars but iron comes partly from stars of intermediate mass (Section~\ref{subsec:Beyond MS}; Section~\ref{subsubsec:Primary elements from different stars}), then iron should be more deficient than oxygen in the outermost stars. The hypothesis of a gradient in the IMF of massive stars would predict the opposite trend in relative abundances. Current data do not detect any gradients in relative abundances, but oxygen itself has not been studied and nor have the faint outer regions of elliptical galaxies.

It is quite possible that all of the processes discussed above were effective in producing abundance gradients in spheroidal systems, so clear choices among the theories are not to be expected.

\subsection{The Metallicity--Mass Relation for Elliptical Galaxies}
\label{subsec:The Metallicity--Mass Relation for Elliptical Galaxies}

The correlation between metallicity and mass (color and luminosity) of elliptical galaxies has been explained in several ways, of which two will be reviewed here. These each involve dynamical effects during galaxy formation, resulting in less complete conversion of the protogalactic gas to stars, and so to a smaller final mean stellar metallicity, in smaller systems. One could, of course, invoke differences in the IMFs of galaxies as a function of their mass, but there is no independent evidence for a trend of the required form.

\subsubsection{Supernova-driven winds}
\label{subsubsec:Supernova-driven winds}

Star formation and chemical enrichment are cut off in a protogalaxy if the residual gas is lost, and a possible loss mechanism is a galactic wind energized by supernova explosions. Galactic winds were first analyzed for a steady-state case by \citet{Johnson1971GalacticWinds} and \citet{Mathews1971GalacticWinds}; similar analyses have been made for nuclear bulges of spirals by \citet{Faber1976HWinds} and for bulge-disk systems by \citet{Bregman1978GalacticSequence.}. A galaxy sustains a steady-state wind if the supernova rate divided by the rate of supply of gas (from evolving stars) gives the gas enough energy to escape from the galactic potential well. For protogalaxies, we are interested not in the steady state, but in conditions for the initiation of a wind that can remove essentially all of the residual gas. \citet{Larson1974b} discussed possible effects of supernovae in heating the gas, and adopted a simple approximation as the condition for its removal: the residual gas is assumed to be lost suddenly when the total heat input from supernovae has provided the gas with the escape energy, assuming uniform conditions throughout the protogalaxy. This approximation is plausible enough to suggest how the loss condition scales with the relevant parameters, but there are unavoidably large uncertainties in the astrophysics involved so the results are not very secure. The scaling can be derived as follows.

\medskip

Let $E$ be the thermal energy imparted to the ISM by supernovae when a unit mass of stars is formed; $E$ is proportional to the fraction of stars that become supernovae, to the mean kinetic energy of material ejected in a supernova explosion, and to the efficiency with which this energy is transferred to the ISM as heat. (The last factor is the most uncertain). As an approximation, let $E$ be treated as a constant, despite finite stellar lifetimes and complicated effects of the clumpiness of the ISM, its chemical composition, etc. Let us consider a spherical protogalaxy of mass $M$ that has formed a mass $M_{\rm s}$ of stars and has residual gas mass $M_{\rm g} = M - M_{\rm s}$. The condition for gas to escape can be written
\begin{equation*}
    \rm Potential \: energy \: of \: gas = Energy \: from \: supernovae,
	\label{eq:eq5.energy}
\end{equation*}
i.e,
\begin{equation}
    K \frac{GMM_{\rm g}}{R} = EM_{\rm s},
	\label{eq:eq5.4}
\end{equation}
where $K$ depends on the density distribution in the galaxy and will be assumed constant as another simplification. Large elliptical galaxies are observed to be more tightly bound than small ones, so a greater fraction of their gas must be converted to stars before the condition (\ref{eq:eq5.4}) is satisfied; therefore, their surviving stars have a greater mean metallicity. Other consequences of this scenario are that the more massive galaxies collapse more extensively before star formation is cut off, so they are predicted to have more condensed nuclei and steeper metallicity gradients than smaller galaxies (Section~\ref{subsubsec:Dissipative collapse}). These trends are observed, lending support to this type of origin for the increase of metallicity with mass.

The form of the metallicity--mass relation can be accounted for using the same approximate model. Let the initial mass--radius relation for protogalaxies have the form
\begin{equation}
    M \propto R^{\alpha},
	\label{eq:eq5.5}
\end{equation}
so Equation~(\ref{eq:eq5.4}) can be written
\begin{equation}
    M^{1 - \frac{1}{\alpha}} \left( M - M_{\rm s} \right) \propto M_{\rm s}.
	\label{eq:eq5.6}
\end{equation}
Asymptotic equations for the mean metallicity of stars can be derived from very general considerations: the mass of metals synthesized and ejected is $yM_{\rm s}$, so at early stages of evolution when $M \simeq M_{\rm g}$, we have approximately
\begin{equation}
    Z_{\rm s} \propto \frac{yM_{\rm s}}{M_{\rm g}} \simeq \frac{yM_{\rm s}}{M}, \: \: \left( Z_{\rm s} \ll y \right).
	\label{eq:eq5.7}
\end{equation}
At late stages, Equation~(\ref{eq:eq3.21}) predicts that in all cases where mass is conserved,
\begin{equation}
    Z_{\rm s} \rightarrow y, \: \: {\rm as} \: \: M_{\rm s} \rightarrow M.
	\label{eq:eq5.8}
\end{equation}
The results from numerical collapse models verify these relations in cases of interest here. Substituting Equation~(\ref{eq:eq5.7}) into (\ref{eq:eq5.6}), we find the stellar metallicity--mass relation,
\begin{equation}
    Z_{\rm s} \propto M_{\rm s}^{\frac{\alpha - 1}{2 \alpha -1}}, \: \: \left( Z_{\rm s} \ll y \right).
	\label{eq:eq5.9}
\end{equation}
The tentative empirical relation (\ref{eq:eq5.2}) is obtained if $\alpha = 1.5$, which agrees fairly well with the observed mass--radius relation for elliptical galaxies if one considers how the stellar system must swell (to conserve energy) when the gas is lost \citep{Tinsley1978b}. According to Equation~(\ref{eq:eq5.8}), the power-law relation between $Z_{\rm s}$ and $M_{\rm s}$ must level off at large masses, with $Z_{\rm s} \rightarrow y$ in the limit when essentially all the original material is converted to stars; this behavior agrees with the levelling of the color--magnitude relation at the magnitudes of the brightest cluster galaxies.

The critical parameter $E$ can plausibly have a value that would give the right scale for the $Z_{\rm s}$--$M_{\rm s}$ relation \citep{Larson1974b}, but its value is very uncertain so the success of this theory must be considered tentative. The interaction between supernovae and the ISM could, in fact, be so weak as to drive a wind in only the very smallest protogalaxies.

\subsubsection{Bursts of star formation in merging subsystems}
\label{subsubsec:Bursts of star formation in merging subsystems}

Since the largest elliptical galaxies are the most metal-rich, a natural hypothesis is that chemical enrichment accompanied the \emph{growth} of galaxies by successive mergers among small subsystems. As noted in Section~\ref{subsec:Galaxy_Formation}, gaseous protogalaxies probably consisted of many dense lumps, so it is only a change of viewpoint to consider these as merging subsystems rather than as a collapsing unit. Moreover, extrapolations backward in time from the incidence of strongly interacting galaxies in sight today suggest that collisions and coalescence were common processes in the early lives of galaxies \citep{Toomre1977MergersConsequences, Vorontsov-Velyaminov1977NewGalaxies}. A property of colliding galaxies most relevant to chemical evolution is that they often appear to be undergoing intense bursts of star formation induced by the dynamical disturbance (Section~\ref{subsec:Colors of Peculiar Galaxies}), so it is reasonable to assume that star formation was caused in the past by coalescence of subsystems in a protogalaxy. A qualitative model of chemical enrichment by this process has been proposed by \citet{Tinsley1979StellarGalaxies}: elliptical galaxies form by a hierarchical sequence of mergers among subsystems, starting from small unenriched gas clouds; a burst of star formation occurs at each merger, so at each stage of growth the fraction of the total mass in stars increases and the mean metallicities of stars and gas increase. In this picture, the final mass of an elliptical galaxy is determined by the total mass of the small pieces initially present in its surroundings. When these have all been mopped up, efficient star formation stops. Any residual gas may get swept away if the system is moving through an ambient IGM, or possibly blown out in a wind; if it remains bound to the system, it could settle to a disk and leave the ``elliptical galaxy'' as the central bulge of a spiral.

The resulting $Z_{\rm s}$--$M_{\rm s}$ relation depends on the \emph{efficiency} of star formation as a function of the mass of the system (i.e., the system that has been built after a given number of mergers), where efficiency is defined as the mass of stars formed (in a given burst) per unit mass of gas. An approximately power-law relation between $Z_{\rm s}$ and $M_{\rm s}$ can be obtained only if the efficiency increases with the total mass of the system, i.e., with successive mergers. For example, a relation
\begin{equation}
    {\rm Efficiency \: of \: star \: formation \propto (Total \: mass)}^{p},
	\label{eq:eq5.10}
\end{equation}
where $p$ is a constant, leads to
\begin{equation}
    Z_{\rm s} \propto M_{\rm s}^{\frac{p}{1 + p}}, \: \: \left(Z_{\rm s} \ll y \right),
	\label{eq:eq5.11}
\end{equation}
with the usual limit $Z_{\rm s} \rightarrow y$ when all the gas is consumed. The relation (\ref{eq:eq5.10}) can be justified qualitatively by considerations of gas compression during collisions and mergers of subsystems. To reproduce the tentative empirical relation (\ref{eq:eq5.2}), Equation~(\ref{eq:eq5.11}) needs $p = 1/3$, which is consistent with the compression arguments. Equation~(\ref{eq:eq5.11}) results from (\ref{eq:eq5.10}) independently of such details as the mass distribution of merging pieces, and it can be understood as follows: Equation~(\ref{eq:eq5.7}) is true in any models with mass conservation (including here conservation of the total mass of merging pieces), while Equation~(\ref{eq:eq5.10}) gives, dimensionally,
\begin{equation*}
    \frac{M_{\rm s}}{M_{\rm g}} \propto M^{p},
	\label{eq:eq5.mass_ratio}
\end{equation*}
so that
\begin{equation}
    M_{\rm s} \propto M^{1 + p} \: \: {\rm when} \: M_{\rm g} \simeq M \: \: \left( M_{\rm s} \ll M \right);
	\label{eq:eq5.12}
\end{equation}
Equation~(\ref{eq:eq5.11}) then follows from (\ref{eq:eq5.7}) and (\ref{eq:eq5.12}). The power law is again predicted to level off, with $Z_{\rm s} \rightarrow y$ at high masses, according to Equation~(\ref{eq:eq5.8}).

As a theory for the origin of the metallicity--mass relation, this model has the advantage of invoking processes that can be studied in nearby interacting galaxies, but it remains to be seen whether the structural properties of elliptical galaxies are fully consistent with its dynamical implications.

\subsubsection{Mergers of stellar systems}
\label{subsubsec:Mergers of stellar systems}

The color--magnitude (metallicity--mass) relation for elliptical galaxies is apparently affected in a way that has nothing to do with chemical evolution: central cluster galaxies accrete their neighbors, by the process of dynamical friction. There is no star formation during these mergers, because the galaxies involved are ellipticals or S0s with almost no gas. Thus the growth in luminosity is not accompanied by chemical enrichment, and it can make the growing system bluer because the surrounding galaxies that it accretes are generally smaller than the central giant. Galactic cannibalism by dynamical friction was first proposed by \citet{Ostriker1975AnotherGalaxies}, and later papers (e.g. \citealp{Hausman1978GalacticClusters}, and references therein) have developed its implications for cosmological tests, the origin of core--halo structure of cD galaxies, the luminosity function of galaxies in clusters, and the color--magnitude relation itself. The process obviously tends to make the color--magnitude relation turn over toward bluer colors at the bright end. This effect has been proposed as a test for the occurrence of cannibalism in clusters, but the results are not unambiguous because there is an intrinsic upper limit, $Z_{\rm s} \rightarrow y$, to the average stellar metallicity in the models discussed above, that leads to a flattening of the relation anyway. Strong evidence that galaxies in the centers of clusters \emph{do} merge with each other is given by the lumpy appearance of the central supergiant (cD) members of some clusters; the lumps are interpreted as recently swallowed galaxies, and the timescale for them to merge into a smooth system is generally $< 10^{9} \: \rm yr$ \citep{Gunn1977ConcludingRemarks}.

\subsection{The Intergalactic Medium and Gas Lost from Galaxies}
\label{subsec:The Intergalactic Medium and Gas Lost from Galaxies}

Loss of interstellar gas from galaxies can both affect their own evolution, as discussed for example in Section~\ref{subsec:The Metallicity--Mass Relation for Elliptical Galaxies} above, and be a significant source of metal-enriched IGM.

\subsubsection{Loss of metals from galaxies}
\label{subsubsec:Loss of metals from galaxies}

The mass of metals lost from an elliptical galaxy can be estimated by the following argument, which is independent of the method of gas loss. The mass of metals ever made by stars in the galaxy is ${\sim} yM_{\rm s}$ (by the definition of the yield, Equation~\ref{eq:eq3.12}), and the mass of metals presently in its stars is $Z_{\rm s} M_{\rm s}$, so the mass lost to the IGM at some stage is ${\sim} (y - Z_{\rm s}) M_{\rm s}$. This reasoning was used by \citet{Larson1975GasGalaxies} to predict a substantial metal-rich IGM in clusters of galaxies, and a number of models with similar results have been advanced since the iron X-ray line was discovered. An essentially model-independent estimate can be made as follows.

Let $\phi(M_{\rm s})$ be the mass function of elliptical galaxies in a cluster. Then the total mass of metals they have supplied to the ISM is
\begin{equation}
    M_{Z1} = \int \left[ y - Z_{\rm s} \left( M_{\rm s} \right) \right] M_{\rm s} \phi \left( M_{\rm s} \right) \ dM_{\rm s},
	\label{eq:eq5.13}
\end{equation}
where $Z_{\rm s} (M_{\rm s})$ is a function derivable from the color--magnitude relation. In practice, $M_{\rm s}$ is expressed in terms of luminosity, and $\phi(M_{\rm s})$ is obtained from the luminosity function. The value of $y$ should be taken as the maximum $Z_{\rm s}$ of an elliptical galaxy, which is hard to obtain since the extensive outer regions that are probably metal-poor are seldom observed; setting $y$ equal to the mean metallicity of local stars (a little under $\rm Z_{\odot}$) is equivalent if elliptical galaxies have the local IMF. In a calculation equivalent to the one just outlined, \citet{Tinsley1979StellarGalaxies} found that a cluster of elliptical galaxies would contain a mass ${\sim} (2 - 5) \: \rm M_{\odot} Z_{\odot}$ of intergalactic metals per solar luminosity.

This is a very significant quantity of metals. For example, about $1/3$ of the luminosity of the Coma cluster is due to its elliptical galaxies, so they would provide a mass ${\sim} 1 \: \rm M_{\odot} Z_{\odot}$ of metals per solar luminosity of the cluster, corresponding to ${\sim} 0.1 \: \rm Z_{\odot}$ of metals per unit mass of galaxies (the ordinary stellar mass). If the bulges of spiral and S0 galaxies also lost their metals due to the IGM, this estimate could be doubled, but some of those metals may be in the disks (cf. Section~\ref{subsubsec:Metals from the young halo}). Iron can be considered as a representative metal in this calculation, so we predict ${\sim} 1 \times \rm (Fe / H)_{\odot} \times M_{\odot}$ of iron in the IGM per solar luminosity of the cluster. The actual mass of iron is quite uncertain, and could be equal to the predicted amount.

\subsubsection{Overall gas loss from galaxies}
\label{subsubsec:Overall gas loss from galaxies}

The total mass of gas lost from elliptical galaxies is a less predictable quantity, depending on gas flows within the galaxies and gain or loss of gas during the time of star formation. Nevertheless, some estimates are interesting.

\medskip

In order of magnitude, almost any model will predict a mean metallicity of the lost gas that exceeds the mean metallicity of stars, since the gas has the composition of the last and most metal-rich stars formed; i.e., $Z_{\rm i} \gtrsim Z_{\rm s}$. The mass of gas lost is therefore $M_{\rm gi} = M_{\rm Zi} / Z_{\rm i} \lesssim M_{\rm Zi} / Z_{\rm s}$. With $Z_{\rm s} \lesssim y \sim \rm Z_{\odot}$ for the mean of a typical cluster of galaxies, we therefore expect that $M_{\rm gi} \sim M_{\rm Zi} / \rm Z_{\odot}$, very roughly. This implies that the elliptical galaxies in the Coma cluster have ejected ${\sim} 1 \: \rm M_{\odot}$ of gas per solar luminosity of the cluster, which is at the lower end of the range of estimates of the cluster gas content, from X-ray data (Section~\ref{subsec:Data2}). Various specific calculations using the models discussed in Sections~\ref{subsubsec:Supernova-driven winds}~and~\ref{subsubsec:Bursts of star formation in merging subsystems} lead to similar results within a factor ${\sim} 3$. It is therefore uncertain whether gas loss from ellipticals could be the entire source of gas in a cluster like Coma, or whether some of the IGM is simply primordial material that was never in a galaxy. (In the latter case, the intergalactic metals could nevertheless have been supplied by galaxies).

An interesting comment on the origin of the cluster IGM has been made by \citet{Ostriker1977a}: the distribution of morphological types of galaxies in clusters like Coma differs from the field distribution in having a much smaller fraction of spirals, many more S0s, and somewhat more ellipticals \citep{Oemler1974TheClusters}. If one ``corrects'' the cluster galaxies by adding disk matter until the overall ratio (disks)/(elliptical galaxies + bulges) equals that of the field, the mass of extra disk matter required is ${\sim} 50\%$ of the (ordinary) mass of galaxies in the cluster. This in turn is a significant fraction of the mass of cluster IGM. \citet{Ostriker1977a} therefore proposes that some of the IGM is material that would have been made into disks in a less dense environment, but instead was swept up into a hot ambient IGM. This idea ties in with several scenarios for the formation of disk galaxies, in which the disk is made from diffuse gas that is accreted after the formation of a spheroidal component. For example, in the model of \citet{Ostriker1975GalacticHalos}, a significant fraction of the disk is shed by halo stars; and in the picture of \citet{Tinsley1979StellarGalaxies}, the disk forms from diffuse gas after denser pieces have merged to make the bulge.

\subsubsection{Ejection from evolving stars in elliptical galaxies}
\label{subsubsec:Ejection from evolving stars in elliptical galaxies}

The stellar population in elliptical galaxies (and S0 galaxies and the bulge--halo components of spirals) is predominantly very old, and the light of these galaxies is dominated by red giants. It is almost certain that such stars lose a few tenths of a solar mass, between the MS turnoff and the end of their lives, to die as white dwarfs of typically $0.7 \: \rm M_{\odot}$ (Section~\ref{subsubsec:Solar mass}). This mass has been included in the total mass loss considered above, but it is interesting to calculate also the present \emph{rate} of mass loss by stars in elliptical galaxies.

\medskip

For an analytical estimate, let us assume that all the stars in the system formed at the same time, $t = 0$. Let $M_{0}$ be the mass of stars formed, and let $\phi(m)$ be the IMF; the mass of stars formed in the mass interval $(m, \: m + dm)$ is therefore
\begin{equation}
    n(m) \ dm = M_{0} \phi(m) \ dm,
	\label{eq:eq5.14}
\end{equation}
by Equation~(\ref{eq:eq2.1}). Now imagine these stars peeling off the MS and dying soon afterward as they reach their lifetimes $\tau_{\rm m}$. The number of stars dying per unit time is clearly
\begin{equation}
    D(t) = n \left( m_{\rm t} \right) \left| \frac{dm}{d \tau_{\rm m}} \right|_{\tau_{\rm m} = t},
	\label{eq:eq5.15}
\end{equation}
where $m_{\rm t}$ is the turnoff mass ($\tau_{\rm m} = t$). The stellar mass--lifetime relation can be approximated by a power law,
\begin{equation}
    \frac{m}{m_{1}} = \left( \frac{\tau_{\rm m}}{\tau_{1}} \right)^{-\theta},
	\label{eq:eq5.16}
\end{equation}
where $\tau_{1}$ is the lifetime of a fiducial mass $m_{1}$ and $\theta \simeq 0.25$ in the mass range of interest ($m_{\rm t} \sim 1 \: \rm M_{\odot}$). It is convenient to use a power-law IMF, Equation~(\ref{eq:eq2.3}), normalized to $\phi(m_{1}) \equiv \phi_{1}$; masses in only the small range ${\sim} 0.5 - 1 \: \rm M_{\odot}$ are relevant to the following calculation, so this IMF may be a reasonable approximation even if a single power law would not apply to all masses. The ejection rate can be obtained by multiplying $D(t)$ by $(m_{\rm t} - w_{\rm m})$, the mass lost per star with remnant mass $w_{\rm m}$, with the result
\begin{equation}
    E(t) = M_{0} \phi_{1} \theta \frac{m_{1}}{\tau_{1}} \left( m_{\rm t} - w_{\rm m} \right) \left( \frac{t}{\tau_{1}} \right)^{-1 + \theta x}.
    \label{eq:eq5.17}
\end{equation}
Since $(m_{\rm t} - w_{\rm m})$ changes slowly with time and $\theta x$ is probably only a few tenths, this expression shows that the ejection rate $E(t)$ varies approximately as $t^{-1}$. In Section~\ref{subsubsec:The stellar mass loss rate relative to luminosity}, an analytical expression is derived for the luminosity of stars in this model, and it is shown that Equation~(\ref{eq:eq5.17}) leads to a ratio of ejection rate to integrated blue luminosity of the population,
\begin{equation}
    \frac{E}{L_{\rm B}} \sim 0.02 \: \rm M_{\odot} \ L_{B \odot}^{-1} \ Gyr^{-1}
    \label{eq:eq5.18}
\end{equation}
at a present time ${\sim} 10^{10} \: \rm yr$.

\medskip

Most elliptical galaxies have a mass of neutral hydrogen that is less than 0.1 times their luminosity, in solar units, and many better studied ellipticals have less neutral hydrogen than 0.01 times their luminosity \citep{Knapp1978HGalaxies}. \citet{Faber1976HWinds} argue that significant amounts of ISM cannot be hiding in elliptical galaxies in ionized or molecular form. Thus the ejection rate given by Equation~(\ref{eq:eq5.18}) would provide more than the observed amount of gas in a few Gyr, or even in less than $1 \: \rm Gyr$. Possible fates for this gas have been thoroughly discussed by \citet{Faber1976HWinds}; they note that star formation at the rate in Equation~(\ref{eq:eq5.18}) would be detectable (unless only low-mass stars form), so they conclude that the gas ejected from stars is being continually lost from the galaxies. On the other hand, \citet{Oemler1979TypeStars} argue that star formation at the rate required to use up this gas could have escaped detection in most ellipticals, and could account for their supernova rate.

\subsection{Abundance Gradients in Disks}
\label{subsec:Abundance Gradients in Disks}

Abundance gradients in disk stars and gas cannot be fully accounted for by gradients in the gas fraction (Section~\ref{subsec:Data2}), so it is of interest to see whether dynamical processes analogous to those discussed in Section~\ref{subsec:Abundance Gradients in Spheroidal Systems}, for spheroidal systems, could be responsible. A gradient in the IMF could again be invoked, but this mechanism will not be discussed further.

\begin{figure*}
	\includegraphics[width=0.97\textwidth]{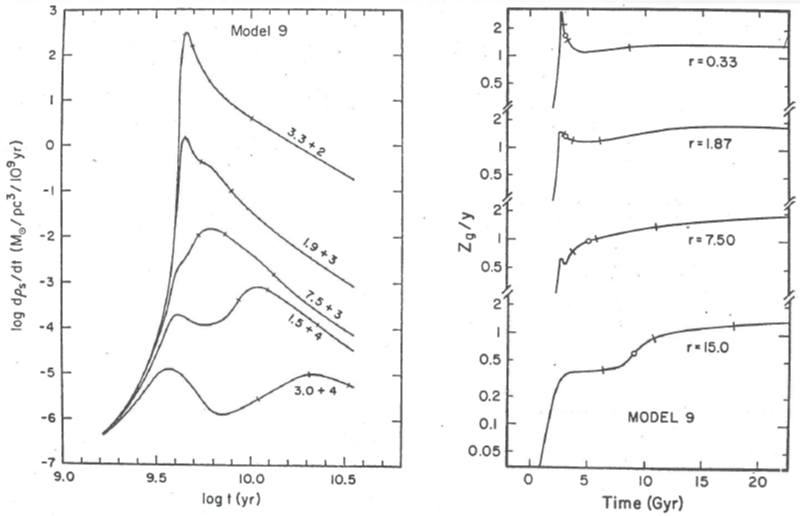}
    \caption{(a) Star formation rates at several radii in the equatorial plane of a collapse model for the formation of a disk galaxy \citep{Larson1976a}. The radius in pc is marked on each curve, and the three ticks indicate the times at which star formation is $10\%$, $50\%$, and $90\%$ complete (relative to the final mass of stars) at that radius. (b) Metal abundances in the gas (relative to the yield) in the equatorial plane of the same model \citep{Tinsley1978ChemicalDisks}. In this Figure, the radii are given in kpc, ticks have the same meaning as before, and open circles denote the time of maximum gas density at each radius.}
    \label{fig:fig12}
\end{figure*}

\subsubsection{Effects of infall}
\label{subsubsec:Effects of infall}

The idea that disks of galaxies form by accretion, incorporating metals from the young halo, has been suggested by dynamical models and supported by the metallicities of stars in the Solar neighborhood (Section~\ref{subsec:Effects of Galaxy Formation}). The properties of the accretion process that affect chemical evolution are the metallicity of infalling gas ($Z_{\rm f}$) and the ratio of SFR to infall rate ($\psi / f$). If there is a radial gradient in these quantities, there must be a corresponding metallicity gradient in the disk. In particular, the metallicity of the gas at any time tends to a value given by Equation~(\ref{eq:eq4.6}), $Z \rightarrow ky + Z_{\rm f}$, where $k = (1 - R) \psi / f$. Thus the disk gas has about this metallicity at the present time, in regions where infall is at all effective, i.e., where $k$ is not so large that $Z$ takes too long to approach its asymptotic value; the stars in turn reflect the metallicity of the gas at the time when they formed.

In the dynamical models studied by \citet{Tinsley1978ChemicalDisks}, the value of $\psi / f$ decreases outward in the disk at all times, because star formation is less efficient at low densities; $Z_{\rm f}$ is negligible at late times at all radii, but it has a significant negative gradient at early stages when metals from the young halo (central bulge) were most important near the center. Figure~\ref{fig:fig12}~(a) illustrates the SFR versus time at several radii in the equatorial plane of one of these models, and Figure~\ref{fig:fig12}~(b) gives the corresponding metallicities in the gas. At small radii, most stars form early from metal-rich infalling gas, while the outer regions experience star formation on a much longer timescale and are still relatively metal-poor. The gas at the present time thus has a negative metallicity gradient due mainly to the gradient in $\psi / f$, while the gradient in stellar abundances is due partly to the early gradient in $Z_{\rm f}$. The sizes of the model gradients are comparable to observed values, so this very schematic model may possibly be showing some effects that occur in real disk galaxies.

Generalizing these results, we can conclude as in Section~\ref{subsec:Effects of Galaxy Formation} that the chemical properties of disks could plausibly be strongly affected by gas flows that constitute the formation of the disk itself.

\subsubsection{Effects of radial gas flows}
\label{subsubsec:Effects of radial gas flows}

Radial inflow of gas in disks possibly occurs as a result of transfer of angular momentum by viscosity, loss of angular momentum from the gas to spiral or bar-like density waves, and other mechanisms \citep[e.g.][]{Kalnajs1978AObservations, Ostriker1977b}. These processes are rather speculative, since the inflow in many models could be a numerical artifact, but it is interesting to see how chemical evolution could be affected by flow velocities of a plausible magnitude.

The metals are concentrated inward by this process, as by gaseous dissipation in a collapsing spheroidal system (Section~\ref{subsubsec:Dissipative collapse}). Let us consider an annulus of a galaxy between radii $r$ and $r + \delta r$, measured in the disk. The chemical evolution of this annulus can be studied in the instantaneous recycling approximation, using Equations~(\ref{eq:eq3.17})~and~(\ref{eq:eq3.19}). Let $M_{\rm g}$ and $\psi$ in those equations be replaced by $2 \pi r M_{\rm g} \delta r$ and $2 \pi r \psi \delta r$, respectively, where $M_{\rm g}$ and $\psi$ now denote the corresponding surface densities; let $f$ be replaced by the net rate of inflow into the annulus, i.e., $F(r) - F(r + \delta r) = -(\partial F / \partial r) \delta r$, where $F$ is a flow rate (in $\rm M_{\odot} \ yr^{-1}$) with a positive sign for outward motion; and let $Z_{\rm f} f$ be replaced by the net rate of inflow of metals, which is $Z(r) F(r) - Z(r + \delta r) F(r + \delta r) = -Z (\partial F / \partial r) \delta r - (\partial Z / \partial r) F \delta r$. The equations then reduce to
\begin{equation}
    \frac{\partial M_{\rm g}}{\partial t} = -(1 - R) \psi - \frac{1}{2 \pi r} \frac{\partial F}{\partial r},
    \label{eq:eq5.19}
\end{equation}
and
\begin{equation}
    M_{\rm g} \frac{\partial Z}{\partial t} = y(1 - R) \psi - \frac{1}{2 \pi r} \frac{\partial Z}{\partial r} F.
    \label{eq:eq5.20}
\end{equation}
Equation~(\ref{eq:eq5.20}) shows that the radial flow is consistent with a steady-state abundance gradient,
\begin{equation}
    \frac{\partial (Z / y)}{\partial r} \sim 2 \pi r \frac{(1 - R) \psi}{F},
    \label{eq:eq5.21}
\end{equation}
which is negative if the flow is inward. The flow causes $Z$ to change on a timescale
\begin{equation}
    \tau_{\rm F} \sim \frac{2 \pi r^{2} M_{\rm g}}{\lvert F \rvert}.
    \label{eq:eq5.22}
\end{equation}
$F$ can be expressed in terms of the flow velocity $v$, where $\lvert F \rvert = $ (mass of gas in the annulus)/(time for gas to flow across $\delta r$) $= 2 \pi r M_{\rm g} \lvert v \rvert$. The timescale for radial flow to be effective is thus
\begin{equation}
    \tau_{\rm F} \sim \frac{r}{\lvert v \rvert},
    \label{eq:eq5.23}
\end{equation}
and the corresponding gradient can be written
\begin{equation}
    \frac{\partial (Z / y)}{\partial \ \ln(r)} \sim \frac{\tau_{\rm F}}{\tau_{*}},
    \label{eq:eq5.24}
\end{equation}
where $\tau_{*} \equiv M_{\rm g} / (1 - R) \psi$ is the timescale for star formation to use up the gas. These relations show that rapid inflow, with a timescale less than that for star formation, quickly obliterates any radial metallicity gradient, while slow inflow can lead to a significant one.

Substituting values of $\psi$, $M_{\rm g}$, and $r$ for the Solar neighborhood, it is found that the interstellar abundance gradient (Section~\ref{subsec:Data2}) is consistent with inflow at a few $\rm km \ s^{-1}$, carrying a flux ${\sim} 1 \: \rm M_{\odot} \ yr^{-1}$; the timescale for the gradient to change is a few Gyr. There is no strong evidence for the occurrence of systematic gas flows of this magnitude in the Galaxy, but nor can they be ruled out.

\citet{Sanders1977ChemicalGalaxy} has suggested that the deep minimum in the surface density of gas in the Galaxy, in an annulus between $0.6$ and $4 \: \rm kpc$, could be due to inflow into the central $600 \: \rm pc$, where the total quantity of ISM is enough to fill the depleted annulus. If so, inflow could perhaps be fueling the strong star formation at the Galactic center.

\section{Approaches to Photometric Evolution}
\label{sec:Approaches to Photometric Evolution}

Evolution of stars in galaxies affects not only their chemical compositions, but also their integrated luminosities, colors, and spectra. Photometric and chemical evolution can be studied separately, because they depend largely on complementary properties of a galaxy: the colors at a given time are governed strongly by the current rate of star formation relative to its past average value, whereas the chemical composition depends mainly on the integrated past star formation relative to the gas content and on the ratio of the SFR to gas flow rates. In studying photometric evolution, we can ignore the effects of ISM, except in correcting colors for any reddening or gaseous emission, and we can avoid assumptions relating the SFR to the gas supply. Of course, a complete understanding of the properties of a galaxy would include the relations among its history of star formation, gas content and gas flows, and chemical composition (cf. Figure~\ref{fig:fig1}), but more can be learned by tackling pieces of the puzzle separately first.

\subsection{Aims and Methods}
\label{subsec:Aims and Methods}

Models for the stellar population of a galaxy address three related questions: what types of stars are present, what history of star formation produced them, and what was the population and its photometric properties in the past? The answers to these questions have many applications, such as interpreting in terms of star formation the correlations between photometric and morphological properties of galaxies, and predicting changes on cosmological timescales.

Methods of constructing population models can be divided into three categories: ``population synthesis'' with no explicit evolution, evolutionary models, and analytical approximations.

\subsubsection{Population synthesis}
\label{subsubsec:Population synthesis}

This approach is to find the ``best'' mixture of stars to match the colors of the galaxy under study. The inferred distribution of stars in the HR diagram then contains information on their past formation rate and the IMF, and it is often possible to judge the mean chemical composition and even to detect minor components of high or low metallicity. The procedure is to observe the colors of the galaxy and a variety of nearby stars, generally including narrow-band photoelectric colors and indices giving the strengths of spectral features that are sensitive to stellar temperature, luminosity, or composition. Then synthetic colors are computed for various mixtures of stars and compared with the galaxy colors. The search for an optimal mixture can be made in many ways, ranging from trial-and-error to elaborate computer algorithms; the method generally used, quadratic programming, was introduced to the field by \citet{Faber1972QuadraticSynthesis.}. Because of observational errors, and because the available nearby stars do not include all types in the galaxy under study, a perfect fit is seldom found, and the solution that (formally) minimizes the errors is not necessarily the most plausible. The choice of a ``best'' synthetic population must therefore be based on imposed astrophysical constraints; these include such requirements as a smooth MS luminosity function, and a distribution of subgiants and giants that could plausibly arise from evolution off the MS. The lack of an objectively defined best fit means that the final solution depends strongly on the imposed constraints, as emphasized by \citet{Williams1976PopulationGalaxies}.

There are often several astrophysically acceptable synthetic populations that match the galaxy colors equally well but correspond to significantly different histories of star formation. An example of such ambiguity appears in models for elliptical galaxies, reviewed by \citet{Faber1977ThePopulations}. All studies agree that \emph{most} of the light of these galaxies, from blue through infrared wavelengths, comes from an old stellar population with a distribution in the HR diagram like an old open cluster plus an extended giant branch (cf. Figure~\ref{fig:fig5}). However, such models almost always fail by a few percent to account for the light around $3500~\si{\angstrom}$ (e.g. $U - B$ is predicted to be too red by ${\sim} 0.1 \: \rm mag$), so they are lacking some hot stellar component that is present in the real galaxies. To date it has been impossible to determine whether the hot stars are a few upper-main-sequence stars, implying ongoing star formation, or a minor population of old objects such as horizontal-branch stars or blue stragglers (which can be seen in the color--magnitude diagram for the old cluster M67, Figure~\ref{fig:fig5}). Obviously, it would be very interesting to know if typical elliptical galaxies really are still making stars at a slow residual rate! For the central bulge of M31, which is optically indistinguishable from an elliptical galaxy, there are broadband colors down to $1550~\si{\angstrom}$, but even these data have not resolved the ambiguity \citep{Wu1980TheM81}. A new avenue has been opened by a demonstration that the integrated light of stars in a nuclear bulge region of our own galaxy matches exactly the integrated light of comparable regions of spirals and ellipticals \citep{Whitford1978SpectralGalaxies}. The brighter stars are individually observable in the Galactic bulge, so a star-by-star synthesis of their contribution to the light is possible. Perhaps in the end this approach will tell whether an old population alone can account for the ultraviolet light.

\medskip

Another general problem is that even the best defined regions in the HR diagram cannot be interpreted uniquely in terms of a past history of star formation. The models are insensitive to many details of the IMF and SFR, for two basic reasons:

\begin{enumerate}
  \item the integrated light of galaxies is dominated by regions of the HR diagram that depend theoretically on rather few parameters of star formation; and
  
  \vspace{1mm}
  
  \item some types of stars, such as red giants, may have evolved from a wide range of MS masses (Section~\ref{subsubsec:1 - 4 Msun}), so they cannot be traced uniquely back to an initial mass and time of star formation.
\end{enumerate}

The second of these problems is avoided in the evolutionary method described next, but the first remains and will be discussed below.

\subsubsection{Evolutionary models}
\label{subsubsec:Evolutionary models}

This approach relies primarily on stellar evolution theory to suggest allowable populations, as follows. Theoretical tracks (or isochrones) of stars in the HR diagram are used to compute the stellar population that would arise, at a given age, from a given SFR and IMF, with a given chemical composition; the integrated colors are then calculated, using observed colors of stars in appropriate parts of the HR diagram, and the results are compared with the colors of the galaxy under study. The aim is to derive from a series of models the SFR, IMF, age, and composition(s) that best match the galaxy, and thereby to learn not only about its present stellar population but also about its past history and past photometric properties.

In practice, stellar evolution is not well enough understood for fully theoretical models to be reliable. The main problems are related to late stages of evolution, including  particularly the giant branches in old stellar populations, whose effects on models for elliptical galaxies are reviewed by \citet{Faber1977ThePopulations}. These problems are alleviated by using statistical studies of nearby giants to provide semi-empirical evolutionary tracks (Section~\ref{subsubsec:Solar mass}), and by allowing the most uncertain types of stars to be present in numbers that are treated as adjustable parameters. This method thus closely resembles some non-evolutionary population syntheses in which the constraints are chosen to represent stellar evolutionary tracks \citep{OConnell1976GalaxyEllipticals}.

The evolutionary approach has several advantages. The best established aspects of stellar evolution theory are incorporated, so the resulting population is a \emph{possible} one as far as can be determined. Uncertainties cannot be formally calculated, but from trials with a variety of assumptions one can estimate subjectively the allowable range of parameters. Often this range is small enough to lead to useful conclusions about the past history of star formation, and to predictions of photometric changes of cosmological interest. For example, it is possible to determine the slope of the IMF in elliptical galaxies closely enough to be sure that their integrated luminosity declines with increasing age (Section~\ref{subsec:Evolution of a Single Generation of Stars}).

Uncertainties in the conclusions from this method arise partly from uncertainties in stellar evolution, and partly from the intrinsic insensitivity of integrated colors to many parameters of interest -- a problem found earlier with population syntheses. Two parts of the HR diagram tend to dominate the integrated light, as illustrated spectroscopically by the work of, e.g., \citet{Morgan1957AGalaxies} and \citet{Morgan1969OnGalaxies}: B stars on the upper main sequence, and late G through early M giants. These dominant regions are extended out to O stars in ultraviolet light and to late M giants in the infrared. If young stars are absent, low-mass giants dominate at visual and longer wavelengths, so the colors depend much more on stellar evolution than on the IMF or past SFR; at shorter wavelengths, however, turnoff stars are seen so the colors give some information on the age of the system. If young stars are present, the light at short wavelengths is dominated by OB stars, whose relative numbers depend on the IMF and whose total numbers (relative to red stars) depend on the ratio of the present SFR to its integrated past value. Stars with lifetimes from a few times $10^{8} \: \rm yr$ to just below the age of the system (usually A and F stars) contribute relatively little light, so there is little information on either their part of the IMF or the detailed time-dependence of the SFR. In Section~\ref{subsec:UBV Colors of Normal Galaxies}, models will be discussed that illustrate the dominance of the upper main sequence and/or low-mass giants, depending on the SFR.

\medskip

Programs for constructing evolutionary models have been described by \citet{Tinsley1968EvolutionGalaxies, Tinsley1972a, Tinsley1978b}, \citet{Searle1973TheGalaxies}, \citet{TinsleyGunn1976a}, and \citet{Larson1978StarGalaxies}. The mechanical details are far less troublesome than the input ``data'' representing stellar tracks, and it is easy to obtain numerical accuracy far exceeding the astrophysical certainty of the calculations. There are two types of this technique.

\begin{enumerate}
  \item The first method is to supply the computer with evolutionary tracks in the HR diagram for stars with a series of discrete masses, or with isochrones for a series of discrete ages; separate stellar data are used for each chemical composition of interest. Then, for a given IMF and SFR, the calculation yields the numbers of stars on a large grid of points in the HR diagram, as a function of the age of the system.
  
  \vspace{1mm}
  
  \item The second method uses the first type of program once only for each IMF and composition, to give the integrated colors at a series of ages of a model whose SFR consists of a single initial burst. These are then regarded as the colors of ``generations'' of stars with a given age (and IMF and composition).
\end{enumerate}

A model with any prescribed SFR can then be treated, at each age, as the sum of such generations in proportions given by the SFR. The number of generations whose properties must be combined to obtain the integrated colors of any model is much smaller than the number of points in the HR diagram that are referred to directly in the first method, so the second approach is more economical. In either method, it is clearly possible to add arbitrary numbers of stars of undetermined evolutionary status, in the spirit of population synthesis.

While models with Solar metallicity can rely on nearby stars to provide colors and semi-empirical evolutionary tracks, there is no such convenient sample for other compositions. In making models for non-Solar metallicities, it is often most convenient to change the ``standard'' models differentially, rather than starting from scratch with tracks and colors for each set of stars. \citet{Faber1973VariationsGalaxies} first used the metallicity effects discussed in Section~\ref{subsubsec:Initial composition} to estimate differential changes in the integrated colors of elliptical galaxies, as a function of metallicity, and her methods have been adapted by others subsequently. Recent results for elliptical galaxies have been cited in Section~\ref{subsec:Data2}. The calculations of metallicity effects in integrated light are still much less secure than one would like, and there is a need for more basic work on stellar evolution and atmospheres at non-Solar compositions, including non-Solar abundance ratios among elements heavier than helium \citep{Faber1977ThePopulations}.

\subsubsection{Analytical approximations}
\label{subsubsec:Analytical approximations}

Some of the results from evolutionary models can be understood qualitatively using analytical approximations. These have proved particularly tractable for models in which all the stars form in a single initial burst, which is a first approximation to the population in elliptical galaxies. Such models will be considered next.

\subsection{Evolution of a Single Generation of Stars}
\label{subsec:Evolution of a Single Generation of Stars}

Many numerical models designed to match detailed photometry of elliptical galaxies have shown that nearly all the light at visual and longer wavelengths can be accounted for by a very old population, with a turnoff near the Sun's position on the MS. The metallicities of the dominant stars appear to be within a factor of two of Solar in wide-aperture observations of giant ellipticals, although their centers may be more metal-rich and small ellipticals are metal-poor (Section~\ref{subsec:Data2}). Reviews by \citet{vandenBergh1975StellarGalaxies} and \citet{Faber1977ThePopulations} cover the history and recent status of this subject, and a few subsequent developments have been referred to in Section~\ref{subsec:Aims and Methods}. The implications of a predominantly very old population for the evolution of elliptical galaxies are best understood using analytical approximations.

\subsubsection{Content and luminosity}
\label{subsubsec:Content and luminosity}

Let us consider a single generation of stars, formed with total mass $M_{0}$ in a short burst (as in Section~\ref{subsubsec:Ejection from evolving stars in elliptical galaxies}), with a fixed chemical composition near Solar. The population evolves by peeling off the MS, as can be visualized from Figures~\ref{fig:fig2}~and~\ref{fig:fig3}.

The IMF will be taken to be a power law, normalized to $\phi(m_{1}) \equiv \phi_{1}$, where $m_{1}$ is the turnoff mass at a fiducial time $\tau_{1}$. The power-law approximation need only hold over a small mass interval, since the light at present comes almost entirely from stars between $0.4 \: \rm M_{\odot}$ and turnoff, and the turnoff mass at ages of interest, ${\sim} 5 - 20 \: \rm Gyr$, lies in the small range ${\sim} 0.9 - 1.2 \: \rm M_{\odot}$.

At a time $t$ after star formation, the MS stars present have masses from the lower limit at formation, $m_{\rm L}$, up to the turnoff mass $m_{\rm t}$, which is given by substituting $\tau_{\rm m} = t$ in Equation~(\ref{eq:eq5.16}). Thus the number of dwarfs with masses in the interval $(m, \ m + dm)$ is, by Equation~(\ref{eq:eq2.3}),
\begin{equation}
\begin{medsize}
    n_{\rm d} (m) \ dm = M_{0} \phi(m) \ dm = M_{0} \phi_{1} \left( \frac{m}{m_{1}} \right)^{-(1 + x)} \ dm, \: \: m_{\rm L} \leq m \leq m_{\rm t}.
	\label{eq:eq6.1}
\end{medsize}
\end{equation}
Stars slightly more massive than $m_{\rm t}$ are present as giants, and their total number is the number of stars that were on the MS with lifetimes between $t$ and $t - \tau_{\rm g}$, where $\tau_{\rm g}$ is the duration of post-MS evolution for masses ${\sim} m_{\rm t}$. (The term ``giants'' is used loosely here to mean all post-MS stars; the analysis can easily be modified to refer to any portion of post-MS evolution). The number of giants is therefore
\begin{equation}
    n_{\rm g} (t) = M_{0} \phi(m_{\rm t}) \left| \frac{dm}{d \tau_{\rm m}} \right|_{\tau_{\rm m} = t}; \: \: \tau_{\rm g} = M_{0} \phi_{1} \theta \frac{m_{1}}{\tau_{1}} \tau_{\rm g} \left( \frac{t}{\tau_{1}} \right)^{-1 + \theta x}.
	\label{eq:eq6.2}
\end{equation}

The luminosity of individual dwarfs in the mass range of interest can be approximated by a power law,
\begin{equation}
    \ell_{\rm d} (m) = \ell_{1} \left( \frac{m}{m_{1}} \right)^{\alpha},
	\label{eq:eq6.3}
\end{equation}
where $\alpha \simeq 5$. For giants, an average luminosity $\ell_{\rm g}$ is defined so that the product $\ell_{\rm g} \tau_{\rm g}$ gives correctly the integrated light output during post-MS evolution. The values of $\ell_{1}$, $\alpha$, and $\ell_{\rm g}$ of course depend on the wavelength interval of interest, and so do the results below relating to luminosities. (For bolometric light, the product $\ell_{\rm g} \tau_{\rm g}$ is proportional to the amount of nuclear energy used, but it has no such interpretation in restricted wavelength bands).

The integrated luminosities and masses of dwarfs and giants can now be derived from Equations~(\ref{eq:eq6.1})~--~(\ref{eq:eq6.3}) and Equation~(\ref{eq:eq5.16}). It will be assumed in the integrals that $m_{\rm L} \ll m_{1}$. The total mass of dwarfs at time $t$ depends critically on whether the slope of the IMF $(x)$ is less than or greater than 1:
\begin{subequations}
 \begin{empheq}[left={ M_{\rm d} (t) = \int\limits_{m_{\rm L}}^{m_{\rm t}} m n_{\rm d} (m) \ dm = \empheqlbrace\,}]{align}
  & \frac{M_{0} \phi_{1} m_{1}^{2}}{x - 1} \left( \frac{m_{\rm L}}{m_{1}} \right)^{-x + 1}, \: \: \: \, x > 1, \label{eq:eq6.4a} \\
  & M_{0} \phi_{1} m_{1}^{2} \ \ln \left( \frac{m_{\rm t}}{m_{\rm L}} \right), \: \: \: \: \: \; x = 1, \label{eq:eq6.4b} \\
  & \frac{M_{0} \phi_{1} m_{1}^{2}}{1 - x} \left( \frac{t}{\tau_{1}} \right)^{-\theta (1 - x)}, \: \: x < 1. \label{eq:eq6.4c}
 \end{empheq}
\end{subequations}
Giants have a total mass ${\sim} m_{\rm t} n_{\rm g} (t)$, and one can quickly verify that the mass ratio of giants to dwarfs is greatest in the case $x < 1$, and is at most ${\sim} \tau_{\rm g} / t \sim 0.1$; the contribution of giants to the total mass will therefore be neglected. The integrated luminosity of dwarfs is
\begin{equation}
    L_{\rm d} (t) = \int_{m_{\rm L}}^{m_{\rm t}} \ell_{\rm d} (m) n_{\rm d} (m) \ dm = \frac{M_{0} \phi_{1} m_{1} \ell_{1}}{\alpha - x} \left( \frac{t}{\tau_{1}} \right)^{-\theta (\alpha - x)},
	\label{eq:eq6.5}
\end{equation}
on the assumption $x < \alpha$, which is justified below. Finally, the integrated luminosity of giants is
\begin{equation}
    L_{\rm g} (t) = \ell_{\rm g} n_{\rm g} (t) = M_{0} \phi_{1} \theta \frac{m_{1}}{\tau_{1}} \ell_{\rm g} \tau_{\rm g} \left( \frac{t}{\tau_{1}} \right)^{-1 + \theta x}.
	\label{eq:eq6.6}
\end{equation}

The above relations will be used to derive some interesting properties of this single generation of stars.

\subsubsection{Remnants of dead stars}
\label{subsubsec:Remnants of dead stars}

There may be a significant dark mass in the form of remnants of stars initially above $m_{\rm t}$, especially if the IMF has a fairly shallow slope so these stars were relatively numerous. Although it is probably a very poor approximation to extrapolate the IMF to high masses with the slope $x$ used near $1 \: \rm M_{\odot}$, the equations will be written to show how the contributions of remnants can be estimated in the simplest cases. (These results can easily be modified to allow for a variable slope). In this approximation, it will be assumed that all remnants have the same mass $w$, and that all stars above $m_{\rm t}$ are dead. Then the total mass of remnants is $w$ times the number of stars formed with masses between $m_{\rm t}$ and the upper limit $m_{\rm U}$:
\begin{equation}
    M_{\rm w} (t) = w \int_{m_{\rm t}}^{m_{\rm U}} M_{0} \phi(m) \ dm = \frac{M_{0} \phi_{1} m_{1} w}{x} \left( \frac{t}{\tau_{1}} \right)^{\theta x},
	\label{eq:eq6.7}
\end{equation}
assuming $m_{\rm U} \ll m_{\rm t}$ and $x > 0$. The relative mass of remnants is potentially greatest if $x < 1$, and then Equation~(\ref{eq:eq6.4c}) shows that $M_{\rm w} / M_{\rm d} \sim w / m_{\rm t}$, which could be close to unity. This result is obviously strongly dependent on the assumption of a single power law for the whole IMF, which would exaggerate the mass of remnants if, for example, elliptical galaxies have a curved IMF like the function in the Solar neighborhood (Figure~\ref{fig:fig4}). It may be concluded that dead remnants could possibly affect the total mass by a factor ${\sim} 2$, which cannot be predicted with any confidence from constraints on the slope of the IMF at turnoff.

\subsubsection{The ratio of giants to dwarfs in the light}
\label{subsubsec:The ratio of giants to dwarfs in the light}

Some spectral features in the integrated light of elliptical galaxies depend sensitively on the relative amounts of light contributed by giant and dwarf stars at the feature wavelength. Examples are an iron hydride band at $0.99 \: \rm \upmu m$, known as the Wing-Ford band, which \citet{Whitford1977ThePopulations.} has found to be extremely strong in late dwarfs but weak in late giants; and a carbon monoxide band at $2.2 \: \rm \upmu m$, studied especially by \citet[][and earlier papers cited therein]{Frogel1978PhotometricGalaxies}, which has the opposite behavior, being much stronger in late giants than in late dwarfs. Since the light of elliptical galaxies at those wavelengths must be dominated by late-type stars, the galaxies should show a weak FeH band and a strong CO band if giants outshine dwarfs, and vice versa. As the following analysis shows, the relative luminosities of giants and dwarfs give important information on the slope of the IMF, which in turn affects many other properties of elliptical galaxies including the rate of evolution of total luminosity; it is the significance of this effect for cosmological tests (Section~\ref{subsubsec:Evolution of luminosity and the Hubble diagram}) that has motivated much of the analysis of spectral features.

Equations~(\ref{eq:eq6.5})~and~(\ref{eq:eq6.6}) together give an approximate expression for the relative luminosities of giants and dwarfs:
\begin{equation}
    G(t) \equiv \frac{L_{\rm g} (t)}{L_{\rm d} (t)} = \theta (\alpha - x) \frac{\ell_{\rm g} \tau_{\rm g}}{\ell_{1} \tau_{1}} \left( \frac{t}{\tau_{1}} \right)^{\theta \alpha - 1}.
	\label{eq:eq6.8}
\end{equation}
An alternative expression is obtained by substituting Equations~(\ref{eq:eq5.16})~and~(\ref{eq:eq6.3}):
\begin{equation}
    G(t) = \theta (\alpha - x) \frac{\ell_{\rm g} \tau_{\rm g}}{\ell_{\rm d} \left( m_{\rm t} \right) t}.
	\label{eq:eq6.9}
\end{equation}
The term $\ell_{\rm g} \tau_{\rm g}$ is the amount of energy radiated (at a given wavelength) by a star of approximately turnoff mass after it leaves the MS, while $\ell_{\rm d} \left( m_{\rm t} \right) t$ is the energy radiated during MS evolution. Thus Equation~(\ref{eq:eq6.9}) says that the value of $G$ in bolometric light is, in order of magnitude, equal to the ratio of nuclear fuel consumed after leaving the MS to that consumed on the MS; since stars near $1 \: \rm M_{\odot}$ burn the hydrogen in only $10\%$ of their mass while on the MS but in $70\%$ before they die (Section~\ref{subsubsec:Solar mass}), this fuel ratio is ${\sim} 6$. This high value is the underlying reason why giants can outshine dwarfs in the integrated light of a galaxy, despite their very short lifetimes. Giants tend to be especially dominant at long wavelengths, because most of the energy from the giant branch as a whole comes from red giants.

The fuel burning ratio is not the only factor affecting $G$, however. The term $(\alpha - x)$ in Equation~(\ref{eq:eq6.9}) introduces a dependence on $x$, the slope of the IMF. A larger value of $x$ reduces the contribution of giants simply by reducing the number of stars in the mass range of giants (just above turnoff) relative to those still on the MS. The dependence of $G$ on $x$ is of great practical importance, since it allows spectroscopic criteria to set constraints on $x$. The work of \citet{Whitford1977ThePopulations.} and \citet{Frogel1978PhotometricGalaxies} shows that the red--infrared light of elliptical galaxies is strongly dominated by giants, to an extent that $x$ must be less than 2, and possibly less than 1. These constraints are consistent with the IMF in the Solar neighborhood, which has $x < 1$ in the relevant mass range (Figure~\ref{fig:fig4} and Equation~\ref{eq:eq2.9}).

The infrared spectra of elliptical galaxies set constraints not only on the IMF but also on the relative numbers of M giants of different spectral types that populate the giant branch. As noted in Section~\ref{subsubsec:Evolutionary models}, these numbers are not firmly predicted by stellar evolution theory, so studies of galaxy spectra can add to an understanding of late stages in the lives of low-mass stars. This application of galaxy models is discussed by \citet{Faber1977ThePopulations}, \citet{Tinsley1978b}, and references therein.

\subsubsection{The stellar mass loss rate relative to luminosity}
\label{subsubsec:The stellar mass loss rate relative to luminosity}

An expression for the rate of mass loss from stars has been derived in Section~\ref{subsubsec:Ejection from evolving stars in elliptical galaxies}, but Equation~(\ref{eq:eq5.17}) is not in a useful form for comparing with observable quantities. It is possible to obtain a useful equation for the ejection rate per unit integrated luminosity, because both quantities scale with the populations of stars near turnoff.

From Equations~(\ref{eq:eq6.5})~and~(\ref{eq:eq6.8}), the total luminosity can be written
\begin{equation}
    L(t) = \left[ 1 + G(t) \right] L_{\rm d} (t) = \frac{M_{0} \phi_{1} m_{1} \ell_{1}}{\alpha - x} (1 + G) \left( \frac{t}{\tau_{1}} \right)^{-\theta (\alpha - x)}.
	\label{eq:eq6.10}
\end{equation}
Then, with Equation~(\ref{eq:eq5.17}), the ratio of ejection rate to luminosity is
\begin{equation}
    \frac{E(t)}{L(t)} = \frac{\theta (\alpha - x)}{\ell_{1} \tau_{1}} \frac{m_{\rm t} - w_{\rm m}}{1 + G} \left( \frac{t}{\tau_{1}} \right)^{\theta \alpha - 1},
	\label{eq:eq6.11}
\end{equation}
which shows that the ratio depends only slowly on time. A more useful relation for finding the present ratio is given by substituting Equation~(\ref{eq:eq5.16})~and~(\ref{eq:eq6.3}) to eliminate $\ell_{1}$ and $\tau_{1}$, with the result
\begin{equation}
    \frac{E(t)}{L(t)} = \theta (\alpha - x) \frac{m_{\rm t} - w_{\rm m}}{1 + G} \frac{1}{\ell_{\rm d} \left(m_{\rm t} \right) t}.
	\label{eq:eq6.12}
\end{equation}

This ratio can be estimated for present-day ellipticals as follows. From spectroscopic studies in \emph{blue} light, $G \simeq 1$ (the value of $G$ is greater in red or bolometric light); and approximate values of the other quantities are $\alpha \simeq 5$, $\theta \simeq 0.25$, $m_{\rm t} \simeq 1 \: \rm M_{\odot}$, $w_{\rm m} \simeq 0.7 \: \rm M_{\odot}$, $\ell_{\rm d} \simeq 1 \: \rm L_{\rm B \odot}$, $t \simeq 10 \: \rm Gyr$, $x \simeq 1$. The result from Equation~(\ref{eq:eq6.12}) is then $E / L_{\rm B} \simeq 0.015 \: \rm M_{\odot} \ L_{\rm B \odot} Gyr^{-1}$, of which the significance was discussed in Section~\ref{subsubsec:Ejection from evolving stars in elliptical galaxies}.

\subsubsection{The mass-to-luminosity ratio}
\label{subsubsec:The mass-to-luminosity ratio}

An analytical estimate of $M_{\rm s} / L$ can be made using the mass of stars $M_{\rm s} \simeq M_{\rm d} (t)$ (neglecting the small contribution of giants and the very uncertain contribution of dead remnants), and the total luminosity $L(t)$. From Equations~(\ref{eq:eq6.4a})~--~(\ref{eq:eq6.4c}), it is clear that the result depends strongly on whether $x \lessgtr 1$. Moreover, it depends critically on the assumption that $x$ is constant down to $m_{\rm L}$, since the least massive stars (or sub-stellar objects) can be numerous enough to dominate the mass while contributing negligibly to the light. If $x < 1$, the result from Equations~(\ref{eq:eq6.4c})~and~(\ref{eq:eq6.10}) is
\begin{equation}
    \frac{M_{\rm s}}{L} = \frac{\alpha - x}{1 - x} \frac{1}{1 + G} \frac{m_{\rm t}}{\ell_{\rm d} \left( m_{\rm t} \right)}, \: \: x < 1,
	\label{eq:eq6.13}
\end{equation}
which is proportional to the mass-to-luminosity ratio of turnoff stars. If $x > 1$ (or if $x$ increases from a value below 1 at turnoff to above 1 at smaller masses) Equation~(\ref{eq:eq6.4a}) shows that $M_{\rm s} / L$ increases in proportion to $m_{\rm L}^{-(x - 1)}$, so it is sensitive to a quantity that cannot be determined photometrically.

In all cases, photometric data (star counts, population syntheses, spectroscopic estimates of $x$) yield only a \emph{lower limit} to the true mass-to-luminosity ratio $(M / L)$ of a galaxy, since any amount of mass could be present in hidden form. When the masses of galaxies are determined dynamically, the empirical $M / L$ values often increase to such large values in the outer regions that a large amount of hidden mass must indeed be present \citep[e.g.][]{Spinrad1978HalosRatios}. For this reason, values of $M / L$ determined from population syntheses and equivalent methods are sometimes called ``photometric $M / L$ ratios'' to distinguish them from the ratios of actual mass (defined dynamically) to luminosity.

\subsubsection{Evolution of luminosity and the Hubble diagram}
\label{subsubsec:Evolution of luminosity and the Hubble diagram}

\begin{figure}
	\includegraphics[width=0.47\textwidth]{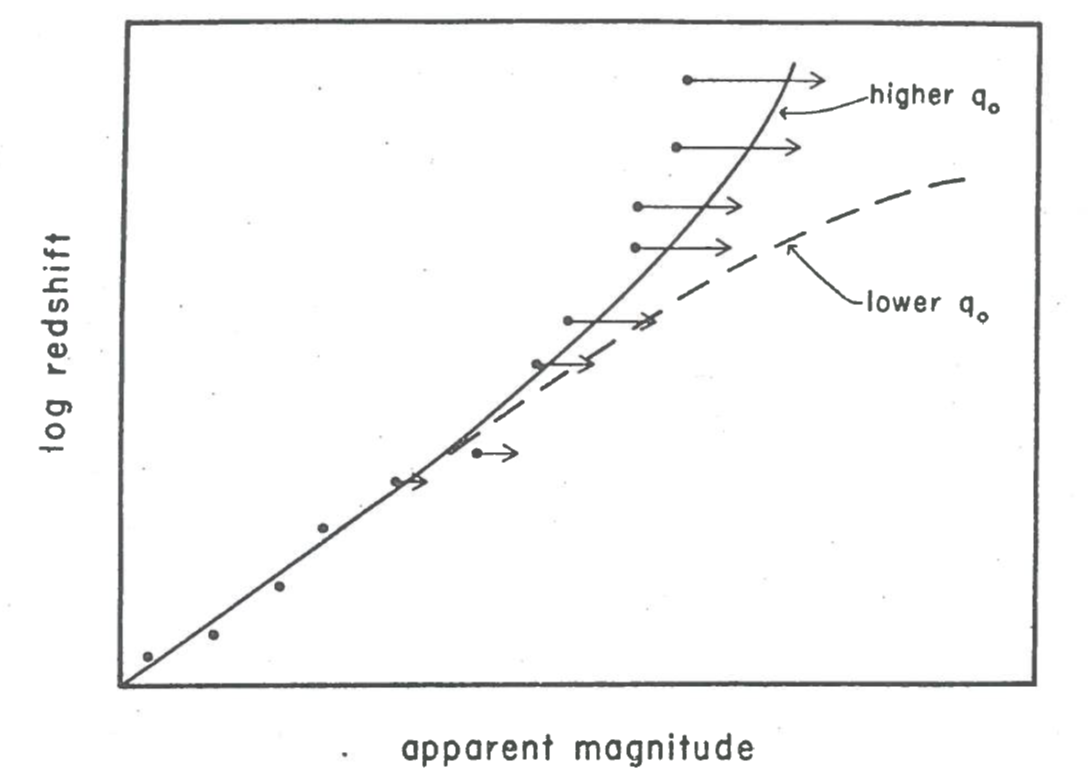}
    \caption{Schematic Hubble diagram showing how both the deceleration parameter ($q_{0}$) and evolution of galaxies affect the departure from linearity. \emph{Lines} are schematic ``theoretical'' curves for two values of $q_{0}$, \emph{dots} are hypothetical data points, and \emph{arrows} indicate qualitatively how they should be corrected if the net effect of evolution is to make distant galaxies intrinsically brighter than nearby ones. More precisely, evolution (in this sense) at the rate of a few percent of a galaxy's luminosity per Gyr makes the true value of $q_{0}$ smaller by about unity than the value inferred from the uncorrected data points.}
    \label{fig:fig13}
\end{figure}

In one of the classic cosmological tests, the Hubble diagram, logarithmic redshifts of galaxies are plotted against against their apparent magnitudes, as illustrated schematically in Figure~\ref{fig:fig13}. For a sample with a well-defined mean absolute magnitude, this diagram can be regarded heuristically as a plot of ``recession velocity'' versus ``distance''. At small redshifts, the regression line is linear with a slope corresponding to Hubble's Law, $\rm redshift \propto distance$. At large redshifts, the deviation from linearity measures the change in the ratio ``velocity'' / ``distance'' with distance itself; since the lookback time (the light-travel time) increases with distance, the curvature of the Hubble diagram thus gives a measure of the past expansion rate of the Universe, and in particular of its deceleration. The deceleration parameter $q_{0}$ can take only positive values in the simplest cosmological models of General Relativity, the Friedmann models, and $1/2$ is a critical value: if $q_{0} > 1/2$, the deceleration is large enough for the expansion eventually to be reversed, but if $0 < q_{0} \leq 1/2$, the Universe will expand forever; if in fact $q_{0}$ is negative, indicating that the expansion is accelerating, more complicated cosmological models are required. Evolution of galaxies enters the picture because the lookback times sampled must be many Gyr for the deceleration to be detectable; the galaxies then had significantly different luminosities, so the ``distance'' parameter, apparent magnitude, cannot be estimated on the assumption of a constant absolute magnitude. The departure of the Hubble diagram from linearity is very sensitive to evolution: if the luminosities of elliptical galaxies grow fainter at a few percent per Gyr, for example, the apparent value of $q_{0}$ (inferred from the shape of the Hubble diagram) exceeds its true value by several tenths. This problem has been discussed by \citet{Humason1956RedshiftsNebulae.}, \citet{Sandage1961a, Sandage1961b}, \citet{Gunn1975SpectrophotometryCosmology}, and \citet{Tinsley1972b, Tinsley1977c}. For an approximate estimate of the evolutionary correction to $q_{0}$, the above analytical equations can be used.

\begin{figure}
	\includegraphics[width=0.47\textwidth]{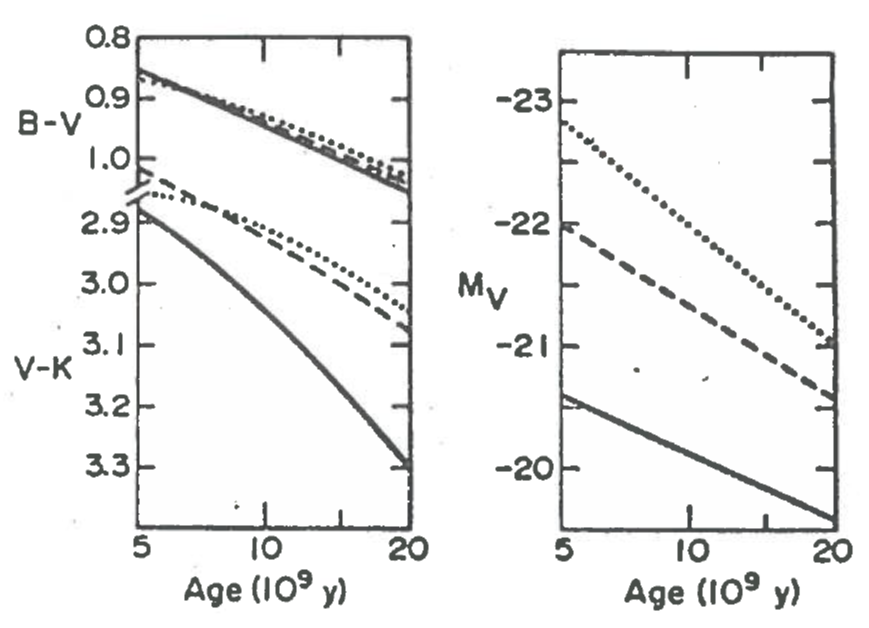}
    \caption{Evolution of colors and magnitudes of single-generation models for the stellar population in elliptical galaxies \citep{TinsleyGunn1976a}. Curves are for three values of the slope of the IMF: \emph{solid lines}, $x = 2$; \emph{dashes}, $x = 1$; \emph{dots}, $x = 0$. Note that if $x$ is small, colors evolve slowly but magnitudes evolve quickly.}
    \label{fig:fig14}
\end{figure}

From Equation~(\ref{eq:eq6.10}), we have
\begin{equation}
    \frac{d(\ln \ L)}{d(\ln \ t)} = -\theta (\alpha - x) + \frac{t}{1 + G} \frac{dG}{dt},
	\label{eq:eq6.14}
\end{equation}
and Equation~(\ref{eq:eq6.8}) can be used to evaluate $dG / dt$. The term $\ell_{\rm g} \tau_{\rm g}$ in the expression for $G(t)$ depends only slowly on time, because giant branch evolution depends only weakly on mass in the relevant range, so only the explicit time-dependence need be considered and that term gives $(t / G) (dG / dt) = \theta \alpha - 1$. Substituting in Equation~(\ref{eq:eq6.14}), we have
\begin{equation}
    \frac{d(\ln \ L)}{d(\ln \ t)} = -\theta (\alpha - x) + \frac{G}{1 + G} (\theta \alpha - 1).
	\label{eq:eq6.15}
\end{equation}
The second term in Equation~(\ref{eq:eq6.15}) is not very important, since $(\theta \alpha - 1)$ is a few tenths and $G / (1 + G)$ lies between 0 and 1. The main term is therefore simply $(-\theta \alpha + \theta x)$, which can be written
\begin{equation}
    \frac{d(\ln \ L)}{d(\ln \ t)} \simeq -1.3 + 0.3 x.
	\label{eq:eq6.16}
\end{equation}
Essentially the same result is obtained for the evolution of luminosity in numerical population models. The examples in Figure~\ref{fig:fig14} show the predicted dependence on the IMF: the rate at which $M_{V}$ gets dimmer is slower in models with a larger value of $x$. Since giants supply most of the light, this behavior is mainly because, when $x$ is large, the giant branch is fed by a more richly populated main sequence as time goes on.

\medskip

In the Hubble diagram, evolution means that departures from linearity are due not only to $q_{0}$ but also to systematic changes in the absolute magnitudes of galaxies (Figure~\ref{fig:fig13}). If the curvature is interpreted without regard to evolution, the result is an apparent value of $q_{0}$ that differs from the true value by
\begin{equation}
    \Delta q_{0} \equiv {\rm apparent \ value - true \ value} \simeq -1.5 \frac{d(\ln \ L)}{d(\ln \ t)}
	\label{eq:eq6.17}
\end{equation}
\citep[e.g.][]{Tinsley1977c}. A first-order estimate, from Equation~(\ref{eq:eq6.16}), is therefore
\begin{equation}
    \Delta q_{0} \simeq 2.0 - 0.4 x.
	\label{eq:eq6.18}
\end{equation}

The slope of the IMF, $x$, emerges as the critical parameter. As discussed in Section~\ref{subsubsec:The ratio of giants to dwarfs in the light}, spectroscopic studies indicate that $x < 2$, and possibly $x < 1$. In the first case, $|\Delta q_{0}| \gtrsim 1$, and in the second case, $|\Delta q_{0}| \gtrsim 1.5$. In either case, the correction for evolution is big enough to make a qualitative difference to the type of cosmology inferred.

Current estimates of the apparent value of $q_{0}$ range from ${\sim} 0$ \citep{Gunn1975SpectrophotometryCosmology} to ${\sim} 1.5$ \citep{Kristian1978TheLine}, the differences being due to unknown sampling and observational effects. Downward corrections of order unity are clearly important in determining whether the true value of $q_{0}$ is greater than $1/2$, less than $1/2$, or even negative.

A value of $x \gtrsim 5$ would be needed to make stellar evolution negligible in the Hubble diagram, but such a steep IMF would make the late dwarfs dominate the infrared light to an extent that is precluded by the giant-dominated spectra of elliptical galaxies. A possible loophole is the following. Equation~(\ref{eq:eq6.16}) depends on the value of $x$ for stars near turnoff, while the infrared spectra depend on the ratio of giants to dwarfs of types K and M; if the IMF were to turn over sharply between ${\sim} 0.5$ and $1 \: \rm M_{\odot}$ (i.e., having far fewer less massive stars that the turnoff slope would predict), one could have both a steep slope at turnoff and a very small contribution from dwarfs to the infrared light. This idea is, of course, completely ad hoc, since the IMF in the Solar neighborhood has $x < 2$ for all masses $< 10 \: \rm M_{\odot}$, and does not cut off above ${\sim} 0.2 \: \rm M_{\odot}$. It is therefore most reasonable to conclude that elliptical galaxies have giant-dominated spectra because the IMF has a fairly shallow slope at turnoff; if so, their luminosity evolves fast enough to make the apparent value of $q_{0}$ exceed its true value by 1 or more.

However, this is not all we need to know to unravel the Hubble diagram. The galaxies used for this test are the central cluster giants, which are believed to grow secularly by cannibalizing their neighbors (Section~\ref{subsubsec:Mergers of stellar systems}). This process could plausibly lead to a growth rate in the total stellar population of several percent per Gyr, with a corresponding increase of luminosity in opposition to the effect just discussed. The dynamical effects cannot yet be calculated accurately enough for a correction to be applied to the Hubble diagram, so this test does not yet give a usefully accurate value of $q_{0}$. The situation is reviewed by \citet{Tinsley1977c}.

\subsubsection{Evolution of colors}
\label{subsubsec:Evolution of colors}

Predictions of color evolution are of interest because they can be tested by observations of distant elliptical galaxies whose ages are several Gyr younger than nearby galaxies.

The colors of a single-generation population become redder with age, if the main course of stellar evolution is peeling off the MS at turnoff and following the red giant branch. The main contribution to the color change is the redward evolution of the turnoff, since giant evolution is insensitive to turnoff mass in the range of interest. Consequently, colors evolve faster if the light is less giant-dominated, i.e. if $x$ is larger, in contrast to the integrated luminosities just discussed. This behavior is illustrated in Figure~\ref{fig:fig14}.

Qualitatively different behavior is predicted if the stars can lose enough mass to become blue horizontal-branch stars, instead of the red ``clump'' giants that normally represent the core helium burning stage of metal-rich low-mass stars (Section~\ref{subsubsec:Solar mass}). It has been suggested that such stars lose mass at a variety of rates, some becoming late red giants and others becoming blue. Numerical models for galaxy populations in which mass loss occurs stochastically on the red giant branch have been studied by \citet{Ciardullo1978TheGalaxies}. Because evolution to a blue position in the HR diagram occurs only if the star has a small mass of envelope left, the fraction of giants becoming blue increases as the turnoff mass decreases. The upshot is that the integrated colors of the model galaxies evolve blueward after ${\sim} 8 \: \rm Gyr$.

Observations of distant elliptical galaxies are ambiguous on this point, as reviewed by \citet{Spinrad1977TheRedshifts}. Some of the most distant central cluster galaxies known, with redshifts ${\sim} 0.6$, have intrinsic colors that are bluer than those of nearby ellipticals, but the distant galaxies were selected on the basis of strong radio emission so they may be atypical. If they are typical, the color change is about that expected according to the type of models that evolve monotonically toward redder colors (e.g. Figure~\ref{fig:fig14}); the lookback time sampled is ${\sim} 4 - 7 \: \rm Gyr$, depending on the cosmological model. Another sample of central cluster galaxies with redshifts up to nearly $0.5$ has no systematic dependence of color on redshift that can be disentangled from the intrinsic scatter \citep{Wilkinson1978SpectralGalaxies}.

Dramatic color differences between nearby and distant galaxy populations in clusters have been discovered by \citet{Butcher1978The295, Butcher1978TheClusters}. Nearby clusters, i.e. those with lookback times $< 1 \: \rm Gyr$, have galaxy populations that are strongly correlated with the cluster morphology: loose, irregular clusters have a large fraction of spiral galaxies, and centrally concentrated, regular clusters have very few spirals and mainly S0 and elliptical galaxies; the brighter galaxies in regular clusters are correspondingly all red. However, in two regular clusters with lookback times ${\sim} 5 \: \rm Gyr$, the bright galaxies are found to have a wide range of colors, including many that are as blue as late-type spiral galaxies. On the assumption that the distant regular clusters represent younger versions of the nearby ones, these very blue galaxies must evolve (in a few Gyr) into red S0s or ellipticals. The color change observed is many times greater than any predictions based on the evolution of single-generation populations, so it is concluded that those galaxies were actively forming stars just a few Gyr ago. Presumably, they are mainly the precursors of S0 galaxies seen in nearby clusters, in which star formation is undetectable.

\section{Colors and Star Formation Rates}
\label{sec:Colors and Star Formation Rates}

The stellar populations in most galaxies are far more complicated than those in ellipticals, because young stars are important contributors to the light. The time-dependence of the SFR is therefore an important parameter in addition to the three quantities (age, IMF, and metallicity) used to characterize old populations, and the latter quantities could also be changing in time. Moreover, the colors of spiral and irregular galaxies are often affected by internal reddening and gaseous emission lines. Despite the complications presented by these galaxies, it is especially interesting to try to understand their photometric properties in terms of histories of star formation. Applications of such studies include explaining correlations between form and photometric properties, finding what physical conditions are conducive to star formation, and searching for young galaxies.

Models for galaxies with ongoing star formation are usually numerical; analytical approximations are cumbersome except in the simplest case of a constant SFR \citep{Tinsley1973AnalyticalGalaxies}. This Section will consider models that study only a few simple properties, mainly just UBV colors. Although more can be learned from spectroscopic details, the UBV system has the advantage of an extensive and homogeneous compilation of galaxy colors in the Second Reference Catalogue of Bright Galaxies (\citealp{deVaucouleurs1976SecondGalaxies.}; to be referred to as \citetalias{deVaucouleurs1976SecondGalaxies.}).

\subsection{UBV Colors of Normal Galaxies}
\label{subsec:UBV Colors of Normal Galaxies}

The UBV colors of a sample of morphologically normal galaxies are shown in Figure~\ref{fig:fig15}; the crosses are all elliptical and S0 galaxies, and the dots are a variety of morphological types, which we consider first. The colors of these galaxies form such a narrow distribution in the two-color diagram that it is tempting to look for one dominant parameter that could vary among galaxies and lead to a one-dimensional set of colors. Because the appearance and spectra of galaxies suggest a progression of star-forming activity, ranging from very active in late-type irregulars to negligible in ellipticals, it is natural to suggest that the color sequence is due to different proportions of young and old stars. Population syntheses and evolutionary models have confirmed this view, and their conclusions can be summarized (with some oversimplification) in a ``standard scenario'' for galaxy evolution: normal galaxies have the same IMF and mean stellar metallicities, and they are of the same age, but they differ in the time-dependence of their SFRs; in particular, the latest (bluest) types of galaxies form stars on a long timescale, while the earliest (reddest) ceased star formation long ago. This hypothesis is obviously inaccurate in detail, but it provides a useful starting point. It will be used to construct a series of ``standard'' galaxy models, whose colors will be compared with observations, and then the effects of factors other than the SFR will be considered in turn.

\subsubsection{``Standard'' models}
\label{subsubsec:Standard models}

\begin{figure}
	\includegraphics[width=0.47\textwidth]{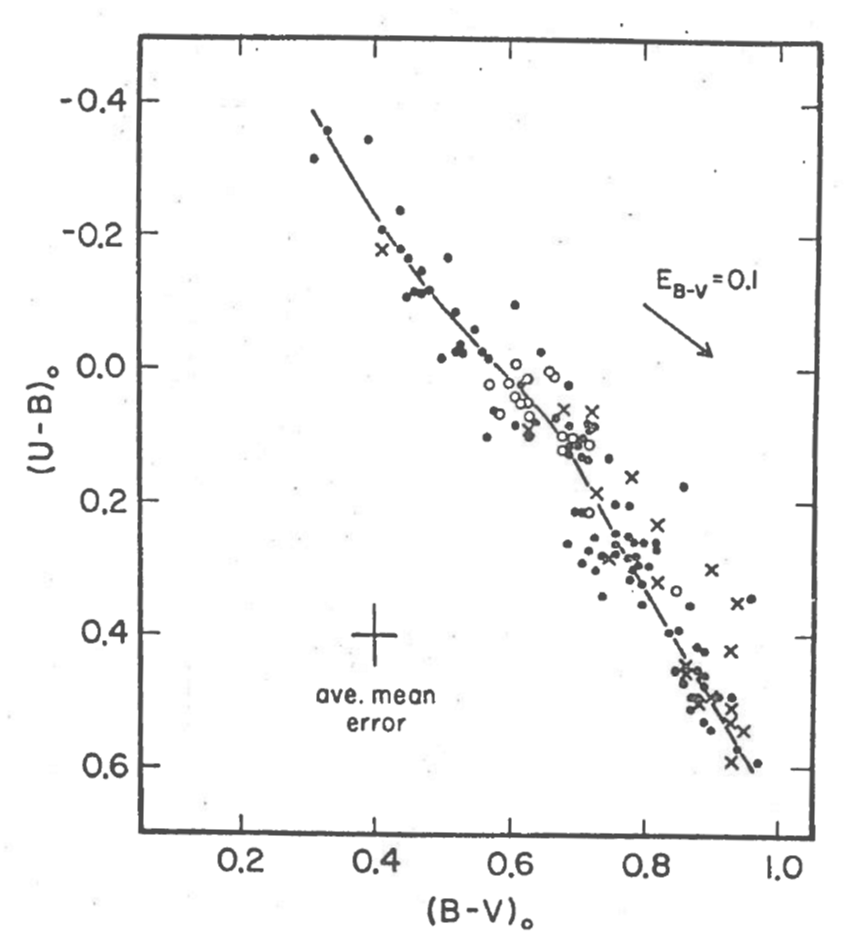}
    \caption{Two-color diagram for morphologically normal galaxies and globular clusters. \emph{Filled circles}: galaxies from the \emph{Hubble Atlas} \citep{Sandage1961c}, excluding peculiars and those with galactic latitudes $|b| < 20^{\circ}$, with corrected colors from the \citetalias{deVaucouleurs1976SecondGalaxies.}; the \emph{error cross} is for this sample, and the \emph{solid line} is its mean locus estimated by eye \citep{Larson1978StarGalaxies}. \emph{Crosses}: E and S0 galaxies in the Virgo cluster, with colors from \citet{Sandage1972AbsoluteColor}, corrected for reddening according to the \citetalias{deVaucouleurs1976SecondGalaxies.} formulae. \emph{Open circles}: galactic globular clusters (excluding those with $E_{B - V} > 0.05$), with colors from \citet{Harris1979GlobularGalaxies}, corrected for reddening according to the \citetalias{deVaucouleurs1976SecondGalaxies.} formulae.}
    \label{fig:fig15}
\end{figure}

\begin{figure}
	\includegraphics[width=0.47\textwidth]{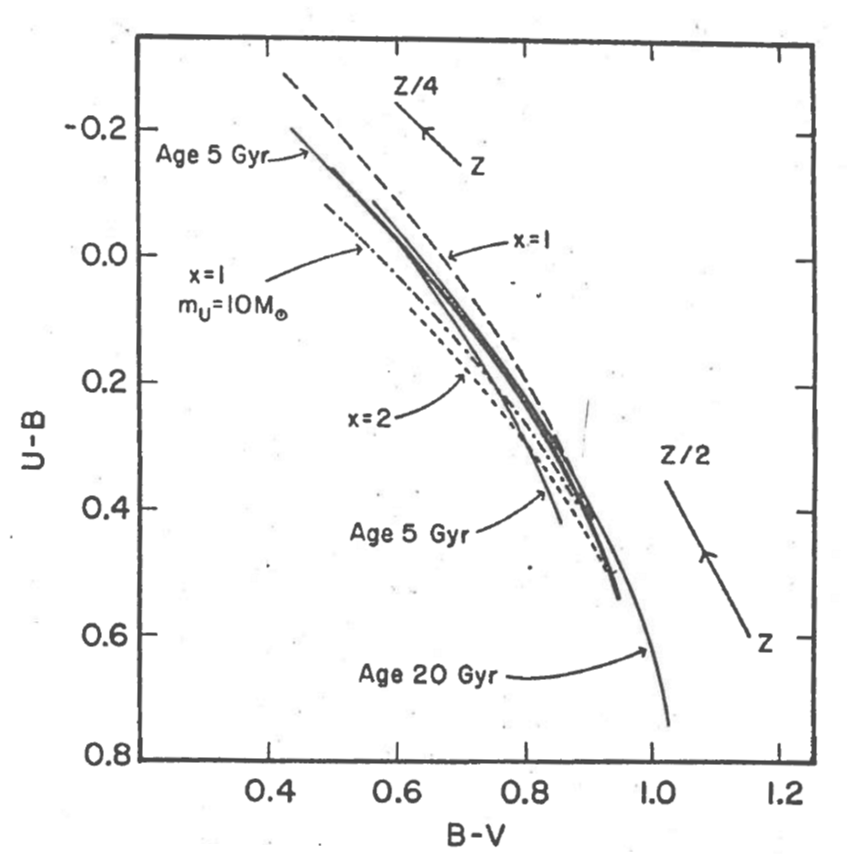}
    \caption{Theoretical two-color diagram for galaxies with monotonic SFRs, Solar metallicity, and the local IMF (Section~\ref{subsec:UBV Colors of Normal Galaxies}). \emph{Heavy line}: ``standard'' models, i.e. those of age $10 \: \rm Gyr$, with SFRs ranging from constant at the top to a single initial burst at the bottom. \emph{Light solid lines}: models differing from the standard set only in age, as indicated. \emph{Dashes}: models differing from the standard set only in having an IMF with a constant slope, as marked. \emph{Dash-dot line}: models differing from the preceding set with $x = 1$ only in having an upper stellar mass limit $m_{\rm U} = 10 \: \rm M_{\odot}$, whereas all other models shown have $m_{\rm U} = 30 \: \rm M_{\odot}$. \emph{Arrows}: approximate estimates of the effect on colors of blue and red galaxies, respectively, of altering the metallicity by the factor indicated.}
    \label{fig:fig16}
\end{figure}

The methods discussed in Section~\ref{subsubsec:Evolutionary models} have been used by various authors to construct models corresponding to the standard scenario; the (typical) results shown here are from \citet{Larson1978StarGalaxies}. Let us consider models with the IMF of the Solar neighborhood, Solar (or old-disk) metallicity, and an age of $10 \: \rm Gyr$; the exact choice of these standard parameters is not critical, as shown below. The models have monotonically decreasing SFRs ($\psi$), ranging from constant to a single initial burst lasting $10^{7} \: \rm yr$. Different series of models have different monotonic functions $\psi(t)$ between these extremes, such as exponential functions, negative powers of time, and combinations of the constant and single-burst models as two components in different proportions. The colors of these series in the UBV diagram all lie very near the locus indicated by a single heavy line in Figure~\ref{fig:fig16}: the model with a constant SFR is at the top of this line, that with an initial burst is at the bottom, and the form of the curve in between is essentially the same for all series with various functional forms for $\psi(t)$. The theoretical locus for these standard models is very close to the observed mean locus for galaxies of different types (line in Figure~\ref{fig:fig15}), so the standard scenario is at least superficially consistent. Two conclusions can be stated.

\begin{enumerate}
  \item \emph{The UBV colors of normal galaxies can in general be accounted for by models with the same age, metallicity, and IMF}; most of the observed colors lie in the range predicted for monotonically decreasing SFRs, within the observational errors. A further conclusion is that late-type galaxies are not necessarily young, even though their appearance and blue-light spectra are dominated by short-lived OB stars; the integrated colors are instead consistent with an underlying population of stars with ages up to many billions of years. These conclusions have been stressed in the context of evolutionary models by \citet{Tinsley1968EvolutionGalaxies}, \citet{Searle1973TheGalaxies}, \citet{Larson1974PhotometricGalaxies, Larson1978StarGalaxies}, and \citet{Huchra1977StarGalaxies}. Caveats and deviations from the norm are discussed below.
  
  \vspace{1mm}
  
  \item \emph{Models with monotonically declining SFRs (and with the same age, metallicity, and IMF) define a one-parameter sequence in the ($U - B$, $B - V$) plane}. Inspection of the models shows that the parameter is the ratio of the present SFR to its average value over the lifetime of the system; or equivalently the SFR per unit mass of stars ($\psi_{1} / M_{\rm s}$ or $\psi_{1} / \overline{\psi}_{1} t_{1}$); or equivalently the inverse of these quantities, a \emph{timescale for star formation} $T_{1} \equiv \overline{\psi}_{1} t_{1} / \psi_{1}$, in the notation of Section~\ref{subsec:IMF}. Galaxies of the latest morphological types are the bluest objects in Figure~\ref{fig:fig15}, and these evidently have the longest timescales for star formation, while the earliest types, which are the reddest, have the shortest timescales; this point will be discussed further in Section~\ref{subsubsec:Relevance to the formation and structure of normal galaxies}.
\end{enumerate}

The one-parameter sequence shows that UBV colors for a given value of $T_{1}$ are almost independent of the functional form of $\psi(t)$, as long as it is monotonic. An unfortunate consequence of this result is that the UBV colors of a galaxy (if near the mean locus in Figure~\ref{fig:fig15}) cannot give any more information about the SFR than the quantity $T_{1}$. In practice, they give less information, because of ambiguities due to possible variations of metallicity, etc., as shown below.

The sensitivity of colors to the single parameter $T_{1}$ is due to the dominance of low-mass giants and/or young OB stars in the light of galaxies, as discussed in Section~\ref{subsubsec:Evolutionary models}. In effect, the contribution of low-mass giants is proportional to the number of long-lived stars ever formed, and the contribution of upper-main-sequence stars is proportional to the present SFR, so their ratio is proportional to $T_{1}$. The integrated colors are insensitive to details of the past SFR because A -- F dwarfs with intermediate lifetimes contribute relatively little light, and because the nature of the giant branch changes little over a wide range of MS lifetimes for the precursor stars. For the same reasons, it is difficult to extract significantly more information about the history of star formation in a galaxy from more detailed photometry than from UBV colors.

\medskip

We next consider some possible problems with the simple one-parameter scenario.

\subsubsection{Possible effects of errors}
\label{subsubsec:Possible effects of errors}

Three systematic discrepancies between the models and data can be seen on comparing Figures~\ref{fig:fig16}~and~\ref{fig:fig15}: the heavy theoretical line lies about $0.05 \: \rm mag$ about the empirical mean locus, some galaxies are bluer than the bluest model, and some are redder than the reddest model. The systematic offset is no more than could be due to errors in the stellar evolution tracks, judged from series of models based on alternative tracks. If this offset is corrected ad hoc by moving the heavy theoretical line downward, there are still some bluer and redder galaxies than predicted. The differences seem to be too big to ascribe to uncertainties in the stellar evolution used, and they cannot be corrected by redefining the ``standard'' age or metallicity, since an improvement at the red end would leave more discrepant galaxies at the blue end, and vice versa. Nor can the ``standard'' IMF be changed, since if the IMF is universal it must be the same as the local function. Therefore, not all of the discrepancies between the heavy line and the data are due to theoretical errors within the framework of the standard scenario.

Although the mean error bar shown for the data in Figure~\ref{fig:fig15} is small, some of the colors may have significantly larger errors due to uncertainties in the reduction. The colors plotted were corrected in the \citetalias{deVaucouleurs1976SecondGalaxies.} on a statistical basis for Galactic and internal reddening, so excessively red and blue galaxies could result from inappropriate corrections in a few cases. A reddening vector (from the \citetalias{deVaucouleurs1976SecondGalaxies.}) is shown in Figure~\ref{fig:fig15}, and it indicates that galaxies away from the ends of the distribution could not be moved far from the mean locus except by extremely large over- or underestimates of their reddening, because the vector happens to lie almost parallel to the mean locus itself.

Emission lines can affect the colors of late-type galaxies, but the estimates made by \citet{Huchra1977StarGalaxies} indicate that morphologically normal galaxies are unlikely to have strong enough gaseous emission for this to be important.

\medskip

In summary, it seems likely that some normal galaxies have colors that are too red or too blue to be accounted for by the standard scenario. Two questions arise: How can the discrepant galaxies be accounted for? And could normal galaxies have significant variations in age, IMF, or metallicity that do not show up on the UBV plot?

\subsubsection{Variations in age}
\label{subsubsec:Variations in age}

Light lines in Figure~\ref{fig:fig16} indicate the effects of allowing ages between 5 and $20 \: \rm Gyr$. The loci for different ages overlap, so most of the galaxies in Figure~\ref{fig:fig15} could have any ages in this range. The extreme colors, however, do depend on age, and the bluest and reddest data points could be accounted for if the ages of galaxies vary by a factor ${\sim} 4$. We shall see that this is not the only possible explanation of those data points, since metallicity effects are probably important.

\subsubsection{Variations in metallicity}
\label{subsubsec:Variations in metallicity}

Effects of different stellar metallicities ($Z_{\rm s}$) can be estimated as outlined in Section~\ref{subsubsec:Evolutionary models}, and some approximate results are indicated in Figure~\ref{fig:fig16}; the slope of the vector for red galaxies is empirical, but the slope for blue galaxies and the length of each vector are uncertain by factors ${\sim} 2$.

As discussed in Section~\ref{subsec:Data2}, the sequence of colors for E--S0 galaxies (crosses in Figure~\ref{fig:fig15}) is regarded as one of metallicity. This sequence closely overlaps the locus of galaxies with different SFRs and ages, so UBV colors alone cannot unambiguously give the SFR parameter ($T_{1}$), age, and $Z_{\rm s}$ for a population of stars. The reddest points in Figure~\ref{fig:fig15} are giant elliptical galaxies, which almost certainly have a mean $Z_{\rm s}$ greater than Solar (in the aperture used for the colors); if the galaxies have some residual star formation, it is undetected to date, but it could affect the colors enough to change the estimated $Z_{\rm s}$ and/or age somewhat.

The bluest points in Figure~\ref{fig:fig15} are small late-type galaxies. It is known from studies of the gaseous emission lines in some such galaxies that they can be significantly metal-poor (e.g. a factor of 4 in the Small Magellanic Cloud; \citealp{Pagel1978AClouds}), so a low $Z_{\rm s}$ could help to make these points very blue. The effects of abundance changes on the colors of blue galaxies are too uncertain to say whether another effect, such as a somewhat younger age, is also required to account for their being bluer than the standard sequence.

\subsubsection{Variations in the IMF}
\label{subsubsec:Variations in the IMF}

To show possible effects of variations in the IMF, Figure~\ref{fig:fig16} includes the loci of UBV colors of models differing from the standard sequence only in their IMF. Two of the variants have constant slopes, $x = 1$ and $x = 2$, while the third has $x = 1$ and an upper limit $m_{\rm U} = 10 \: \rm M_{\odot}$ compared to $30 \: \rm M_{\odot}$ in all other cases\footnote{The models in Figures~\ref{fig:fig16}~and~\ref{fig:fig18} use an IMF slightly different from Equation~(\ref{eq:eq2.9}): the upper limit is $30 \: \rm M_{\odot}$ (except as stated), and the slope is $x = 1.3$ ($\phi \propto m^{-2.3}$) for all $m > 2 \: \rm M_{\odot}$. The UBV colors would be little affected if Equation~(\ref{eq:eq2.9}) itself, with $m_{\rm U}$ taking any value $\geq 30 \: \rm M_{\odot}$, were used \citep[cf.][]{Huchra1977StarGalaxies}.}. For blue galaxies, the local IMF gives colors between those for $x = 1$ and $x = 2$. In general, the colors are redder with a larger value of $x$ or a smaller value of $m_{\rm U}$, since there are relatively few upper-MS stars. Comparisons with Figure~\ref{fig:fig15} show that the variants illustrated are about the largest deviations from the local IMF that one could have without predicting a greater color spread than is observed.

This conclusion applies only to the bluer galaxies, and only to stars $\gtrsim 1 \: \rm M_{\odot}$ that contribute significantly to their light. It is clear from Figure~\ref{fig:fig16} that the UBV colors of redder galaxies ($B - V \gtrsim 0.8$) are very insensitive to the variations of IMF considered. Additional information on the IMF, derived from spectroscopic studies and $M / L$, was discussed in Section~\ref{subsubsec:Other IMF}. Possible departures from the local IMF discussed there are a lack of stars above $10 \: \rm M_{\odot}$ in some early-type spirals, and ubiquitous variations in the fraction of very low-mass objects. In elliptical galaxies, the IMF between ${\sim} 0.4$ and $1 \: \rm M_{\odot}$ cannot be very much steeper than the local function (Section~\ref{subsubsec:The ratio of giants to dwarfs in the light}).

\subsubsection{Relevance to the formation and structure of normal galaxies}
\label{subsubsec:Relevance to the formation and structure of normal galaxies}

To summarize the preceding discussion, every aspect of the standard scenario has been shown to have its weaknesses: the metallicity, IMF, and age are known to vary from one galaxy to another and/or could vary significantly without affecting the locus of normal UBV colors. Nevertheless, it is true that the main parameter causing the progression of colors of morphologically normal galaxies is the timescale for star formation. The average UBV colors of galaxies of different Hubble types lie along the middle of the distribution in Figure~\ref{fig:fig15}, with the latest types at the top and the earliest at the bottom \citep{deVaucouleurs1977QualitativeGalaxies.}. Thus \emph{there is a strong correlation between the structure of galaxies and their timescales for star formation}. Star formation seems to be most efficient in the galaxies with the highest bulge-to-disk ratios, and, among the spirals, it is most efficient in those with the tightly wound spiral arms.

\medskip

An obvious question is whether the shape of a galaxy is a consequence of its efficiency of star formation, or whether, conversely, the timescale for star formation is determined by the structure. Both effects are believed to be present. On one hand, more efficient early star formation leads to a galaxy with a greater bulge-to-disk ratio (Section~\ref{subsec:Galaxy_Formation}): the formation of a spheroidal component requires that the stars in this component formed on a timescale less than that for gaseous dissipation in the protogalaxy. On the other hand, the present structure of a galaxy governs its large-scale dynamics, which in turn has important effects on star formation (Section~\ref{subsubsec:SFR factors}); for example, a more prominent bulge implies a greater surface density, which can lead to stronger gas compression and so to more efficient star formation in the disk. The papers cited in Section~\ref{subsec:Galaxy_Formation} and Section~\ref{subsubsec:SFR factors} give many more detailed discussions, and further references, on the origin of the Hubble sequence of galaxy types and the numerous properties of galaxies that correlate with their forms.

S0 galaxies, which are disk galaxies without spiral structure, are an inscrutable class. Like elliptical galaxies, they have colors consistent with no ongoing star formation (or possibly a little, showing at short wavelengths; Section~\ref{subsubsec:Population synthesis}). There are currently two types of theory on the origin of S0 galaxies: in the first, they are former spirals that have no more star formation because their ISM was lost, either in a collision with another galaxy or by ram-pressure sweeping due to motion through an ambient IGM; in the second type of theory, S0 galaxies had intrinsically very efficient star formation in their disks at early stages. The second type of theory is preferred by some authors because S0 galaxies in isolation and in dense clusters have essentially the same low contents of neutral hydrogen \citep{Faber1976HWinds} and the same distributions of colors \citep{Sandage1978Color-absoluteGalaxies}, and because there are some structural differences between S0s and spirals \citep{Burstein1978TheGalaxies.}. Nevertheless, the first picture is supported circumstantially by the high proportion of S0 galaxies in clusters, especially in their central regions, and especially in clusters with hot IGM \citep{Oemler1974TheClusters, Melnick1977TheClusters}. Arguments based on colors are not decisive, because models in which star formation stopped long ago have similar present colors to those in which it stopped only a few Gyr ago \citep{Biermann1975OnGalaxies}. Whatever mechanism cuts off star formation, the very blue galaxy content of distant, regular clusters strongly suggests that many S0 galaxies were actively making stars only ${\sim} 4 \: \rm Gyr$ ago (Section~\ref{subsubsec:Evolution of colors}). It is especially interesting that some nearby clusters contain ``anemic'' spirals, with weak spiral structure and a subnormal neutral hydrogen content, that have been interpreted as disk systems at a stage of evolution between normal spirals and stripped S0s \citep{vandenBergh1976a}.

\subsection{Colors of Peculiar Galaxies}
\label{subsec:Colors of Peculiar Galaxies}

\begin{figure}
	\includegraphics[width=0.47\textwidth]{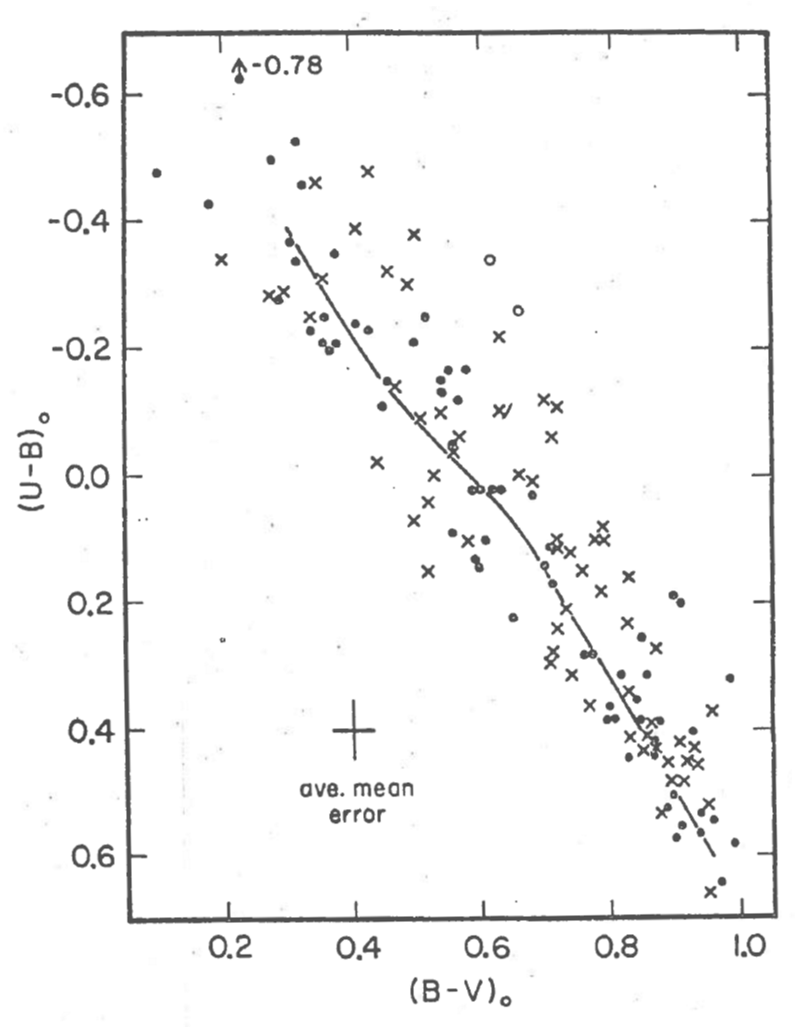}
    \caption{Two-color diagram for galaxies in the \emph{Atlas of Peculiar Galaxies} \citep{Arp1966AtlasGalaxies}, excluding those with galactic latitudes $|b| < 20^{\circ}$, with corrected colors from the \citetalias{deVaucouleurs1976SecondGalaxies.} and other sources cited by \citet{Larson1978StarGalaxies}. \emph{Crosses} denote interacting systems. \emph{Open circles} are two Type~I Seyfert galaxies, whose colors may be affected by non-thermal emission.}
    \label{fig:fig17}
\end{figure}

Galaxies with morphological peculiarities have peculiar colors too, as illustrated in Figure~\ref{fig:fig17}, which is a UBV diagram for systems in the \emph{Atlas of Peculiar Galaxies} \citep{Arp1966AtlasGalaxies}. The width of their color distribution is in striking contrast to the narrow locus of normal galaxies (Figure~\ref{fig:fig15}), and on closer inspection the width turns out to be due almost entirely to interacting galaxies, which are shown as crosses in Figure~\ref{fig:fig17} \citep{Larson1978StarGalaxies}. The implication is that dynamical disturbances have led to an unusual star formation history, so a study of these galaxies and their colors might shed some light on the process of star formation in general.

\subsubsection{Bursts of star formation and blue colors}
\label{subsubsec:Bursts of star formation and blue colors}

If the SFR in a galaxy does not decrease monotonically, colors very different from those of the standard models in Section~\ref{subsec:UBV Colors of Normal Galaxies} can be obtained. The idea of star formation in ``flashes'' or ``bursts'' was introduced by \citet{Searle1972InferencesGalaxies} (see also \citealp{Searle1973TheGalaxies}) to explain the very blue colors of some dwarf irregular galaxies, and it has appeared in other contexts including elliptical galaxies with patches of star formation \citep{vandenBergh1975StellarGalaxies} and the peculiar galaxies discussed here.

\medskip

\begin{figure}
	\includegraphics[width=0.47\textwidth]{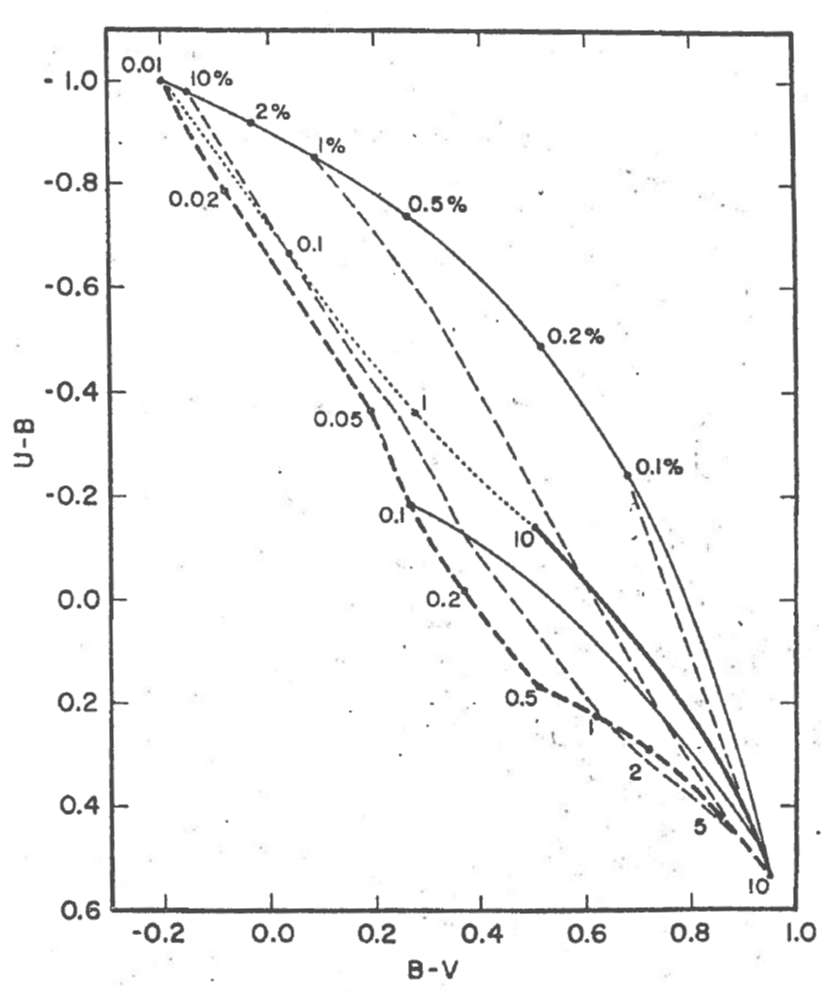}
    \caption{Theoretical two-color diagram showing the colors of young galaxies (\emph{dotted and heavy dashed lines}) and an old red galaxy with bursts of star formation of various strengths and ages (\emph{light solid and dashed lines}). Details of these lines and their labels are explained in Section~\ref{subsubsec:Bursts of star formation and blue colors}. The \emph{heavy solid line} is the locus of standard old models, from Figure~\ref{fig:fig16}.}
    \label{fig:fig18}
\end{figure}

The effects of a burst of star formation on a formerly red galaxy are illustrated in Figure~\ref{fig:fig18}, where the heavy curve is the locus of standard models aged $10 \: \rm Gyr$, from Figure~\ref{fig:fig16}. The colors of younger galaxies are shown in two extreme cases: the dotted line is a model with a constant SFR, evolving through the ages shown (in Gyr), and the heavy dashed line (on the left) is a model whose star formation stopped at $10^{7} \: \rm yr$. The latter curve can be regarded as the evolution of a cluster of stars formed in a period of $10^{7} \: \rm yr$, or equivalently as the colors resulting from stars formed in a burst lasting $10^{7} \: \rm yr$. The light solid lines are the loci of models made of two components:

\begin{enumerate}
  \item a red galaxy aged $10 \: \rm Gyr$, with no ongoing star formation; and
  
  \vspace{1mm}
  
  \item stars formed in a burst of duration $10^{7} \: \rm yr$, seen at ages $10^{7} \: \rm yr$ (upper line) and $10^{8} \: \rm yr$ (lower light solid line).
\end{enumerate}

The numbers along the upper curve are a burst strength parameter, defined as the mass ratio of stars formed in the burst to stars in the old red galaxy. Finally, the light dashed lines represent the evolution of a composite system, from age $10^{7} \: \rm yr$ when the star formation in the burst stops; these are lines of constant burst strength, and they cross the lower light solid line when the age is $10^{8} \: \rm yr$. (The heavy dashed line is the limiting case of a burst of infinite strength, i.e., without any underlying old stars). It can be seen that a burst of star formation in a red galaxy gives colors initially above the normal UBV locus, since the young stars cause an ``ultraviolet excess''; as the burst ages, the colors evolve across the normal locus after nearly $10^{8} \: \rm yr$; then they fall below the normal line and eventually become imperceptibly different from the colors of an undisturbed old galaxy.

Evidently the colors of peculiar galaxies in Figure~\ref{fig:fig17} can be explained by bursts of star formation of various strength and ages in galaxies with various initial colors. Using a series of models like those in Figure~\ref{fig:fig18}, \citet{Larson1978StarGalaxies} found that most of the \citet{Arp1966AtlasGalaxies} galaxies can be accounted for with bursts of strength less than $5\%$ and duration $< 2 \times 10^{7} \: \rm yr$; a few more deviant colors are probably due to observational scatter, internal reddening, non-thermal continuum emission (in two Type~I Seyfert galaxies in the sample), and strong gaseous emission lines, whose effects on colors are discussed by \citet{Huchra1977StarGalaxies}. The strongest bursts of star formation are inferred for galaxies in distorted close pairs, often with bridges or tails, and in apparently single systems with tails and filamentary streamers. Dynamical models for colliding galaxies predict features like these in cases of strong tidal deformation and recent mergers \citep{Toomre1977MergersConsequences}, so it appears that violent dynamical interactions lead to star formation. For this reason, it has been suggested that the stars in elliptical galaxies could have formed in bursts due to collisions and mergers among protogalactic subsystems (Section~\ref{subsubsec:Bursts of star formation in merging subsystems}).

Figure~\ref{fig:fig18} shows that the colors of models with bursts of strength $10\%$ and infinity are very similar. The differences are less than the observational uncertainties for many faint peculiar galaxies, and the theoretical uncertainties in stellar evolution, metallicity effects, etc. In other words, it is not possible to tell from UBV colors alone whether a galaxy is really young or has $90\%$ of its mass in old stars! Some of the galaxies near the upper left in Figure~\ref{fig:fig17} are very chaotic in appearance, and are tempting candidates for truly young galaxies, but in all cases the colors are inconclusive and the appearance could be due to a violent collision or irregularly distributed star formation in an old galaxy.

Colors at longer wavelengths, such as $V - K$, are much more sensitive than $B - V$ to the presence of some old stars that would distinguish between a truly young system and one with a burst strength ${\sim} 10\%$ \citep{Struck-Marcell1978StarRadiation}. Galaxies with very active star formation are often dusty, so red colors could be due to reddening rather than to age; on the other hand, very blue values of $V - K$ would indicate a lack of red giants, and much stronger limits could be put on the mass of any underlying old component. As yet, there are no known nearby galaxies in which the dominant young stellar component could not be masking a significant mass of old stars.

\subsubsection{Highly reddened galaxies}
\label{subsubsec:Highly reddened galaxies}

Some regions of galaxies that are suspected of having intense star formation are extremely dusty, and they show thermal infrared (IR) emission that is interpreted as re-radiation of starlight by the dust. Example of such regions include the centers of M82 and NGC~253, and the dust band around NGC~5128 \citep[e.g.][]{Kleinmann1977InfraredSources, Telesco1978ExtendedA/}. Star formation is indicated by early-type spectra and blue colors in unobscured patches \citep[e.g.][]{vandenBergh1971The82, vandenBergh1978Multi-ColourA}, emission from interstellar molecules \citep[e.g.][]{Whiteoak1978RadioGalaxies}, and the lack of more plausible explanations for the IR emission \citep[e.g.][]{Kleinmann1977InfraredSources}. In NGC~5128, the dust band has an IR luminosity of a few times $10^{10} \: \rm L_{\odot}$ \citep{Telesco1978ExtendedA/}, which rivals the visual luminosity of the entire elliptical galaxy.

If the IR luminosity is assumed to represent the bolometric luminosity of buried stars, an SFR can be estimated \citep{Struck-Marcell1978StarRadiation}: models like those of Section~\ref{subsec:UBV Colors of Normal Galaxies} show that any system with a mass-to-luminosity ratio $M_{\rm s} / L_{\rm bol} < 0.5$ is so dominated by young stars that $L_{\rm bol}$ is almost directly proportional to the SFR; with the local IMF, the relation is
\begin{equation}
    \psi \simeq (0.1 - 0.4) \frac{L_{\rm bol}}{\rm L_{\odot}} \: \rm M_{\odot} \ Gyr^{-1}.
	\label{eq:eq7.1}
\end{equation}
An upper limit to the time for which star formation could have continued at this rate is approximately $M_{\rm s} / \psi$, so the limiting timescale $\tau_{\rm s}$ depends only on $M_{\rm s} / L_{\rm bol}$, according to the relation
\begin{equation}
    \tau_{\rm s} \equiv \frac{M_{\rm s}}{\psi} \sim (3 - 10) \frac{M_{\rm s} / L_{\rm bol}}{\rm M_{\odot} / L_{\odot}} \: \rm Gyr.
	\label{eq:eq7.2}
\end{equation}
Some galactic nuclei have such strong IR emission that $M_{\rm s} / L_{\rm bol}$ is only a few hundredths, so the timescale is only a few times $10^{8} \: \rm yr$. The dust-band region of NGC~5128 is also making stars at a prodigious rate: given a luminosity ${\sim} 10^{10} \: \rm L_{\odot}$, Equation~(\ref{eq:eq7.1}) leads to an SFR of about $2 \: \rm M_{\odot} \ yr^{-1}$, so a respectable disk of stars could be built in just a few times the dynamical timescale of the system. \citet{vandenBergh1975StellarGalaxies} has suggested that NGC~5128 could evolve into an early-type spiral seen edge-on, like the Sombrero galaxy M104.

Since most of the bolometric light comes from massive stars, the above ratios of SFR to $L_{\rm bol}$ could be overestimates if low-mass stars are not forming. For example, in the models discussed, stars above $10 \: \rm M_{\odot}$ contribute $90\%$ of $L_{\rm bol}$, but stars below $1 \: \rm M_{\odot}$ account for $80\%$ of the mass formed into stars. There is no reason to suspect a lack of low-mass stars, however, especially since low-mass protostars (T Tauri stars) are associated with dark clouds in the Milky Way.

These very dusty galaxies with intense star formation lead again to the question of what truly young galaxies would look like. In the absence of dust they would be very blue, like the young models in Figure~\ref{fig:fig18}, but it seems that the regions of galaxies with the most intense bursts are the very reddened ones just described. This question is important in the context of primeval galaxies at large redshifts, i.e. the early stages of most normal galaxies that now have ages ${\sim} 10 \: \rm Gyr$. Starting with the ideas of \citet{Partridge1967AreVisible}, most models for primeval galaxies have assumed that hot young stars would be visible and lead to detectable luminosities (despite the great distances) during a brilliant early burst of star formation \citep[e.g.][]{Meier1976TheGalaxies, Kaufman1977YoungRedshifts, Sunyaev1978ObservableRedshift.}. A model for a very dusty primeval galaxy has been studied by \citet{Kaufman1976PrimevalLuminosities}, and \citet{Sunyaev1978ObservableRedshift.} considered the possibility of substantial radiation from dust. If the very dusty nearby galaxies are the best analogs of truly primeval systems, as suggested by \citet{Larson1976c}, their prospects for detection at optical wavelengths are dim. Observational searches have so far produced null results.

Another factor making primeval galaxies hard to detect could be that most galaxies have such a long timescale for star formation that there is not a very bright phase at an early peak SFR \citep{Tinsley1978a, Tinsley1979StellarGalaxies}. One argument is that most spiral galaxies have colors that correspond to rather long timescales for star formation (Section~\ref{subsec:UBV Colors of Normal Galaxies}), precluding a significant early burst with a corresponding peak in luminosity. Moreover, even elliptical galaxies could form their stars over rather long time intervals if star formation occurs during a series of mergers among subsystems (Section~\ref{subsubsec:Bursts of star formation in merging subsystems}). According to these ideas, late-type galaxies could be fainter in the past than they are now, and early-type galaxies could experience their brightest evolutionary stages only a few Gyr ago. Galaxies in interestingly early evolutionary stages have indeed already been found at moderate redshifts: distant clusters have excess numbers of blue galaxies (Section~\ref{subsubsec:Evolution of colors}; \citealp{Butcher1978The295}), and counts in the field show a large excess of blue galaxies at faint apparent magnitudes \citep{Kron1978PhotometryGalaxies.}. Further studies of these phenomena will surely shed light on the ways in which stars and galaxies have formed during cosmological time.

\section{Conclusion}
\label{sec:Conclusion}

Returning to the outline of galactic evolution in Figure~\ref{fig:fig1}, one can see how much remains to be learned before the jigsaw puzzle will be complete enough for a clear picture to emerge. Essentially every aspect of the subject needs further observational and theoretical study, so galactic evolution will long be a fertile field for research.

\section*{Acknowledgements}

I am grateful to Pierre Demarque, Richard B. Larson, Curtis Struck-Marcell, and Bruce A. Twarog for their suggestions and help in the preparation of this review. The work was supported in part by the National Science Foundation (Grant AST77-23566) and the Alfred P. Sloan Foundation.

%%%%%%%%%%%%%%%%%%%%%%%%%%%%%%%%%%%%%%%%%%%%%%%%%%



\bibliographystyle{mnras}
\bibliography{main.bib} % if your bibtex file is called example.bib

%%%%%%%%%%%%%%%%%%%%%%%%%%%%%%%%%%%%%%%%%%%%%%%%%%

%%%%%%%%%%%%%%%%% APPENDICES %%%%%%%%%%%%%%%%%%%%%

%%%%%%%%%%%%%%%%%%%%%%%%%%%%%%%%%%%%%%%%%%%%%%%%%%

% Don't change these lines
\vspace{5mm}

This paper has been transcribed from a hard copy of Beatrice M. Tinsley's original manuscript into a digital {\TeX}/{\LaTeX} file prepared by Michael J. Greener, a PhD student at the University of Nottingham. If you notice any errors or problems with this transcribed paper, please email Michael at either michael.greener@nottingham.ac.uk or mickgreener@protonmail.com.	% typesetting comment
\label{lastpage}
\end{document}